\documentclass[useAMS,usenatbib]{mn2e}
\usepackage{times}
\usepackage{subfiles}
\usepackage{hyperref}
\usepackage{xkeyval}

   \usepackage{wrapfig}
  \usepackage{gensymb}
  \usepackage[english]{babel}
  \usepackage{arydshln}
     \usepackage[pdftex]{graphicx}
     \usepackage[format=plain,justification=raggedright,singlelinecheck=false,font=small,labelfont=bf,labelsep=space]{caption}
\usepackage[justification=centering]{subcaption} 
     \usepackage{lipsum}
\usepackage{siunitx}
\usepackage[usenames,dvipsnames,svgnames,table]{xcolor}
\definecolor{americanrose}{rgb}{1.0, 0.01, 0.24}
\def\citejap#1{\citeauthor{#1}\ \citeyear{#1}}
\def\fig#1{Fig. \ref{#1}}

\usepackage{url}
\usepackage{sidecap}
\usepackage{amssymb}
\hypersetup{draft}
\newcommand{\MsMh}{M$_{\rm s}/$M$_{\rm h}$}
\title[Assembly bias in GAMA]{Galaxy and Mass Assembly (GAMA): halo formation times and halo assembly bias on the cosmic web}

\author[R. Tojeiro]{
\parbox{\textwidth}{
Rita Tojeiro$^{1}$\thanks{E-mail:rmftr@st-andrews.ac.uk}, 
Elizabeth Eardley$^{1,2}$, John A. Peacock$^{2}$, Peder Norberg$^{3}$, Mehmet Alpaslan$^{4}$, Simon P. Driver$^{1,5}$, Bruno Henriques$^{6}$, Andrew M. Hopkins$^{7}$, Prajwal R. Kafle$^{5}$, Aaron S. G. Robotham$^{5}$, Peter Thomas$^{8}$, Chiara Tonini$^{9}$ and Vivienne Wild$^{1}$
}\vspace{3mm}\\
{}$^{1}$School of Physics and Astronomy, University of St Andrews, North Haugh, St Andrews, KY16 9SS, UK\\
{}$^2$Institute for Astronomy, University of Edinburgh, Royal Observatory, Blackford Hill, Edinburgh EH9 3HJ, UK\\
{}$^3$ICC \& CEA, Department of Physics, Durham University, South Road, Durham DH1 3LE, UK\\
{}$^4$NASA Ames Research Center, N232, Moffett Field, Mountain View, CA 94035, United States\\
{}$^5$ICRAR, M468, University of Western Australia, Crawley, WA 6009, Australia\\
{}$^6$Institute for Astronomy, ETH Zurich, CH-8093 Zurich, Switzerland \\
{}$^7$Australian Astronomical Observatory, PO Box 915, North Ryde, NSW 1670, Australia\\
{}$^8$Astronomy Centre, University of Sussex, Falmer, Brighton BN1 9QH,UK\\
{}$^9$School of Physics, University of Melbourne, Parkville, 3010 VIC, Australia\\
}
\begin{document}


\maketitle
%
\begin{abstract}
We present evidence for halo assembly bias as a function of geometric environment. By classifying GAMA galaxy groups as residing in voids, sheets, filaments or knots using a tidal tensor method, we find that low-mass haloes that reside in knots are older than haloes of the same mass that reside in voids. This result provides direct support to theories that link strong halo tidal interactions with halo assembly times. The trend with geometric environment is reversed at large halo mass, with haloes in knots being younger than haloes of the same mass in voids. We find a clear signal of halo downsizing - more massive haloes host galaxies that assembled their stars earlier. This overall trend holds independently of geometric environment. We support our analysis with an in-depth exploration of the L-Galaxies semi-analytic model, used here to correlate several galaxy properties with three different definitions of halo formation time. We find a complex relationship between halo formation time and galaxy properties, with significant scatter. We confirm that stellar mass to halo mass ratio, specific star-formation rate and mass-weighed age are reasonable proxies of halo formation time, especially at low halo masses. Instantaneous star-formation rate is a poor indicator at all halo masses. Using the same semi-analytic model, we create mock spectral observations using complex star-formation and chemical enrichment histories, that approximately mimic GAMA's typical signal-to-noise and wavelength range. We use these mocks to assert how well potential proxies of halo formation time may be recovered from GAMA-like spectroscopic data.

\end{abstract}

 \newpage
 \section{Introduction}

It is well established that the clustering of {\it galaxies} depends on several of their properties; e.g. star-formation rate (SFR), luminosity, stellar-mass, etc. (e.g., \citealt{Norberg2001, Norberg2002, Swanson2008, Cresswell2009, Ross2011, Zehavi2011,Christodoulou2012,Guo2013}). Within the standard $\Lambda$ Cold Dark Matter ($\Lambda$CDM) paradigm, this behaviour can be simply explained by the fact that haloes of different mass cluster differently. Galaxies with different properties are then assumed to live in haloes of different mass, mimicking the observed dependence of, e.g., clustering with luminosity. The above makes two explicit assumptions: the mass of a halo fully determines its bias, and the mass of a halo drives the properties of the galaxies within it. Both assumptions sit on strong theoretical ground: excursion set formalism predicts a mean bias a function of halo mass (see e.g. \citealt{Mo1996,Zentner2007}), and galaxy formation models have been successfully reproducing many observations relying solely on halo mass since \cite{White1978}.

However, early work on dark matter N-body simulations quickly cast doubt on the first assumption. \cite{ShethTormen2004} showed that dark matter haloes in dense environments assemble earlier, motivating a series of dedicated investigations on the question of whether, at fixed halo mass, haloes that assemble at different times cluster differently. Early work converged impressively fast: at the high end of the halo mass function, the oldest haloes are less biased; at low mass end the trend is reversed and the oldest haloes are more biased.
(e.g.\citealt{Gao2005,Wechsler2006,Dalal2008}). Common to all works is the fact that the effect is seen to be strongest at low halo mass. {\it Halo assembly bias}, therefore, refers to the fact that the bias of dark-matter haloes depends on something other than their mass. Although most commonly discussed in terms of halo formation time or halo concentration, any residual dependence of halo bias beyond halo mass is considered a form of halo assembly bias. \cite{Dalal2008} argued for two different mechanisms acting at high and low mass end; they show that some form of halo assembly bias is unavoidable once the statistics of the peaks of Gaussian random fluctuations are considered (see also \citealt{Kaiser84, Zentner2007}). At low mass, they propose the cause lies with a sub-population of low-mass haloes in high-density regions that stops accreting, increasing the bias of the population of haloes at that mass (see also \citealt{LacernaPadilla12001}). \cite{Hahn2009} (and, more recently, \citealt{Borzyszkowski16}) then linked such interactions to strong tidal fields and a dependence on geometric environment. \cite{Faltenbacher2010} looked at the dynamical structure of haloes within N-body simulations. They found that at fixed halo mass, all highly biased haloes have more isotropic velocity distributions than the least biased haloes, and suggest that it is the manner in which a halo grows that primarily drives assembly bias. More recently, \cite{ChavesMontero2016} report for the first time a detection of assembly bias in a hydrodynamical simulation. Working on the EAGLE simulation \citep{Schaye2015,Crain2015}, they shuffle galaxies in bins of constant halo mass, and find that the clustering amplitude cannot be recovered when information on halo mass alone is kept. Keeping information on the maximum circular velocity of a halo  recovers most of the behaviour seen in the clustering of both EAGLE galaxies and matched catalogues constructed using subhalo abundance matching (SHAM).

The implications of halo assembly bias are numerous. Small- and large-scale clustering have proven to be powerful tools in studies of galaxy evolution. On large-scales, a measurement of the evolution of the large-scale bias can put constraints on the evolution and merger history of massive galaxies (e.g. \citealt{Tojeiro2012, Bernardi2016}). On smaller scales, empirical techniques such as the halo occupation distribution (HOD) model (see e.g. \citealt{Jing1998,Seljak2000,Peacock2000}) or several flavours of abundance matching (see e.g. \citealt{Kravtsov2004,Conroy2006,Cooray2006}) allow one to easily model halo abundance and galaxy populations in haloes in terms of centrals and satellites (e.g. \citealt{Zehavi2005,Zheng2007,Ross2009,Skibba2009,Wake2011}). This type of analysis is powerful and versatile, yielding for example measurements of quenched fractions as a function of redshift and luminosity \citep{Tinker2010} or merger rates as a function of redshift (e.g. \citealt{White2007,Zheng2007,Brown2008,Wake2008}). HOD models assume a mean bias as a function of halo mass and, crutially, assume that the galaxy content of a halo is exclusively determined by its mass. One of the first questions that must be answered, therefore, concerns the applicability of such methods to current datasets. \cite{Zentner2014}, using galaxy mock catalogues with built-in assembly bias, showed that a standard HOD analysis applied to such mocks resulted in significant systematic errors on the inferred parameters. HOD or SHAM modeling are also often used in the fast construction of large sets of mock catalogues in support of large-scale cosmological clustering measurements (e.g. \citealt{delaTorre2013b,Manera2013,Manera2015,White2014}). As far as we are aware, the effect of assembly bias on the derived covariances from such mocks remains unknown. Looking at how assembly bias manifests in velocity space, \cite{Hearin2015a} point out the potential impact on cluster redshift-space distortion measurements, via a dependence of the pairwise-velocity dispersion on halo concentration. Finally, assembly bias has fundamental implications for our understanding of galaxy formation, and is probably directly linked to the observed phenomenon of galactic conformity - the observation that the properties of galaxies within the same halo - and even nearby haloes - are correlated \citep{Hearin2015b}.

The distribution of dark matter in numerical simulations shows a striking and familiar structure of voids, filaments and clusters that characterise the cosmic web; we will refer to this cosmic web classification as geometric environment (GE). The cosmic web comes to be via the influence of tidal forces on small and nearly-uniform density fluctuations in the early Universe, and we note that it is largely distinct from other, traditionally more local, measures of environment such as some estimation of overdensity, which we capture here via halo mass. A halo accretion rate is connected to its geometric environment \citep{Hahn2009}, and strong tidal fields can leave imprints of galactic conformity in the evolution of galaxies \citep{Hearin2015b}. This leaves room for GE to act as a driver of assembly bias and/or galactic conformity. In this paper we answer the question: {\it at fixed halo mass, do the properties of galaxies in different geometric environment reveal a different assembly history of their haloes?} 

Over the last decade, a substantial amount of effort has been spent looking for signs of assembly bias in data. Whilst the work on simulations points to a clear, converged picture, work on data is far from it. Part of the difficulty lies in isolating samples of galaxies at fixed halo mass, or finding robust observational proxies for halo formation time. For example, \cite{Blanton2007} studied the relative importance of group environment and large-scale density on Sloan Digital Sky Survey (\citealt{York2000}, SDSS) data. They found that the position of a halo within the large-scale density field is unimportant, and found no evidence for assembly bias. Similarly, \cite{Tinker2008b} found that a standard HOD model fits galaxies in voids equally well as those in higher density regions, and found that the content of haloes of fixed mass does not vary with large-scale environment. In another non-detection, \cite{Tinker2011} found no residual correlation of the quenched fraction with large-scale density, at fixed halo mass. \cite{Lin2015} looked at the bias of SDSS galaxies as a function of halo mass and specific star-formation time or stellar assembly history. They show a detection of assembly bias when using the group catalogue of \cite{Yang2007} as a proxy for halo mass, but attribute the signal to a difference in halo mass of the two samples, which they confirm using weak lensing profiles. 

Equally, there have been several claims of a detection. \cite{Yang2006} provided what was possibly the first evidence towards assembly bias on data, by cross-correlating SDSS galaxies with groups of different mass. They found that at fixed group mass, bias decreased with the SFR of the central galaxy. A similar conclusion was reached by \cite{Wang2008}, who claim that at fixed group mass, groups with red centrals are more clustered than groups with blue centrals. \cite{Lim2015}, again working on group catalogues within SDSS, was motivate by the work of \cite{Wang2011} to use \MsMh, the ratio of stellar mass of the central galaxy to halo mass, as a proxy for halo formation time. At fixed halo or stellar mass, they report a varying \MsMh with the colour of the central, particularly at low masses: at fixed mass, central galaxies with larger \MsMh are redder and more quenched. \cite{Wang2013} instead focused on specific star-formation rate to detect assembly bias, showing that more passive central galaxies, at fixed stellar mass, are more strongly clustered. More recently, \cite{Zentner2016} have conducted for the first time an HOD analysis with sufficient freedom to allow extra dependencies beyond halo mass. Working on SDSS DR7, they claim the data significantly prefer a model with assembly bias, but the additional parameters themselves are currently poorly constrained.

Several GE estimators have been applied to data from galaxy redshift surveys, with the goal of establishing what the role of GE is in the formation and evolution of dark matter haloes and the galaxies that populate them. In general, once the dependence on local density (large overdensities are more common in knots than in voids) or stellar mass (massive galaxies are more common in denser regions) are accounted for, most authors find no significant effect of the GE on the galaxy or halo properties. For example, working on data from the Galaxy and Mass Assembly survey (GAMA, \citealt{Driver2011}), \cite{Eardley2015} show that the luminosity function of galaxies has no residual dependence on GE. \cite{DarvishEtAl14}, working on the HiZeLS survey at $z\approx 0.84$ found that the mean properties of galaxies are indistinguishable across different GEs, but found an increase in the number density of H$\alpha$ emitters in what they identify as filaments. When controlling for stellar mass, \cite{AlpaslanEtAl16} found that GE left no residual effect on several properties of GAMA galaxies. Focusing on the evolution of dark matter haloes, \cite{BrouwerEtAl16} worked on a 100 sq.deg. overlap between GAMA and the Kilo-Degree Survey (KiDS, \citealt{deJong2013}) to compute stacked weak-lensing profiles. These were weighted by stellar mass to a common stellar mass distribution across all environments and separately controlled for local density; they found that the mean halo mass did not change as a function of geometric environment. Working on dark matter simulations, \cite{AlonsoEtAl14} looked for a dependence of the halo mass function (the number of haloes per halo mass) with GE. Once again, all variations could be attributed to the underlying density field, leaving no room for GE influencing the halo mass function. None of the above papers necessarily rule out assembly bias.

In this paper we look for observational evidence of halo assembly bias using the GAMA survey, focusing on halo formation time and, for the first time, a full geometric classification of the cosmic web. We will look for a dependence of galaxy properties - {\it at fixed halo mass} - that vary with GE and might be indicative of the formation time of the haloes within each GE. While there are different ways in which to numerically quantify GE, here we will use the classifications of \cite{Eardley2015}, who apply the tidal tensor method to the GAMA survey. Computed using the second derivatives of the gravitational potential, the tidal tensor can capture the cosmic web environment by computing the dimensionality of collapse within a region: 3 dimensions for knots, 2 for sheets, 1 for filaments and zero for voids. 

Assembly bias is a wide net, which captures any deviation from halo mass as the sole driver of halo clustering and content. In this paper we focus on formation time as a potential source of assembly bias, and begin with an in-depth analysis of the the semi-analytic model of \cite{Henriques2015}, L-Galaxies, which we use to determine the relationship between galaxy observables and the age of a halo. We explicitly correlate different definitions of halo formation time with observable properties of galaxies, such as star-formation history, mass-weighted age, stellar-to-halo mass ratio and present-day star-formation rate. This analysis equips us to interpret the results we find in GAMA data. Acknowledging that many galaxy observables are difficult to measure robustly in real data, we create GAMA-like spectroscopic mocks from the L-Galaxies model. Our mock observations therefore are based on complex star-formation histories, and go a convincing step beyond the traditionally simplistic mocks used to test spectral and photometric analysis codes. We use these mocks in order to determine how well certain proxies for halo formation time can be recovered from real data, using the full spectral fitting code VESPA (VErsatile SPectral Analyses, \citealt{Tojeiro2007,Tojeiro2009}).

This paper is structured as follows: we present the analysis of the L-Galaxies semi-analytic model in Section 2, and the creation and spectral analysis of mock galaxies in Section 3. Then, using a VESPA analysis on GAMA and SDSS spectroscopy, the cosmic-web classification of \cite{Eardley2015}, the stellar masses of \cite{Taylor2011}, the group catalogue of \cite{Robotham2011} and the \cite{HanEtAl15} weak-lensing calibrations, we study the potential dependence of halo formation time on geometric environment in Section 4. We discuss our findings and conclude in Section 5.

\section{Simulations}\label{sec:L-Galaxies}
 
 \begin{figure*}
\includegraphics[scale=0.46]
                        {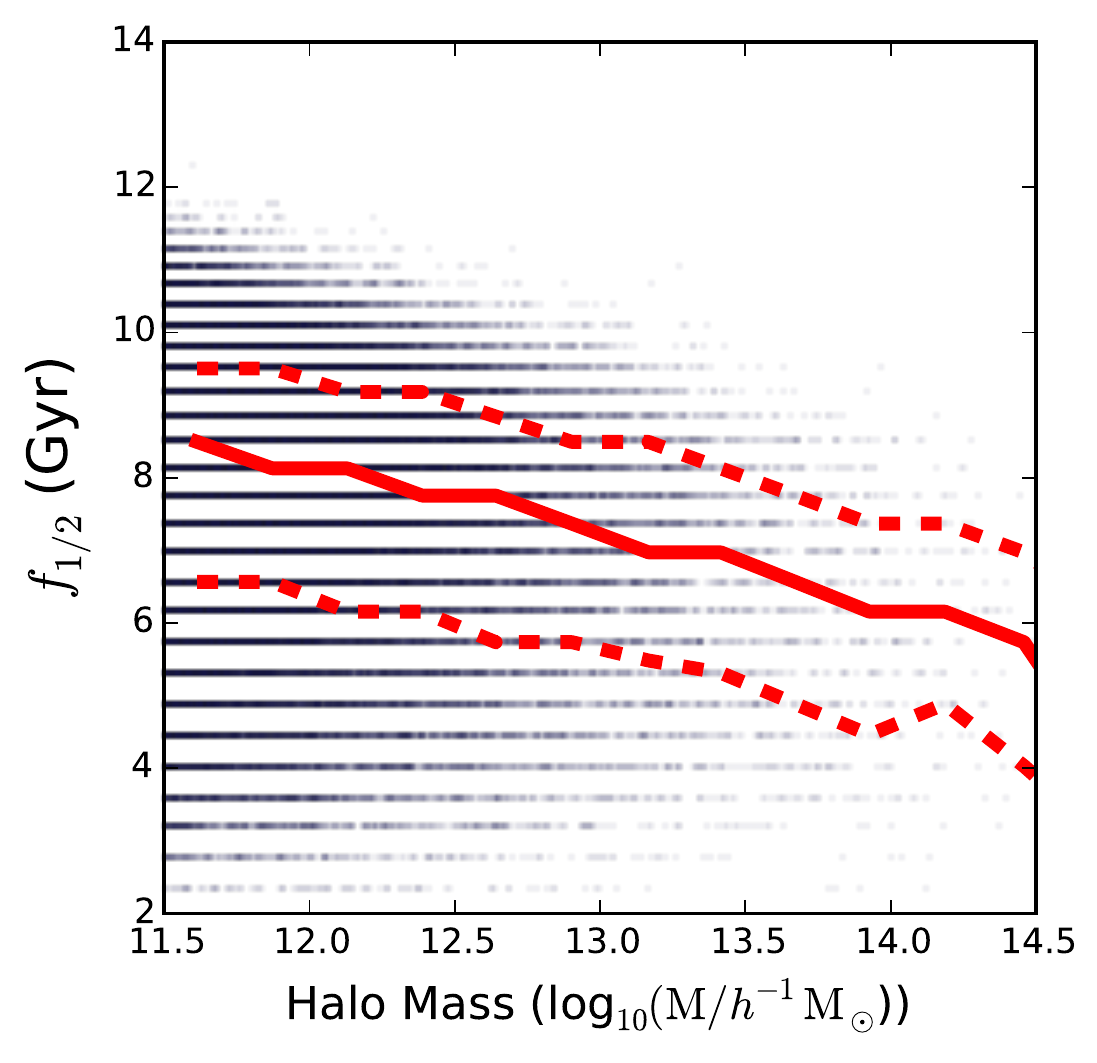}
\includegraphics[scale=0.46]
                        {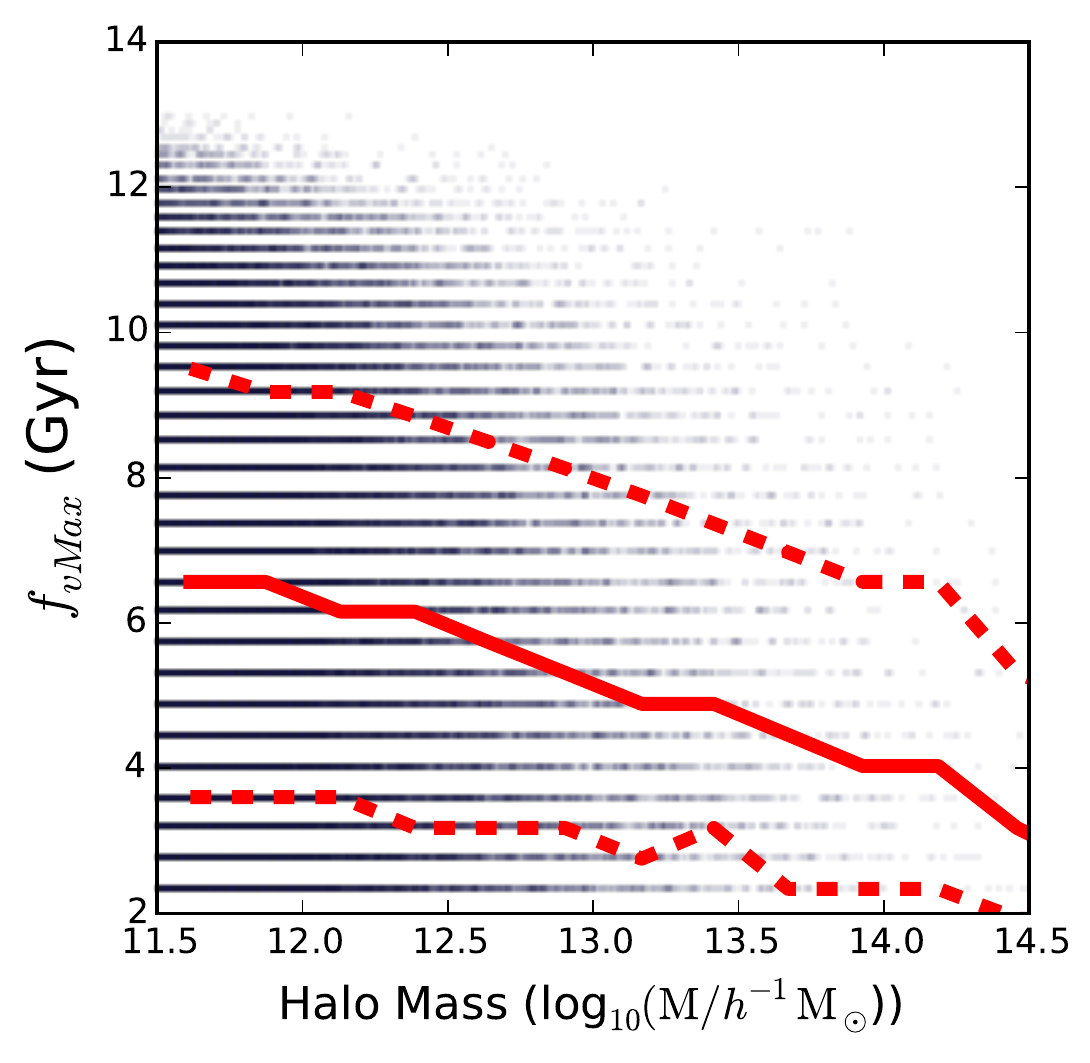}
\includegraphics[scale=0.46]
                        {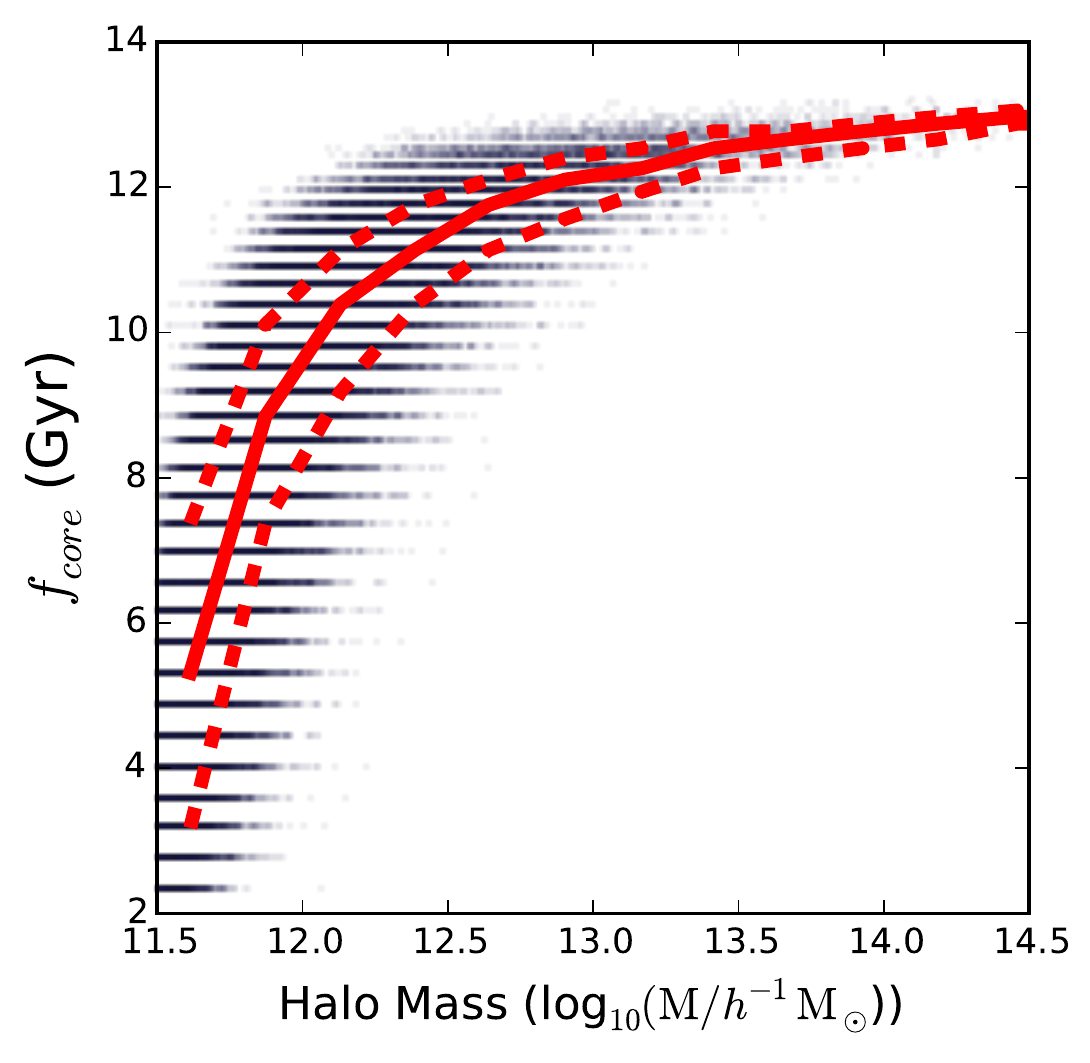}
\caption{Halo formation time vs. halo mass for the 3 definitions of formation time as described in the text (left to right: $f_{1/2}$, $f_{\rm vMax}$, $f_{\rm core}$). Red lines show the median (solid) and 20th and 80th quantiles (dashed) of each formation time (lookback time to the epoch of formation) in bins of host halo mass. It is clear that there are significant differences between the definitions of halo formation times.}
\label{fig:FM}
\end{figure*}
 
We use the Millennium Simulation (\citejap{Springel2005}, \citejap{Lemson2006}), which consists of $2160^3$ particles of mass $8.6\times 10^{8}\,h^{-1}{\rm M_{\odot}}$ in a $500\, h^{-1}$Mpc box from $z=127$ to $z=0$. The original Millennium Simulation uses a WMAP1 $\Lambda$CDM cosmological model with ($\Omega_{\rm m},\Omega_{\rm b},\Omega_{\rm \Lambda}, h, n_{\rm s}, \sigma_{\rm 8})=(0.25, 0.045, 0.75, 0.73, 1, 0.9)$.
 Full details of how the halo merger trees were constructed can be found in \cite{Springel2005} and \cite{deLucia2006}. The simulation data is stored over 64 snapshots and only substructures containing at least 20 self-bound particles are considered. The parent catalogue of dark matter haloes is identified with a standard friends-of-friends (FoF) algorithm with a linking length of 0.2 units of the mean particle separation. Throughout this work we use only haloes of mass $M_{\rm h} > 10^{11.5} h^{-1}{\rm M_{\odot}}$ as the calculation of formation times for lower mass haloes may be affected the resolution of the simulation.
 
 We use the galaxy catalogues generated by the Munich semi-analytic galaxy formation model (\citejap{Henriques2015}). The evolution of dark matter structure within the Millennium simulation is scaled to the Planck cosmology, with parameters ($\Omega_{\rm m},\Omega_{\rm b},\Omega_{\rm \Lambda}, h, n_{\rm s}, \sigma_{\rm 8})=(0.315, 0.0487, 0.685, 0.673, 0.96, 0.829)$. 
 Central galaxies are considered to be those at the position of the most bound particle of the FoF halo.
 We use data from a snapshot corresponding to $z\approx0.15$, chosen to reflect a redshift characteristic of current and future observational surveys.
 
 \begin{figure}
\includegraphics[scale=0.5]{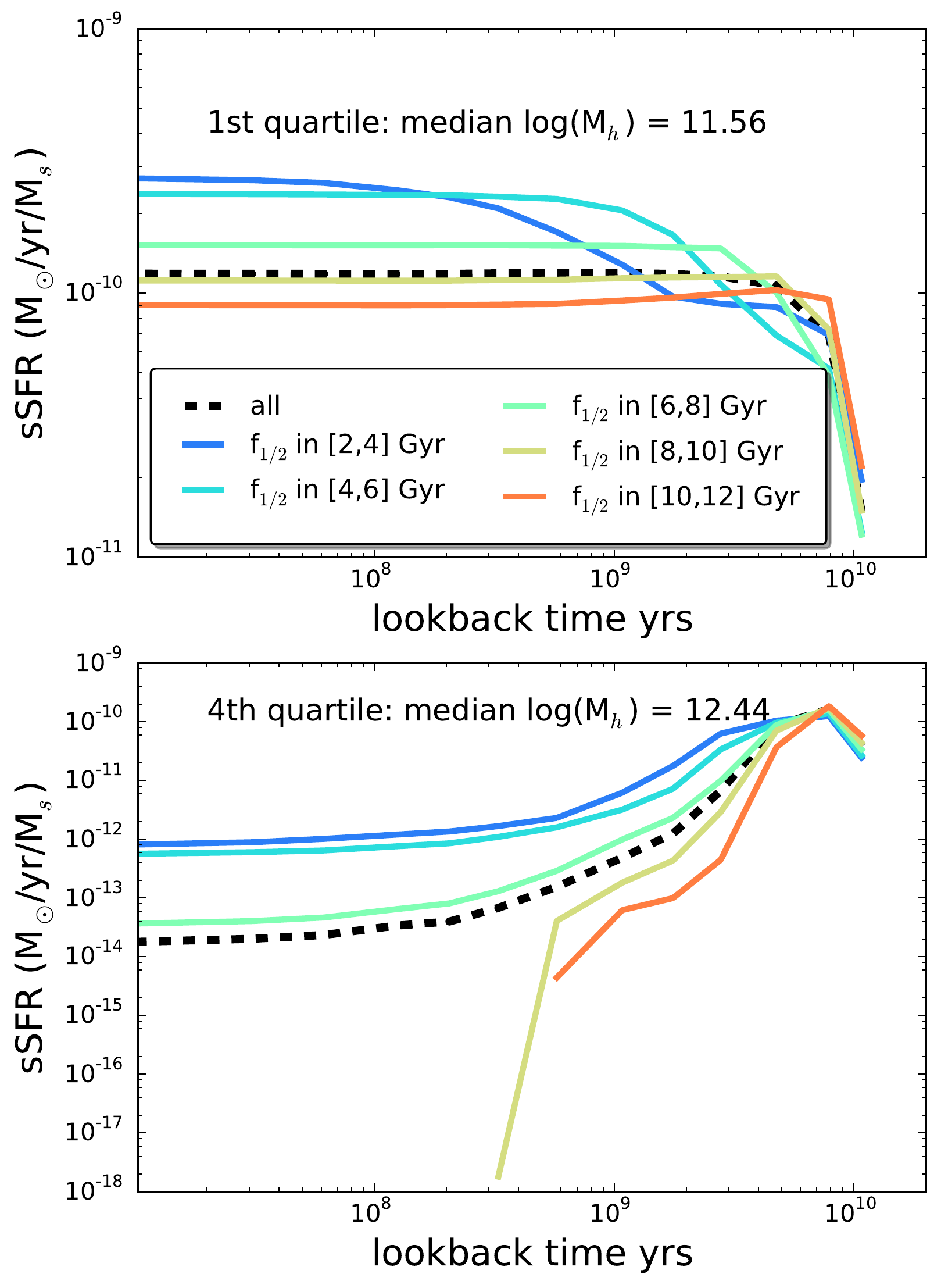}
\caption{The median specific star-formation rate as a function of lookbacktime of the one quarter most and least massive haloes in the L-Galaxies model, as a function of halo age. We plot stellar mass formed per unit time, as a function of lookback time, in units of solar mass formed per year per present-day stellar mass. The different coloured lines show the age of the haloes, as defined by $f_{1/2}$, and the black dashed line shows the median over all halo ages. Halo downsizing is clearly seen: galaxies residing in the most massive haloes formed most of their mass earlier. A clear dependence on halo age is also seen: older haloes host older galaxies. }
\label{fig:SFHs}
\end{figure}

 \subsection{Halo formation times}
 The formation of dark matter haloes is a complex process. There is no clear definition of when a halo can be said to have `formed' and it is difficult to fully characterise a haloes assembly history with one number. The formation time most commonly used in the literature, here $f_{1/2}$, is the time at which the haloes main branch assembled half of its present mass, ${\rm M}_{\rm h}$. However, this widely used definition captures only one aspect of a halo's assembly history and alternate definitions are, arguably, of similar value. \cite{Li2008} present an analysis of 8 different definitions of halo formation times within the Millennium Simulation, finding the definitions to have significant differences. In this work we chose to focus on 3 possible definitions of halo formation time and use the following parameters in order to characterise a halo's assembly history:
  \begin{itemize}
\item $f_{1/2}$: The time at which the halo's main branch assembled half of its present day mass.
  \item $f_{\rm vMax}$: The time at which the halo's virial velocity reaches a maximum.
  \item $f_{\rm core}$: The earliest time at which the halo's most massive progenitor reaches the  fixed mass, ${M_{\rm c}}=10^{11.5}h^{-1}{\rm M_{\odot}}$ (haloes with masses $\sim{M_{\rm c}}$ have the minimum mass-to-light ratio (\citejap{Bosch2003}), and thus are the most efficient in star formation.)
\end{itemize}

Whilst $f_{1/2}$ may be considered to capture the hierarchical nature of halo formation, each of the aforementioned definitions has their own justification; $f_{\rm vMax}$ indicates the time when the halo mass accretion transits from a fast accretion phase to a slow accretion phase and $f_{\rm core}$ indicates when a halo is able to host a relatively bright central galaxy. The dependence of each of these formation times with halo mass within the Millennium Simulation are shown in \fig{fig:FM}. The formation times are shown as lookback time from today, and the (solid) dashed red lines show (median) $20^{\rm th}$ and $80^{\rm th}$ quantiles. The horizontal lines visible in the scattered points of \fig{fig:FM} are due to the discrete formation times possible due to the snapshots used in the Millennium Simulation. The variety in the formation time-halo mass relationship is striking -- whilst $f_{1/2}$ and $f_{\rm vMax}$ are hierarchical in the sense that more massive haloes are seen to form later, formation times defined by $f_{\rm core}$ occur significantly earlier for more massive haloes than for low mass ones. The disparity between these definitions, just 3 of many possible choices of formation times, illustrates the difficulty in fully characterising a halo's assembly history and the importance of carefully considering the meaning of `halo formation time'. In this work we will focus on $f_{1/2}$, but retain all 3 definitions throughout much of the analyses in order to examine the dependence of our results on the choice of definition used. 
    
\subsection{The star formation history}

The first aim of this work is to investigate, quantify and contrast the performance of a number of potential observational proxies of halo formation time. The full star-formation history (taken here to be a non-parametric description of the mass of stars formed as a function of lookback time, as well as their chemical enrichment) encapsulates almost any observable quantity from a galaxy: e.g. ages, current SFRs or stellar mass are simply ways to compress the information in the full star-formation and chemical-enrichment history. Figure~\ref{fig:SFHs} shows the star formation history of central galaxies, in the first quarter of most and least massive haloes, shown separately for haloes of different age. We plot specific star-formation rate, i.e., stellar mass formed per unit time divided by present-time stellar mass. 

The dashed black line shows the median star-formation history for the full sample shown in each panel. Focusing first on this median relationship, downsizing is clearly evident in the L-Galaxies model: galaxies in more massive haloes form their stars earlier. Albeit not shown here, this result is independent of whether they are centrals or satellites. At similar halo mass, however, a clear distinction is seen across haloes of different ages. Older haloes form a greater proportion of their stars early on - i.e., at fixed halo mass older haloes host older galaxies. Although the previous statements hold true at all masses, the way the shape of the SFHs changes as a function of halo age is distinctively different in low and high mass haloes.

\subsection{Simplifying the full star-formation history}

\begin{figure}
\includegraphics[scale=0.45]{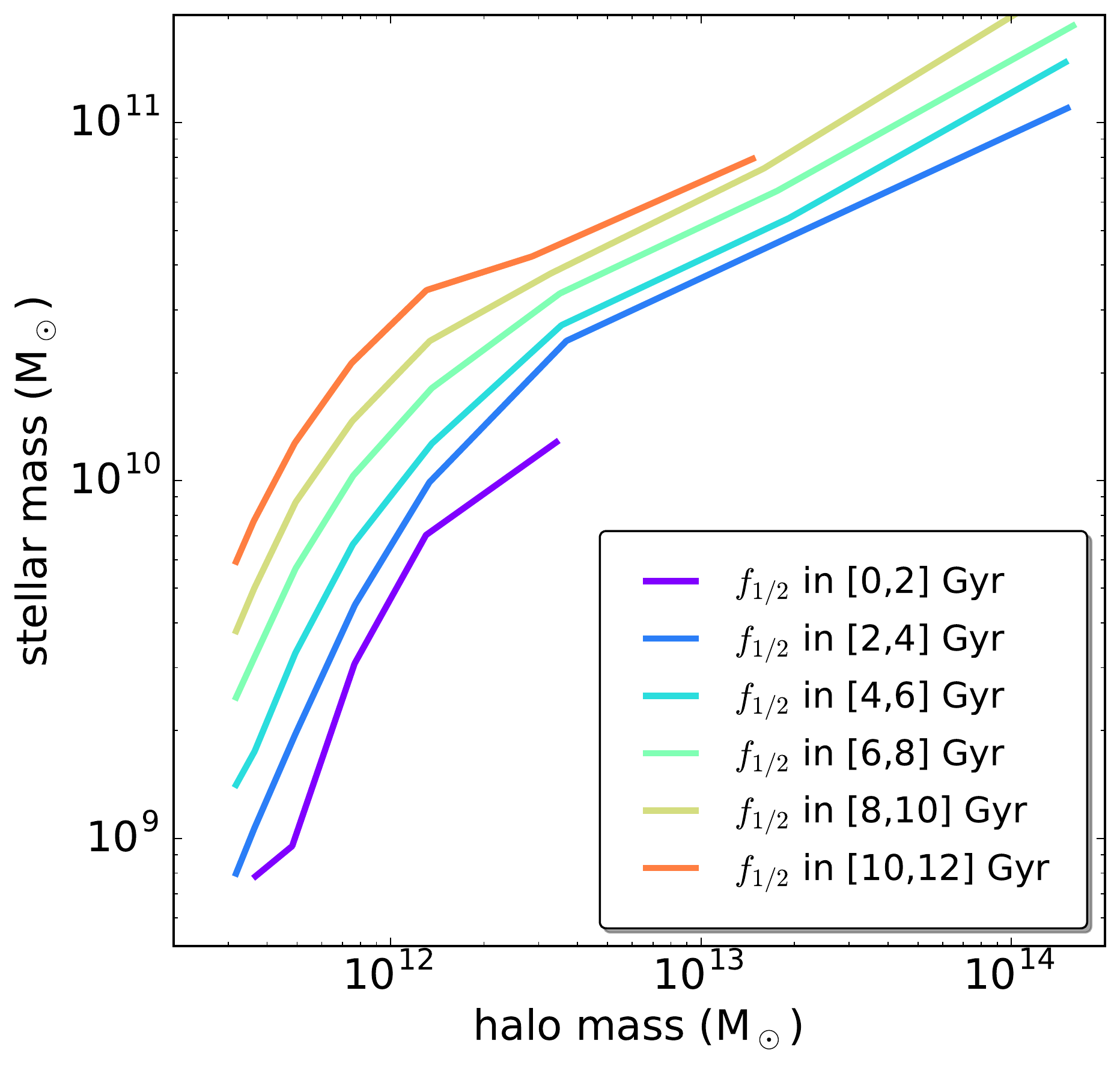}
\caption{The median halo mass to stellar mass relationship for central galaxies residing in haloes of different ages. 
}
\label{fig:MsMh}
\end{figure}

A number of different approaches to observationally estimating halo formation times have been used in the literature. The proxies used are always some simplification of the star-formation history of a galaxy, e.g., SFR \citep{Yang2006}, galaxy colour \citep{Wang2008}, specific SFR \citep{Wang2013}, D4000 \citep{Tinker2011}, luminosity-weighted age \citep{Lacerna2014} and others.
More recently, \cite{Lim2015} use $M_{\rm s}/M_{\rm h}$, the ratio of the central galaxy's stellar mass to the host halo mass, as a proxy for halo formation time. This was motivated by the work of \cite{Wang2011} who found that the sub-structure fraction, $f_{\rm s}=1-(M_{\rm main}/M_{\rm h}$), where $M_{\rm main}$ is the mass of the main subhalo for which $M_{\rm s}$ is a good observational proxy, is tightly correlated with halo formation time. Fig.~\ref{fig:MsMh} shows the stellar-mass to halo-mass relation for central galaxies in haloes of different ages. A form of assembly bias is clearly seen here: at fixed halo mass, central galaxies residing in older haloes are more massive. As the stellar mass is the integral of the SFR over cosmic time, Fig.~\ref{fig:MsMh} is simply a compressed view of Fig.~\ref{fig:SFHs}, showing continuity all halo masses. 

Here we will focus on some of these properties, and additionally we consider $t_{x}$, which may be derived from the SFH of the galaxy. We define $t_{x}$ as the epoch at which $x\%$ of the stellar mass currently in a galaxy had formed, and hence $x$ can take values $0<x<100$. We consider a range of values of $x$ in order to identify the optimal value, that for which $t_x$ is most closely correlated with halo formation time. In calculating $t_x$ from the SFHs, we interpolate across the individual age bins assuming a constant rate of star formation across the bin. In summary, this work considers the following observational proxies for halo formation time:
  \begin{itemize}
 \item SFR: the current star formation rate of the halo's central galaxy;
 \item sSFR: the current star formation rate of the halo's central galaxy, divided by the present-day stellar mass of the galaxy;
 \item MWA: the mass-weighted age of the halo's central galaxy, defined as ${\rm MWA} = \sum t_im_i / \sum m_i$, where the sum is over age bins $i$; $t_i$ is the age in Gyrs and $m_i$ is the mass formed within age bin $i$;
 \item \MsMh: the ratio of the central galaxy's stellar mass to the host halo mass; and
 \item $t_x$: the time at which the central galaxy had formed $x\%$ of its current stellar mass, with $x$ allowed to take a range of values $0<x<100$.
  \end{itemize}

\begin{figure*}
       \includegraphics[scale=0.32]
				{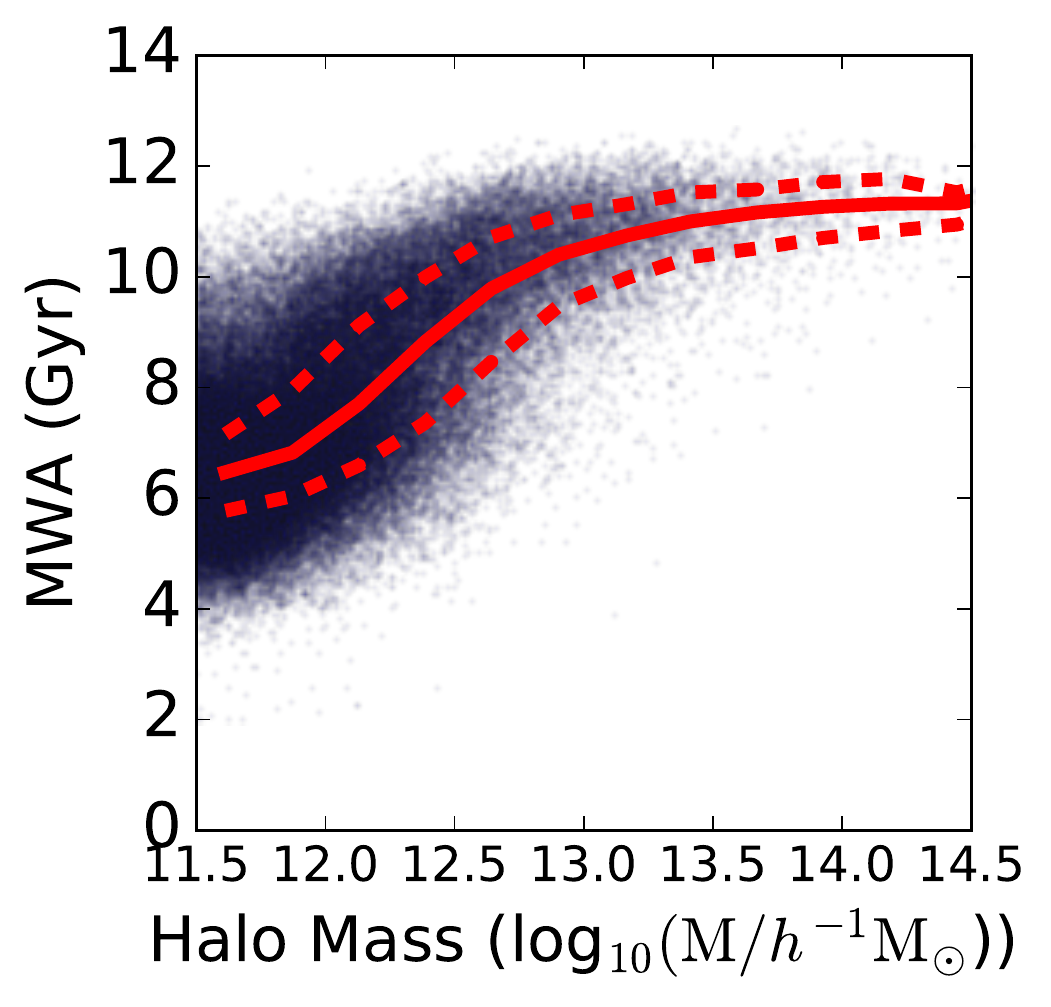}
\hspace{-0.2cm}
       \includegraphics[scale=0.32]
                        {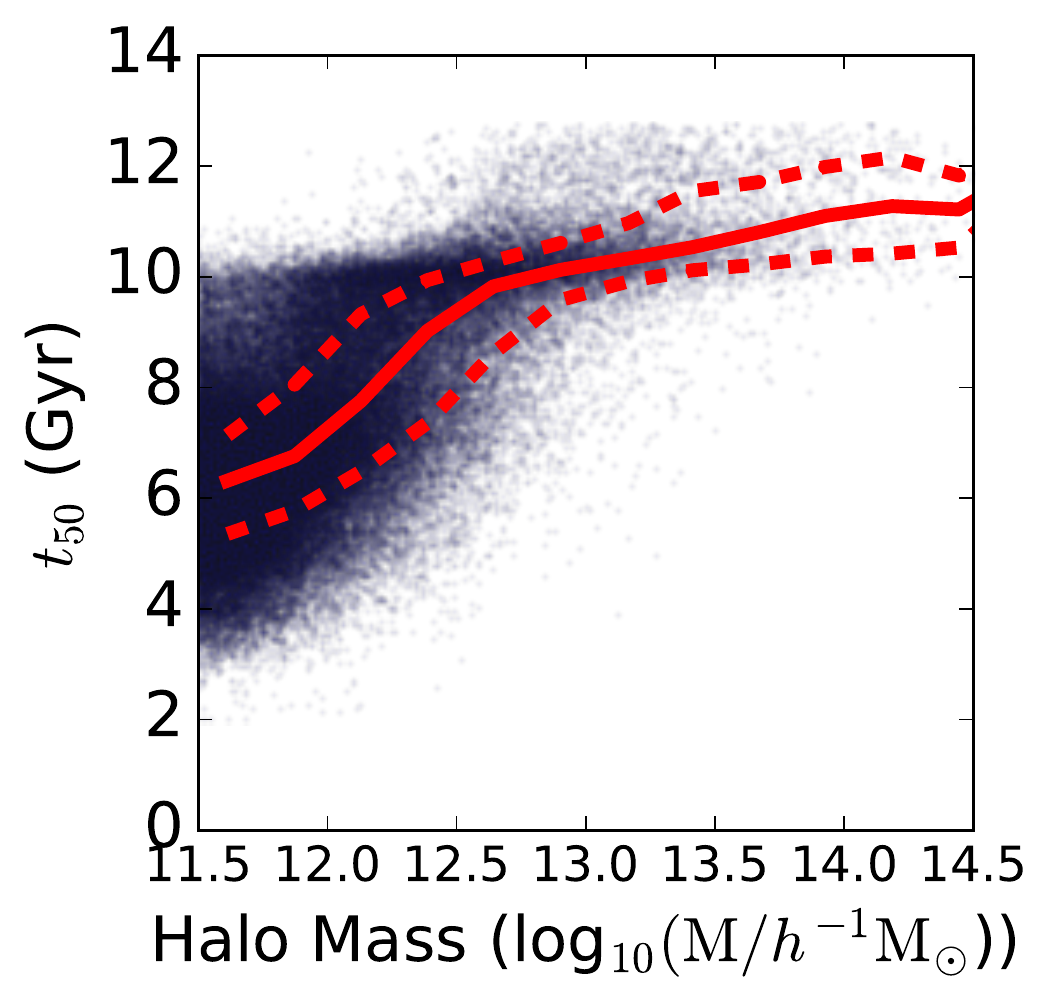}
\hspace{-0.2cm}
       \includegraphics[scale=0.32]
                        {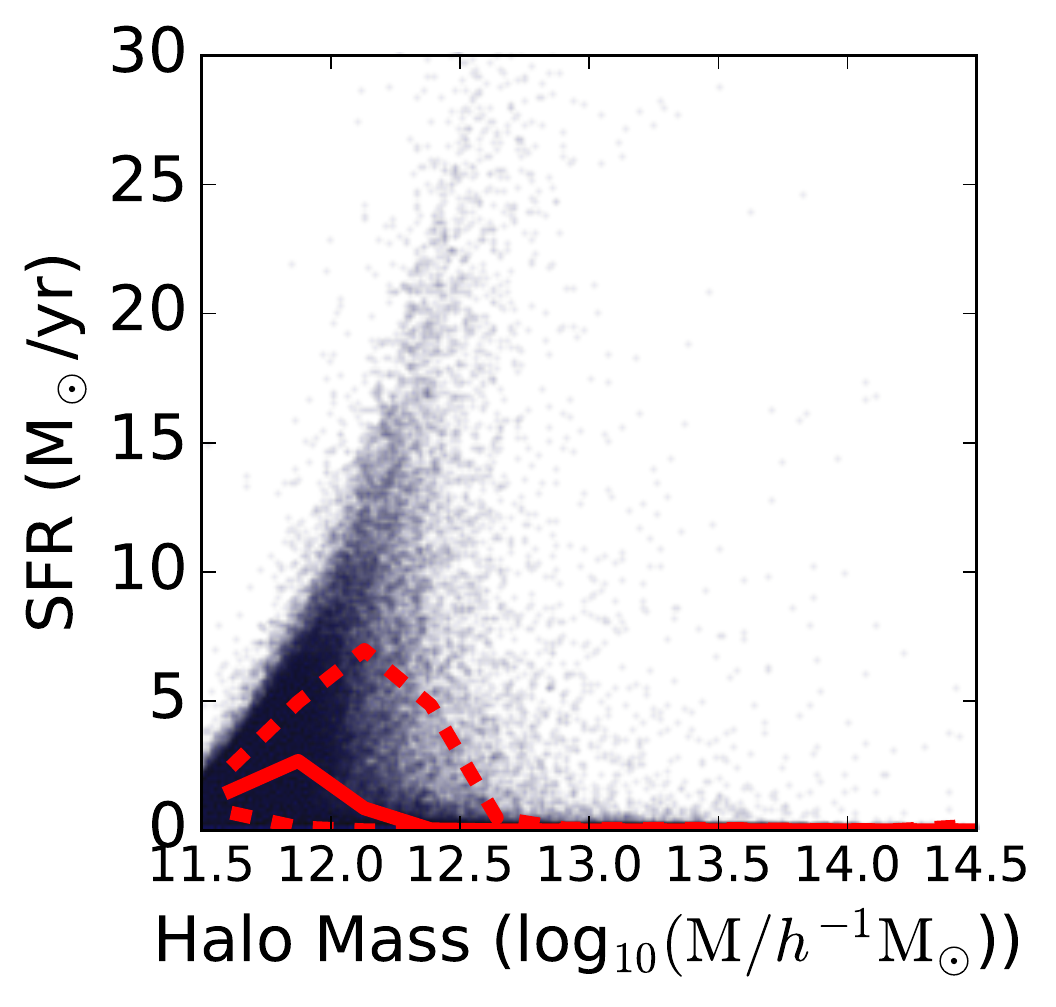}
\hspace{-0.2cm}
       \includegraphics[scale=0.32]
                        {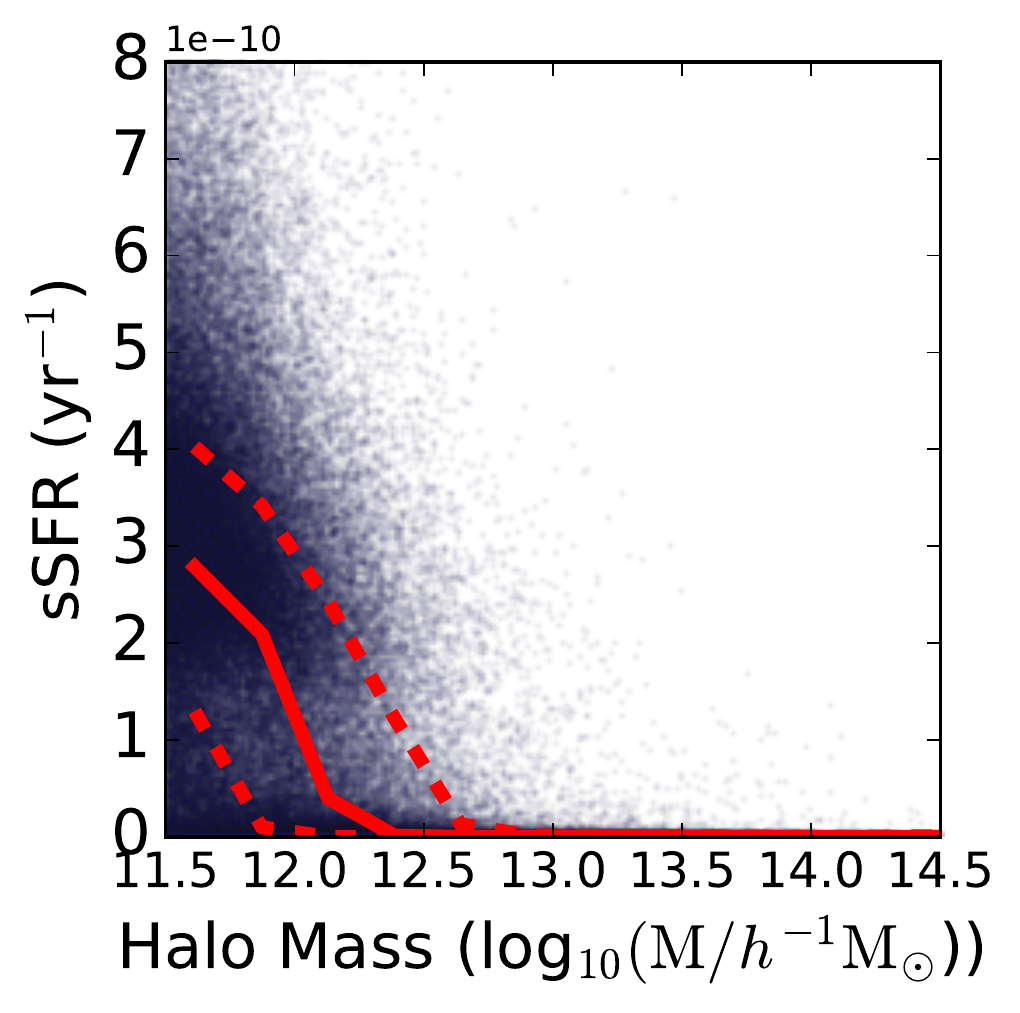}
\hspace{-0.2cm}
       \includegraphics[scale=0.32]                      {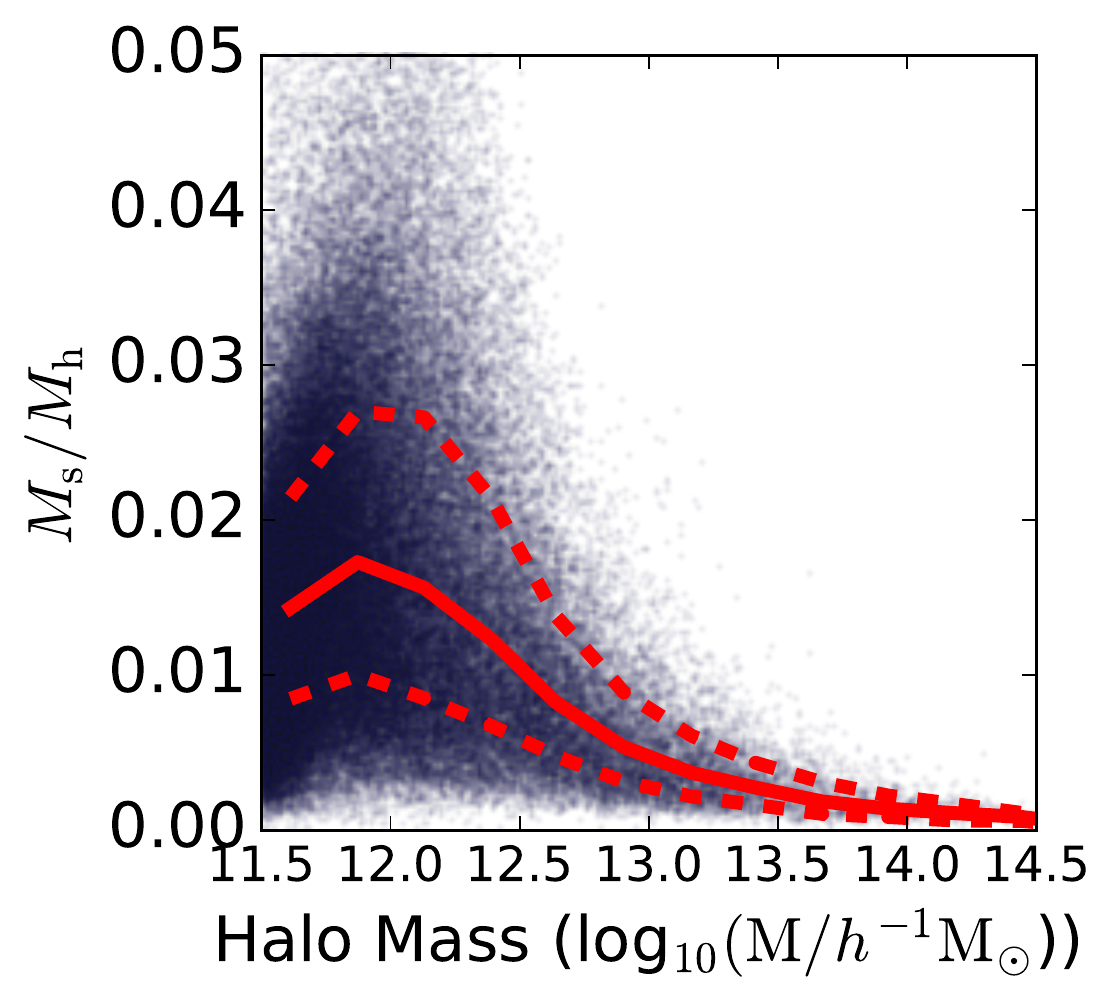}
\hspace{-0.2cm}
\caption{The proxy vs. halo mass relationship for the 5 observational proxies discussed in the text (left to right: mass-weighted age, $t_{\rm 50}$, SFR, specific SFR and $M_{\rm s}/M_{\rm h}$). Red lines show the median (solid) and 20th and 80th quantiles (dashed) of each proxy in bins of host halo mass.}
\label{ProxM}
\end{figure*}

In keeping with the majority of other approaches, we focus on the properties of central galaxies, as these by definition better capture the assembly history of the halo within which they live. All ages and times in this work are converted to look-back time to the epoch in question. \fig{ProxM} shows the relationship of each of these proxies with halo mass for central galaxies. Here we show $t_x$ with $x=50$. 

\begin{figure*}
       \includegraphics[scale=0.347]
                        {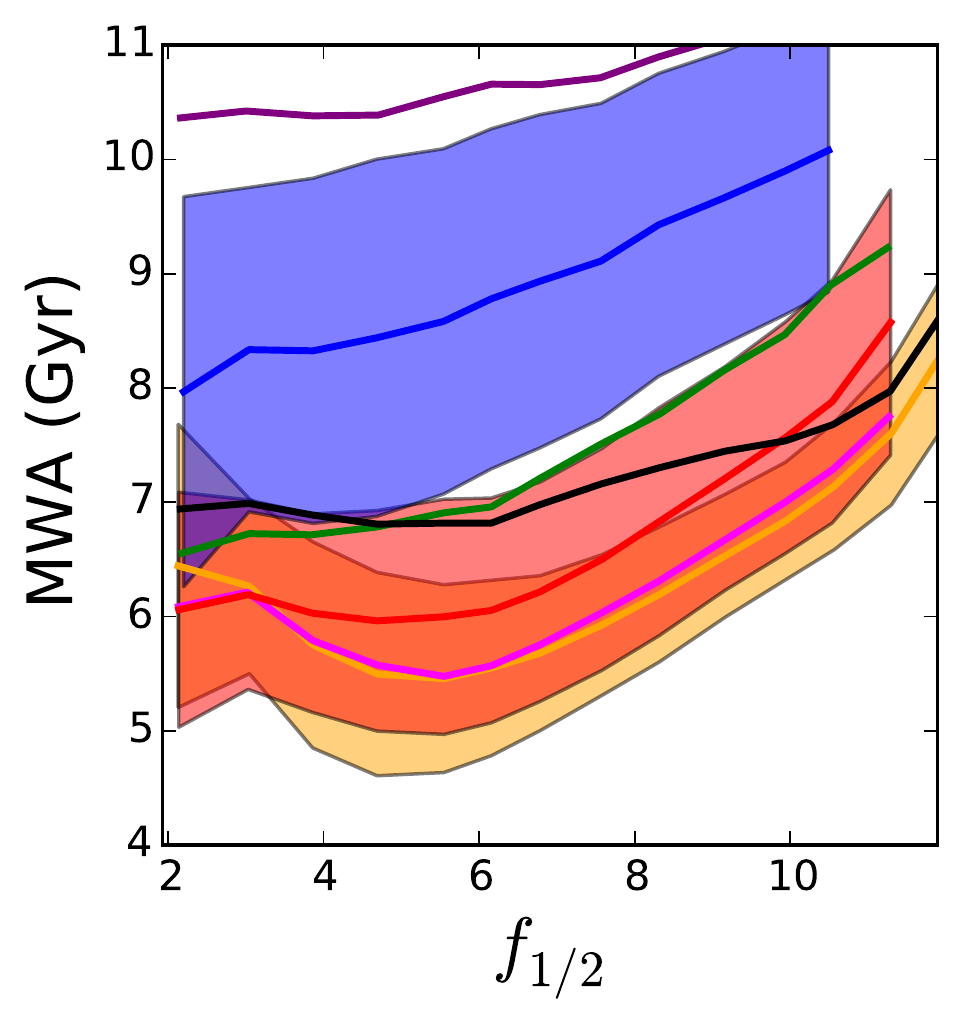}
\hspace{-0.2cm}
       \includegraphics[scale=0.347]
                        {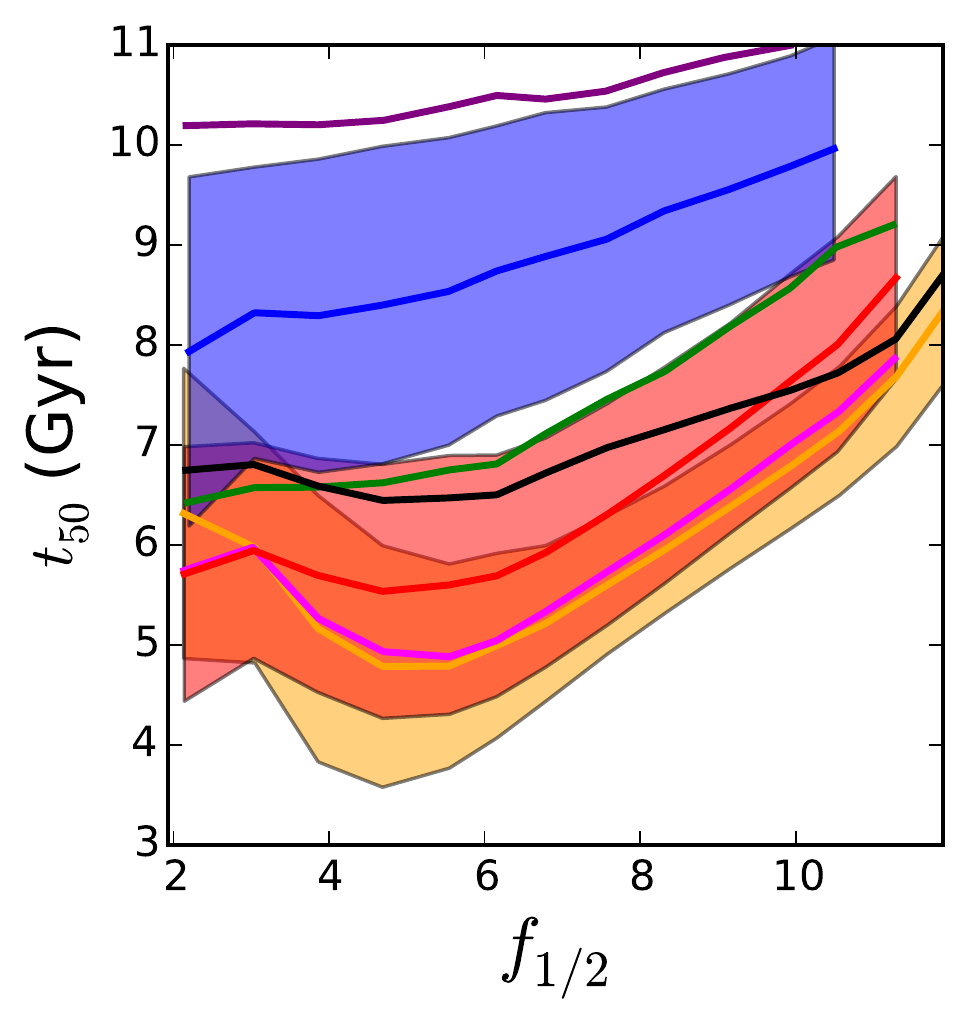}
\hspace{-0.2cm}
       \includegraphics[scale=0.347]
                  {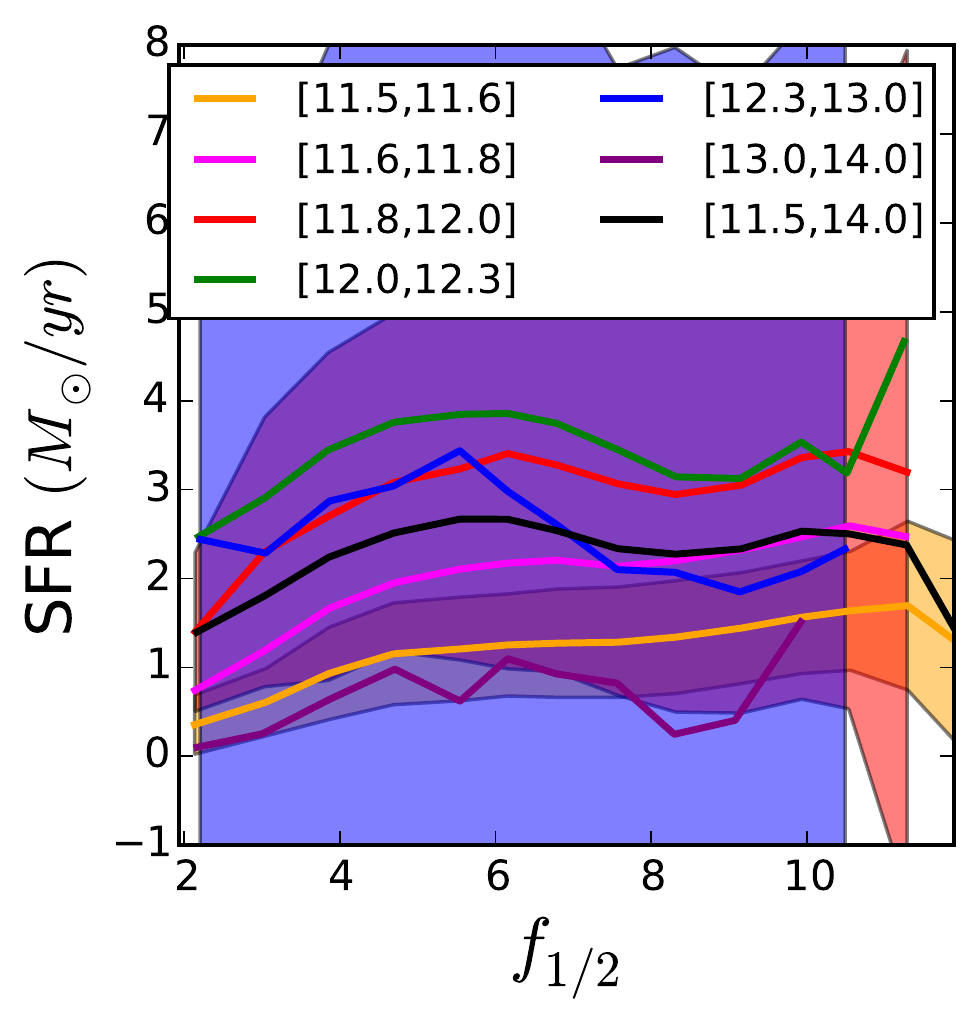}
\hspace{-0.2cm}
        \includegraphics[scale=0.347]
                      {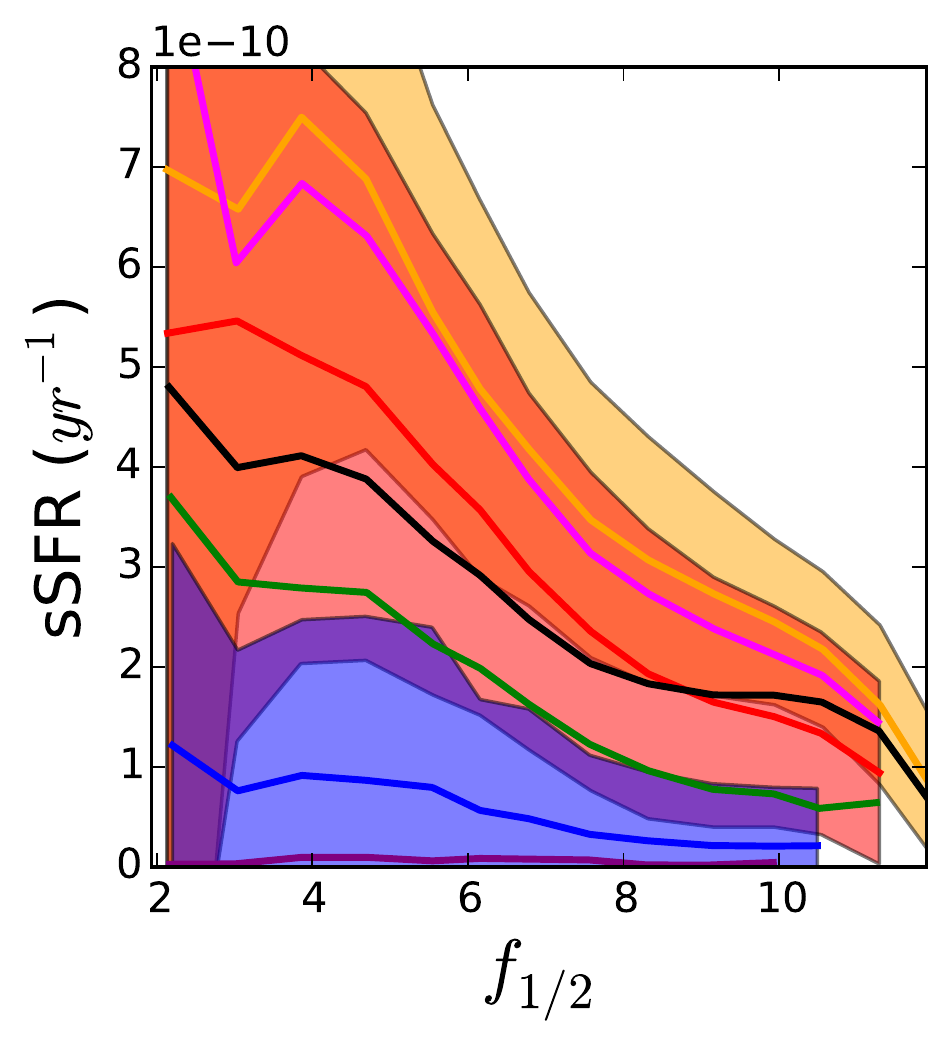}
\hspace{-0.2cm}
       \includegraphics[scale=0.347]
                      {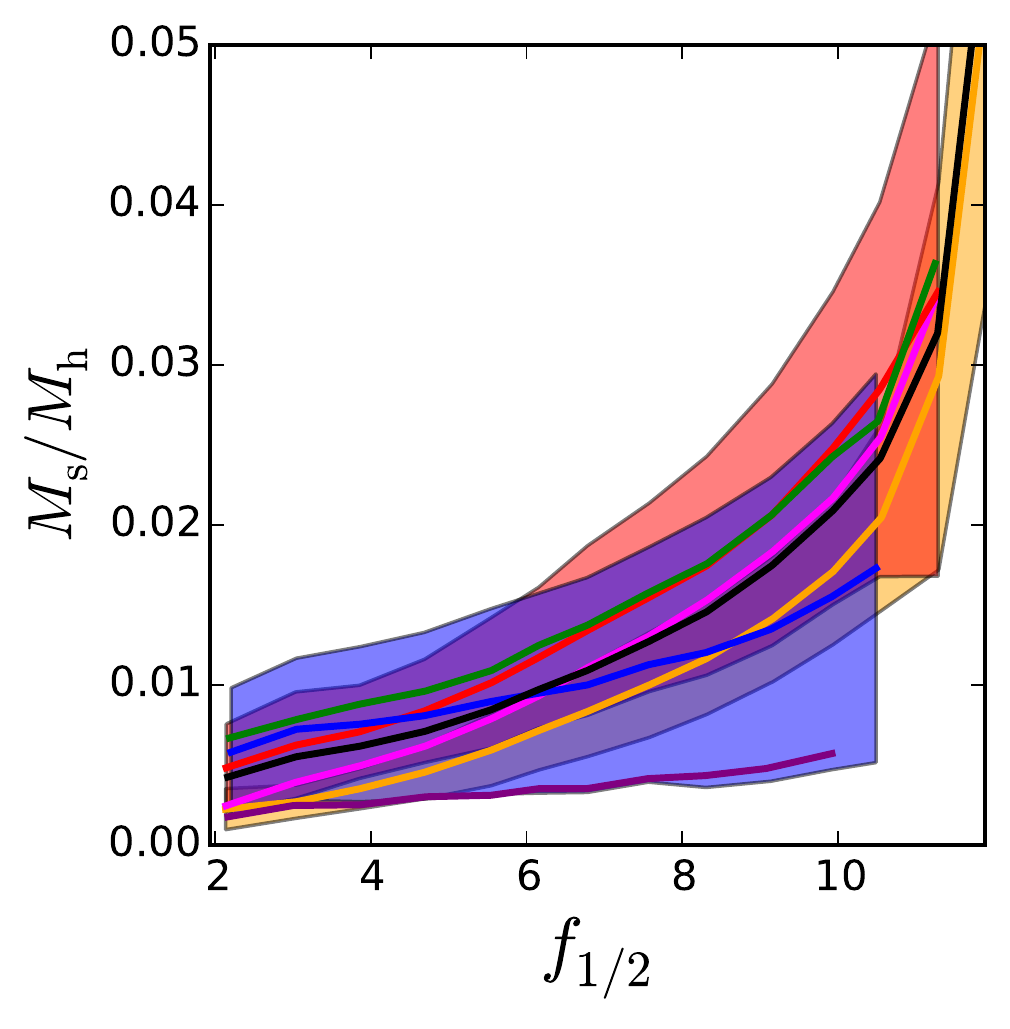}
\hspace{-0.2cm}                         
       \includegraphics[scale=0.347]
                        {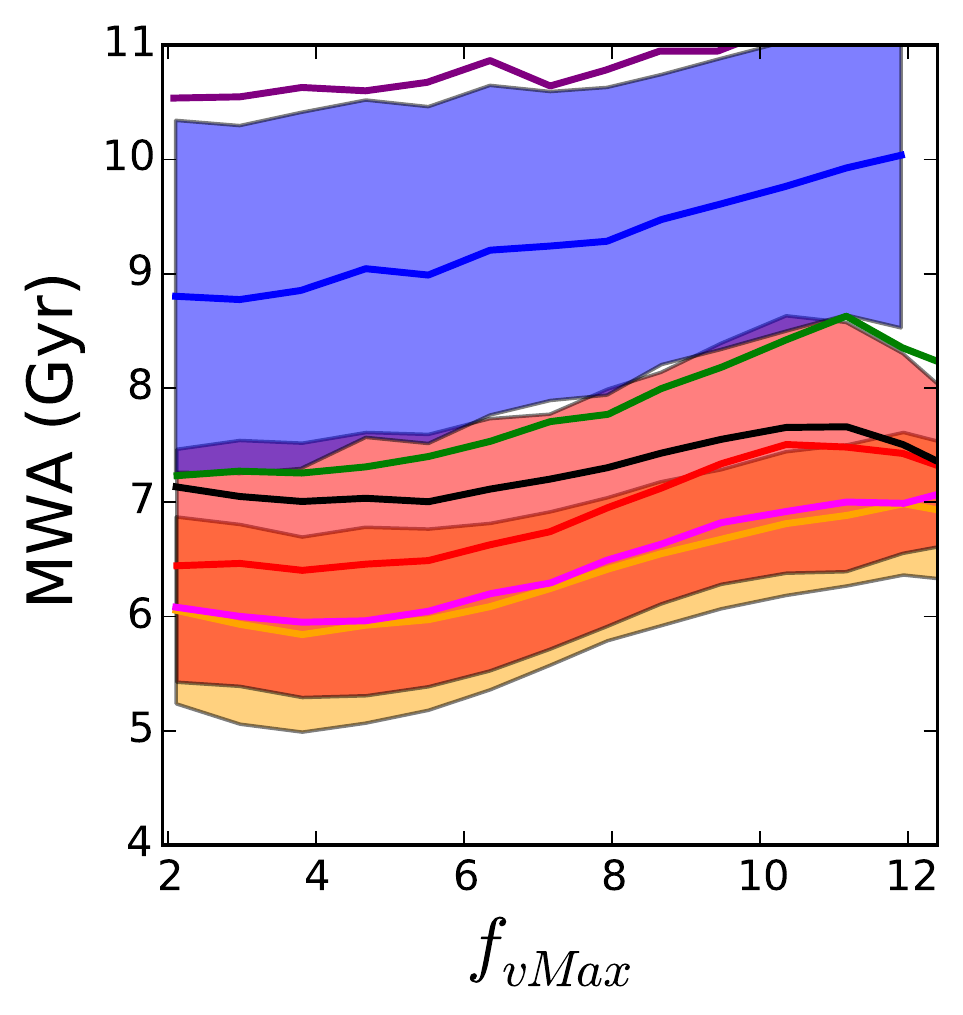}
\hspace{-0.2cm}
       \includegraphics[scale=0.347]
                         {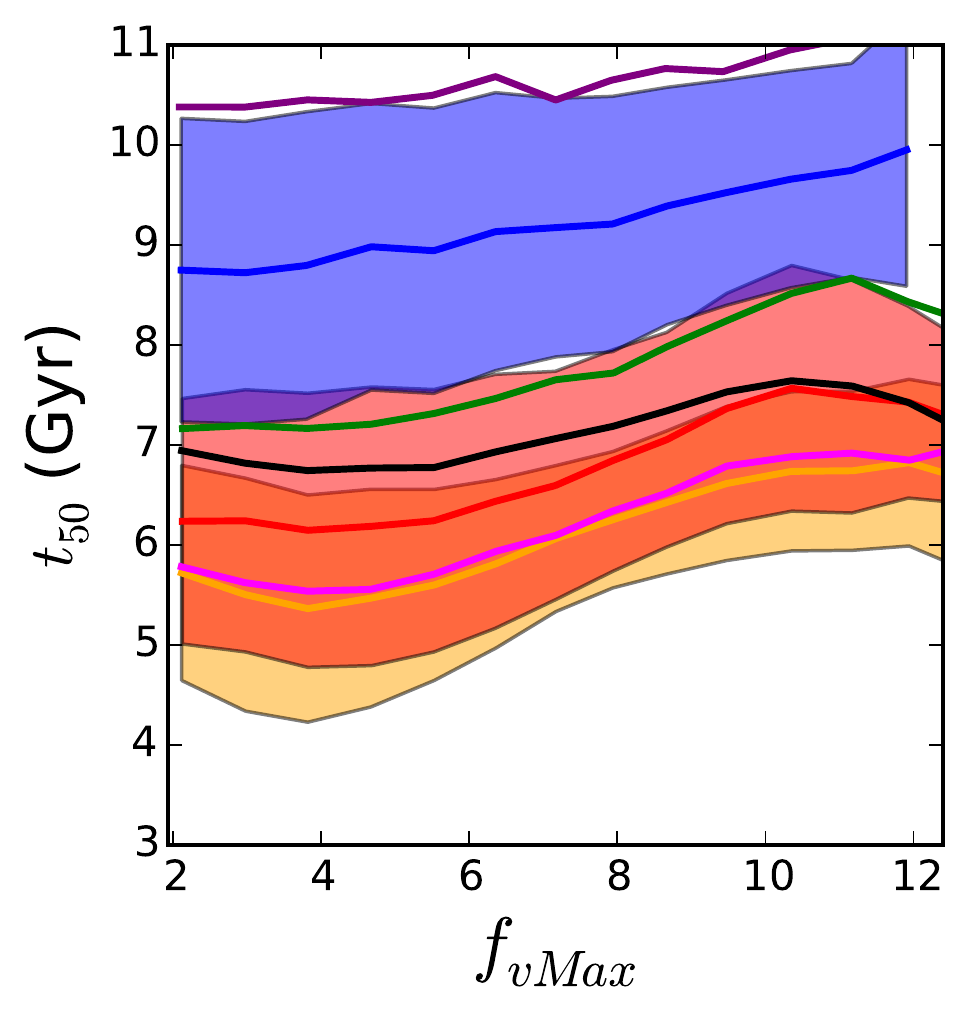}
\hspace{-0.2cm}
       \includegraphics[scale=0.347]
                  {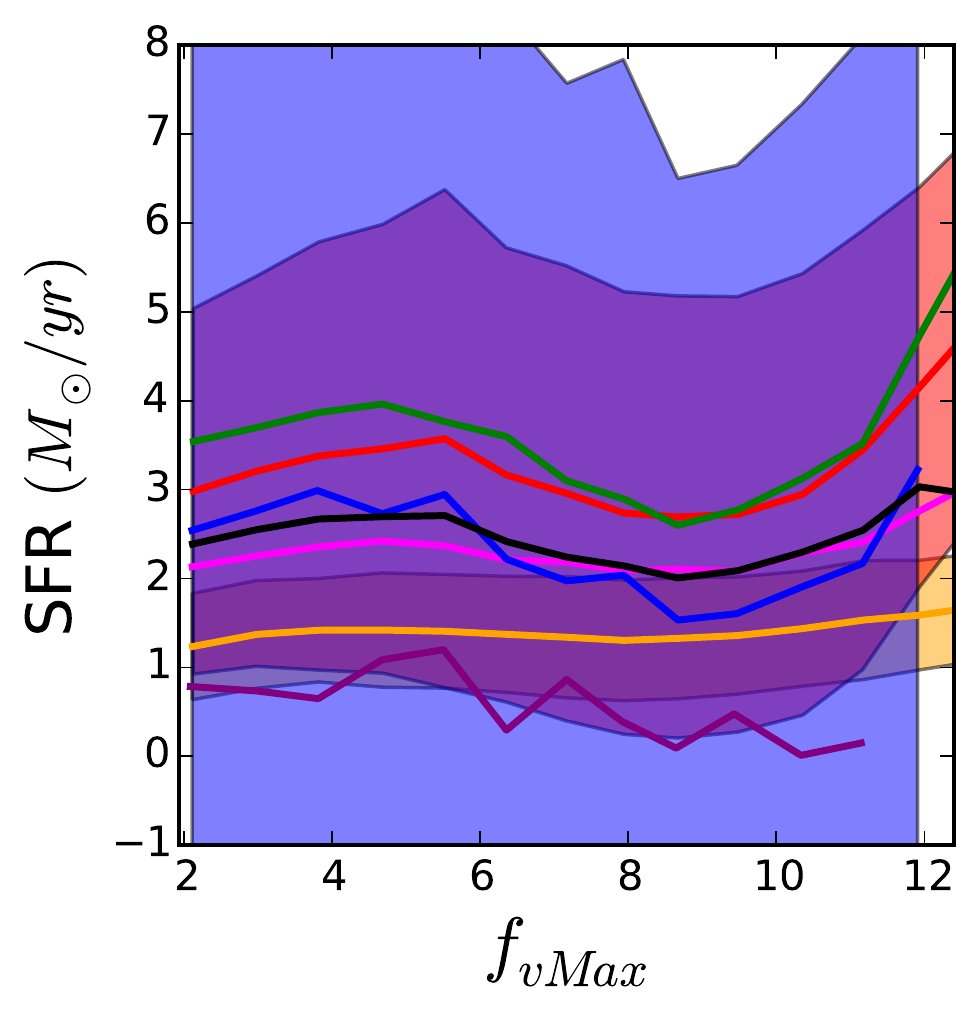}
\hspace{-0.2cm}
       \includegraphics[scale=0.347]
                        {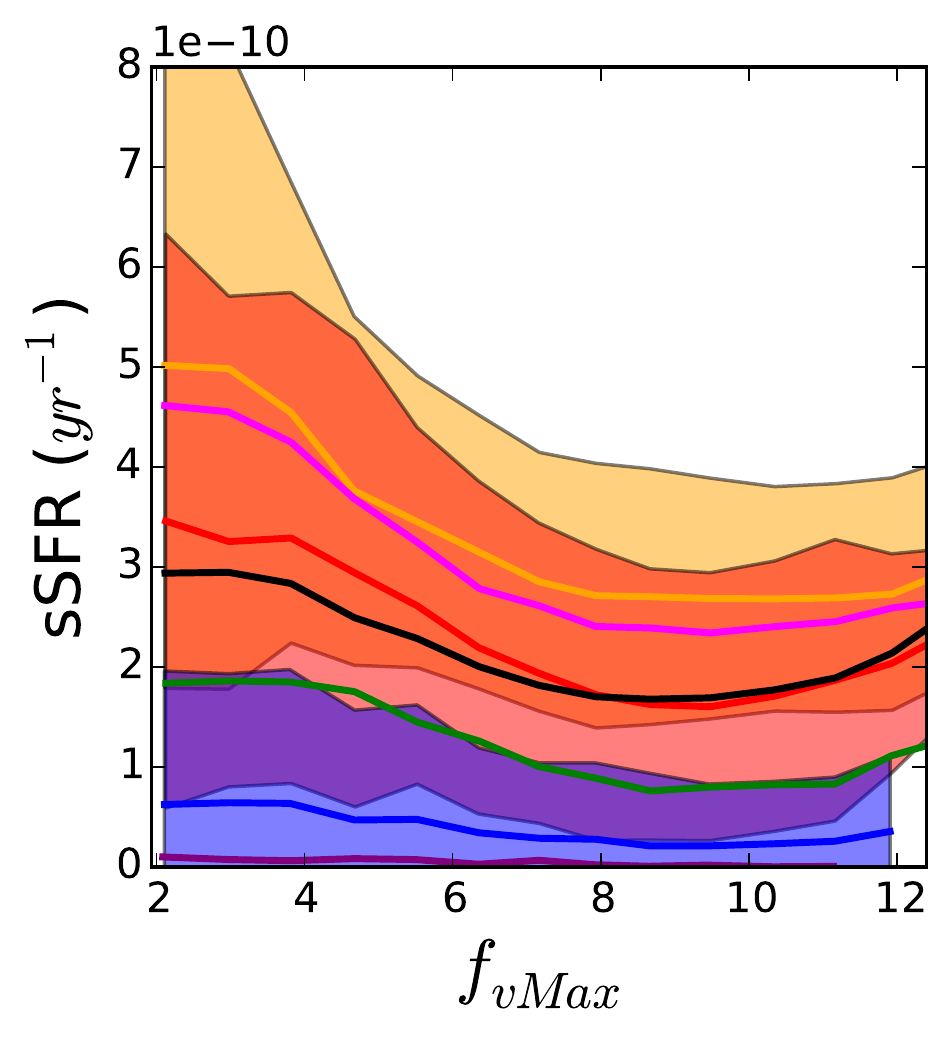}
\hspace{-0.2cm}
       \includegraphics[scale=0.347]
                      {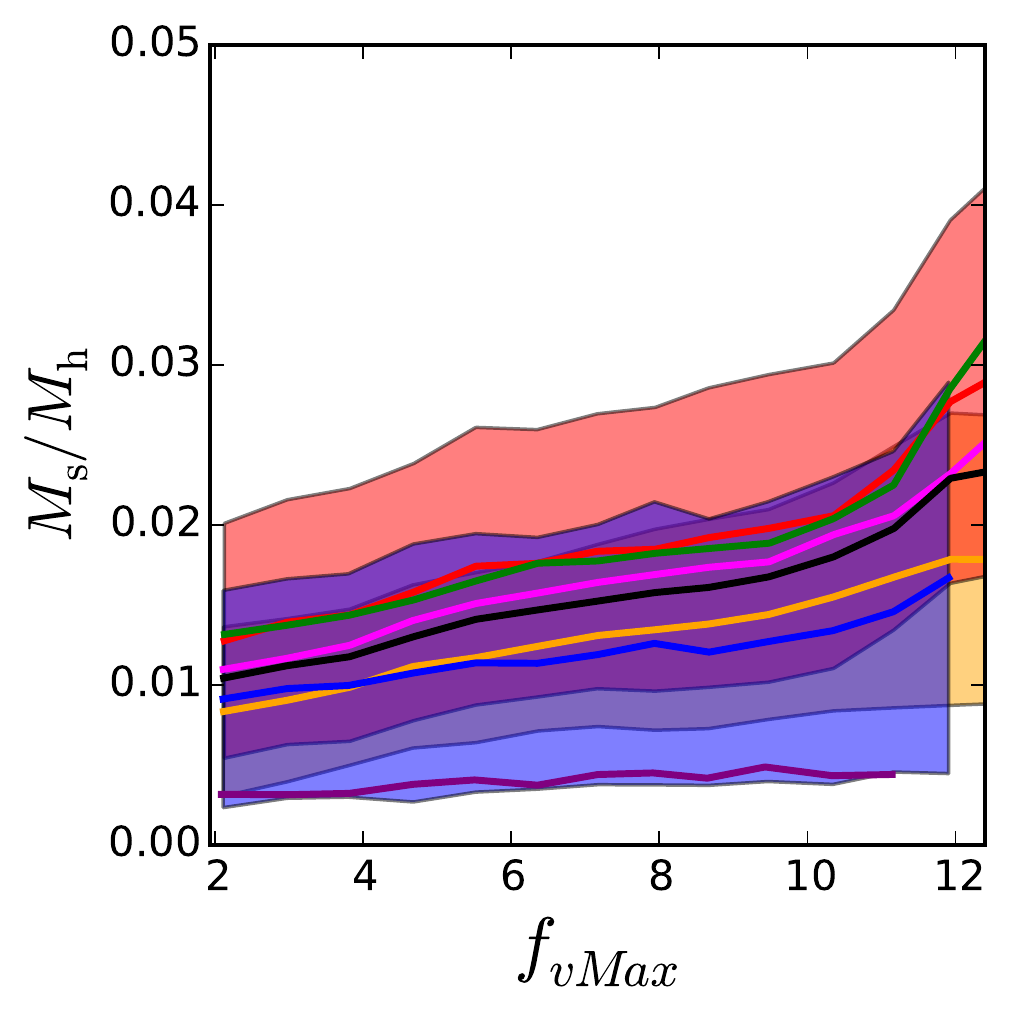}                
\hspace{-0.2cm}                         
       \includegraphics[scale=0.347]
                        {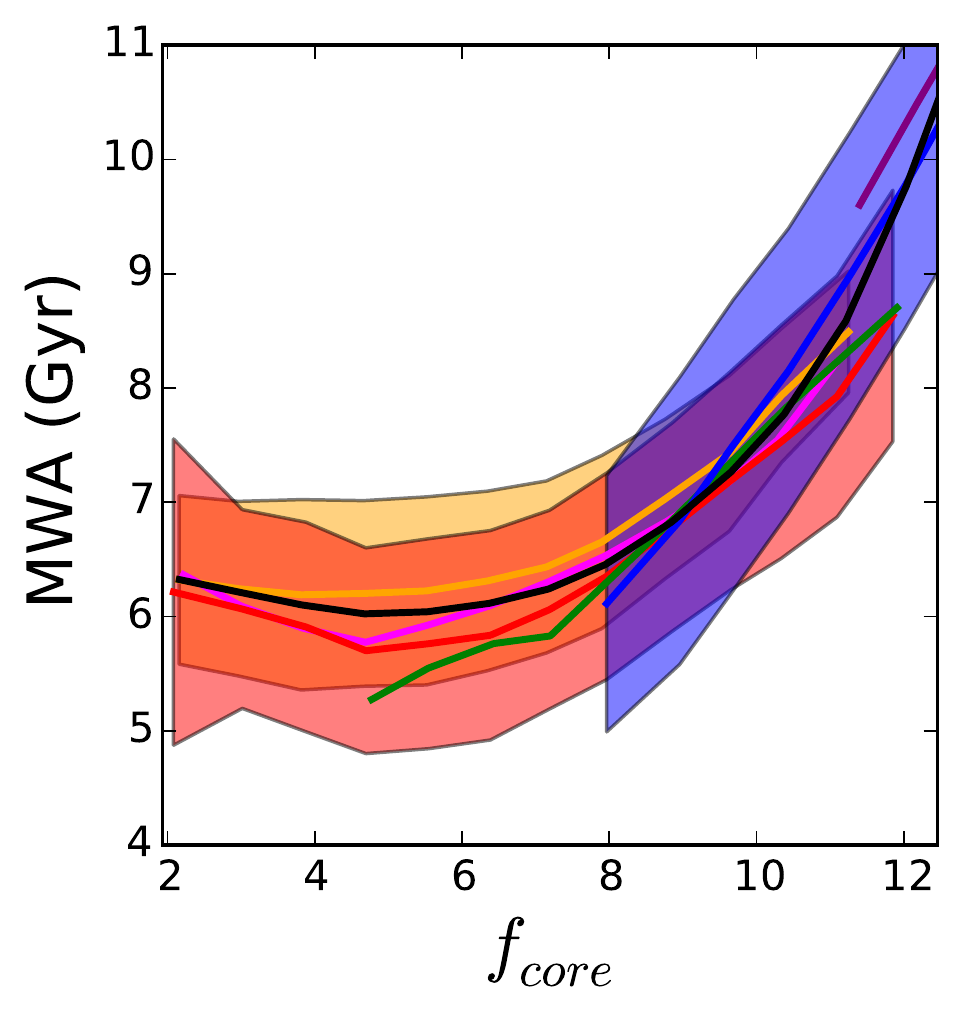}
\hspace{-0.2cm}
       \includegraphics[scale=0.347]
                         {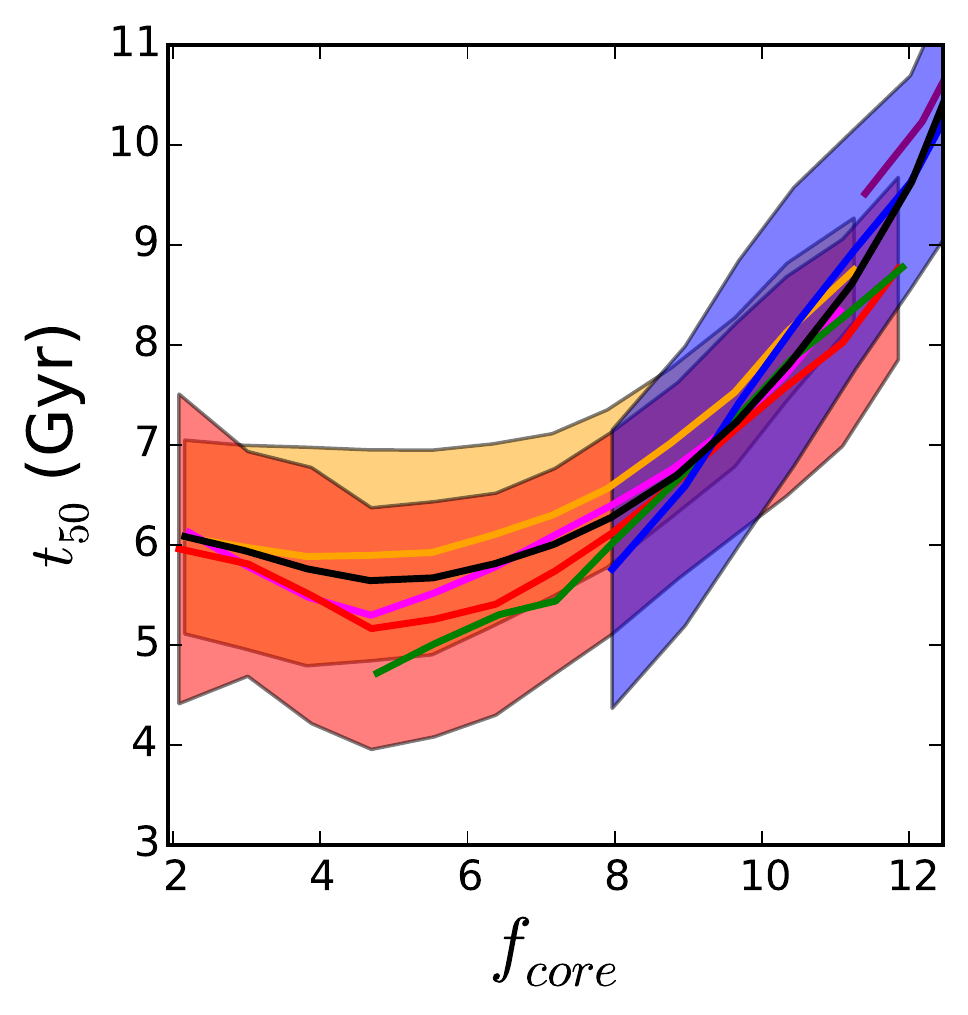}
\hspace{-0.2cm}
       \includegraphics[scale=0.347]
                  {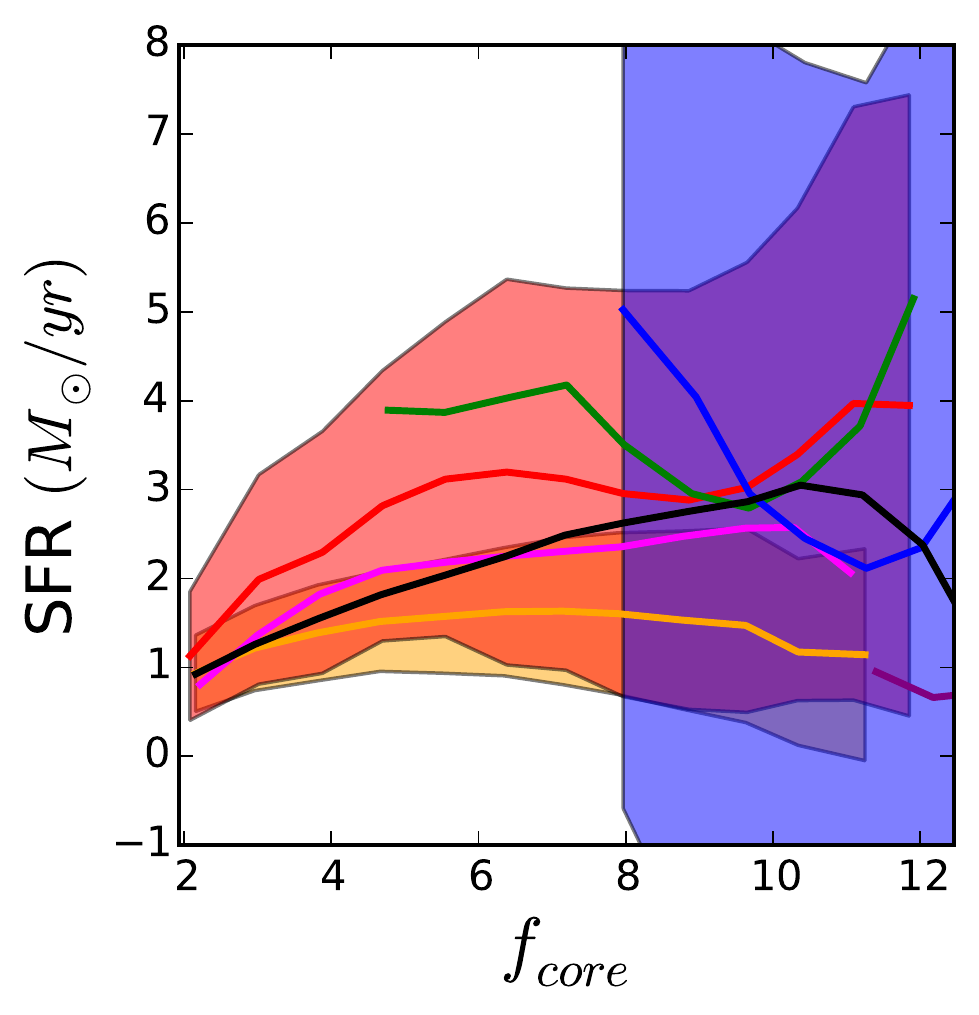}
\hspace{-0.2cm}
      \includegraphics[scale=0.347]
                   {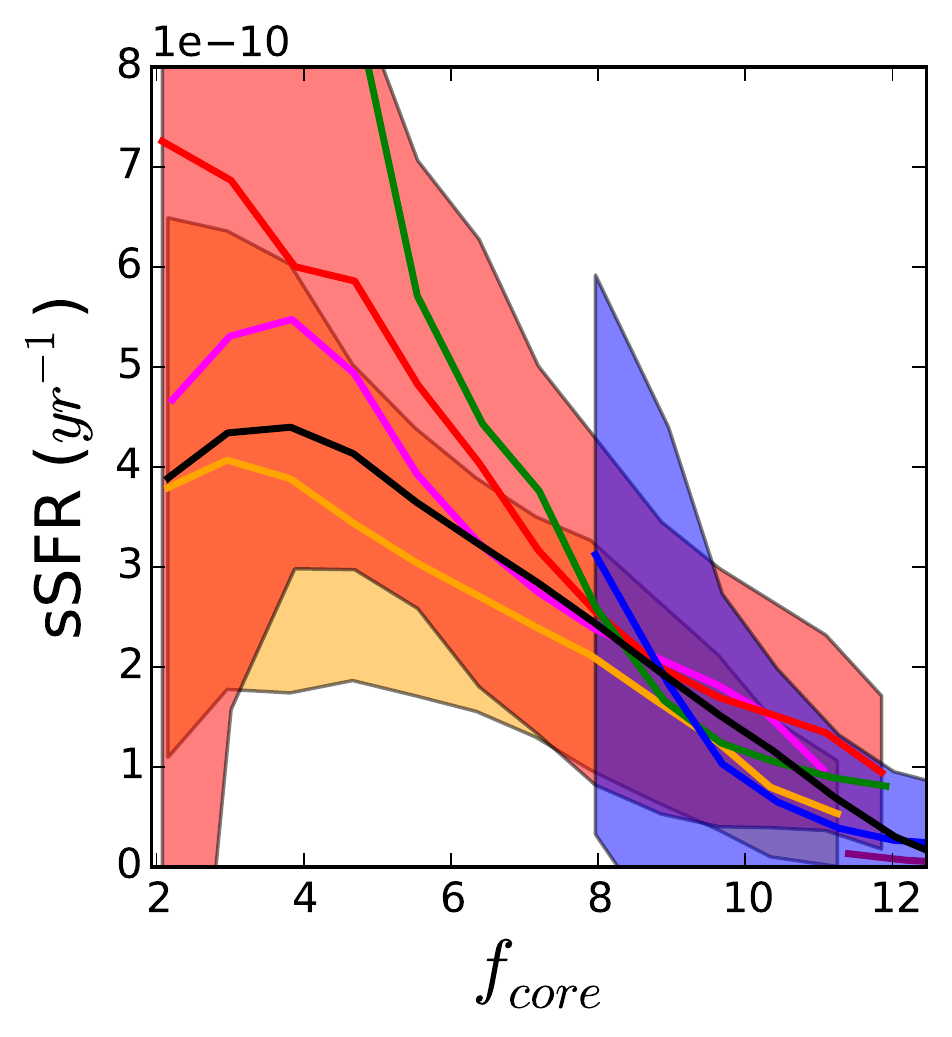}
\hspace{-0.2cm}
   	  \includegraphics[scale=0.347]
                  {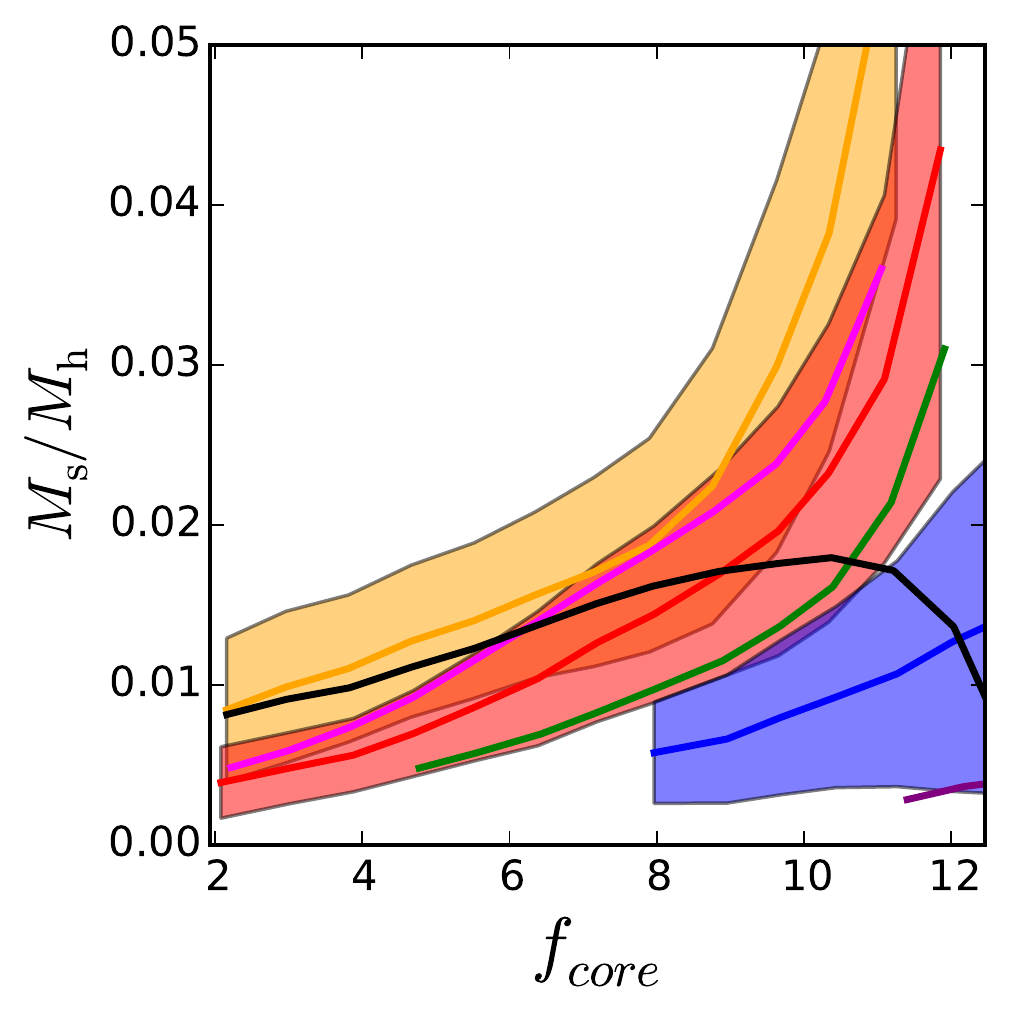}
\caption{The dependence of each proxy (left to right: mass-weighted age, $t_{\rm 50}$, SFR, specific SFR and M$_{\rm s}/$M$_{\rm h}$) on halo formation time. Each row of panels shows one of the three definitions of formation times (from top to bottom: $f_{1/2}$, $f_{\rm vMax}$ and $f_{\rm core}$). Different coloured lines show mean quantities for galaxies in haloes of different mass as described in the legend (written in terms of $\log $M$_h/$M$_\odot$). The shaded regions show the scatter at each halo mass. For clarity, we only show the scatter for three of the halo mass bins shown. }
\label{ProxFs}
\end{figure*}

The relationship between each of the 5 proxies and the $f_{1/2}$, $f_{\rm vMax}$ and $f_{\rm core}$ formation times are shown in \fig{ProxFs}, for different ranges of halo mass. In all panels, the coloured lines show relationships at different halo mass, with the black line showing the mean relation for all halo masses. $f_{1/2}$ and $f_{\rm vMax}$ show qualitatively similar trends, due to the fact they are both hierarchical in nature. $f_{\rm core}$ behaves distinctively differently, and shows little dependence on halo mass when the other two measures of halo age do. This is explained by the fact that $f_{\rm core}$ has a much tighter relationship with halo mass: haloes with a given $f_{\rm core}$ have a small range in halo mass, compared to haloes with a given $f_{1/2}$ or $f_{\rm vMax}$. At fixed formation time, halo mass vastly determines the properties of the central galaxy, and the considerable scatter in the halo mass - formation time relationships (see Fig.~\ref{fig:FM}) then results in larger scatter and halo mass dependence seen in $f_{1/2}$ and $f_{\rm vMax}$, compared to $f_{\rm core}$. 

The first and second columns show measures of stellar age that capture the stellar assembly history at typically intermediate-to-old ages. Although with a large scatter, these quantities do provide some information on halo age, although mostly only at early times. 


The third and fourth columns show current star-formation and specific star-formation rate, which in contrast with the two estimators considered above, are only sensitive to the last few hundred thousand years. We find that SFR is not at all related with halo formation times. Largely this is caused by the fact that SFR is tightly correlated with stellar mass, and there is significant scatter between formation time and stellar mass. For that reason, specific SFR does a significantly better job. Although dependent on halo mass and with a large scatter, sSFR is a reasonable predictor of $f_{1/2}$, especially in low-mass haloes. 

 
The fifth column shows stellar mass to halo mass ratio. As discussed in \cite{Wang2011} and more recently in \cite{Lim2015}, \MsMh shows a convincing and tight correlation with halo age, particularly at low halo mass. 

\subsection{Correlation}
\begin{figure*}
       \includegraphics[scale=0.28]
                        {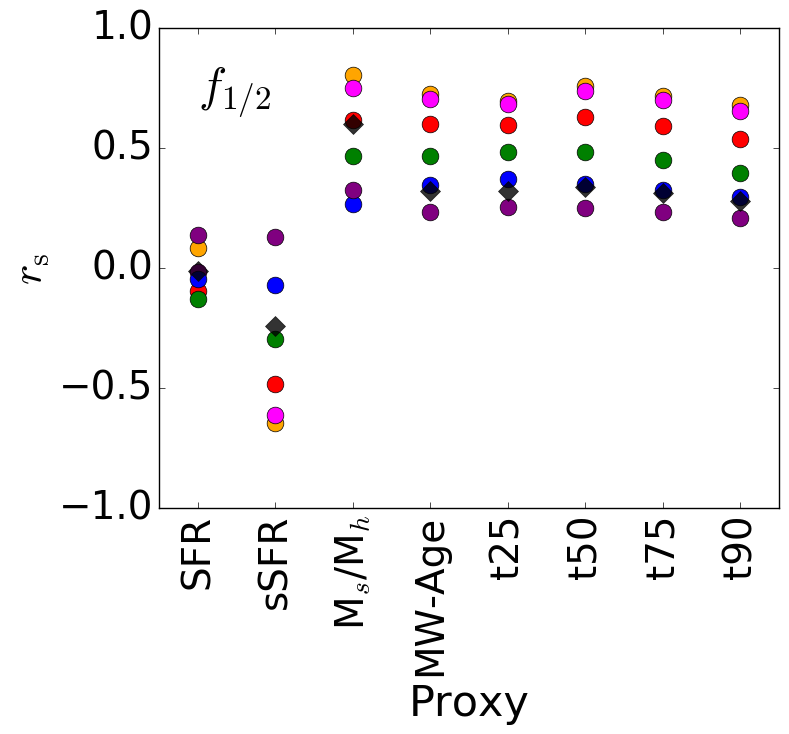}
       \includegraphics[scale=0.28]
                        {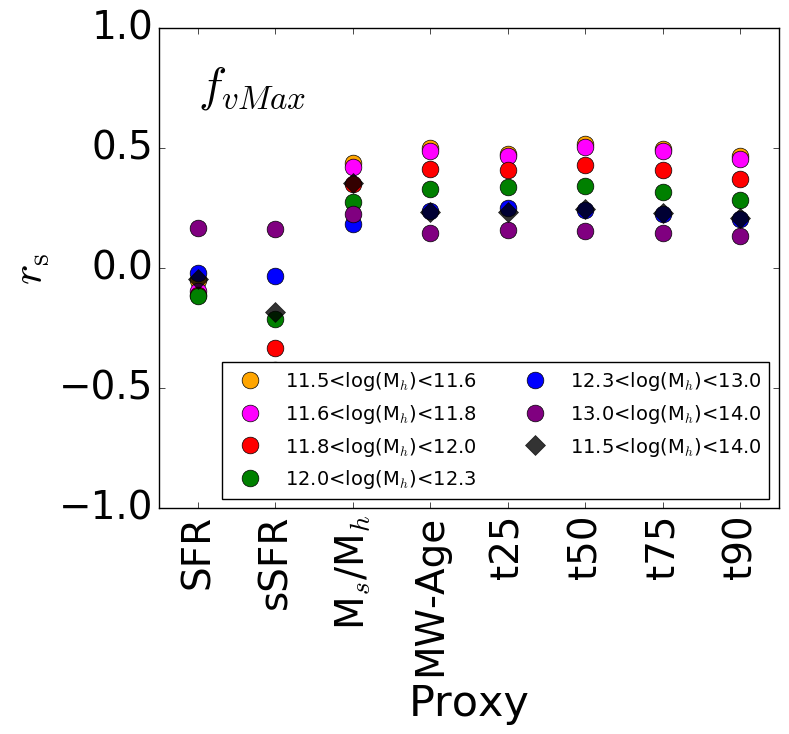}
       \includegraphics[scale=0.28]
                        {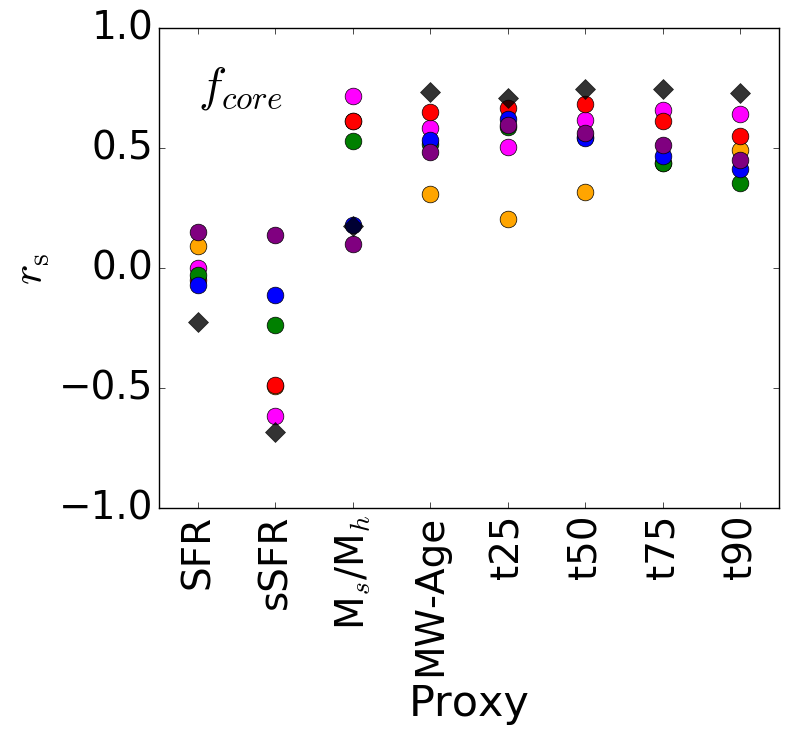}                
                        \caption{The Spearman rank correlation between each proxy and the three halo formation times discussed in the text (left to right: $f_{1/2}$, $f_{\rm vMax}$, $f_{\rm core}$). Coloured circles show results for the sub-catalogues defined by the halo mass ranges given in the key in the left panel. Black diamonds show the correlation coefficient for the full mass range considered here: $11.5<\log{(M/h^{-1}{\rm M_{\odot}})}<14$. }
\label{spearm}
\end{figure*}

In order to quantify the discussion in the previous section, we measure the Spearman rank correlation coefficient between each of the three definitions of formation time and the proxies under consideration. The Spearman rank correlation coefficient is a measure of the statistical dependence between two variables, and how well this relationship may be described by any monotonic function. We chose this measure due to its non-parametric nature, so that complexities of the exact form of the relationship need not be considered. The coefficient, $r_{\rm s}$ is computed from the ranked variables, ($x_i, y_i$) of the sample set ($X_i, Y_i$) of size $n$:

\begin{equation}
r_{\rm s}=1-\frac{6\sum{(x_i - y_i)^2}}{n(n^2-1)}.
\end{equation}

Hence, $r_{\rm s} =1$ or $-1$ equates to the variables being a perfect monotone function of each other. The value of the Spearman correlation coefficients between the three halo formation times and each of the observational proxies, with $t_x$ now shown for a range of values of $x$, are shown in \fig{spearm}. The black diamonds show the correlation coefficients calculated for the full sample, i.e. all haloes with $M_{\rm h} > 10^{11.5} h^{-1}{\rm M_{\odot}}$.

In the case of $f_{1/2}$ and $f_{\rm vMax}$, the formation times of the lower mass haloes are consistently better correlated with all proxies than the formation times of the high mass haloes (or anti-correlated, in the case of sSFR). In contrast, the relationship between $f_{\rm core}$ and the proxies shows a more complicated dependence on halo mass. Due to the smaller scatter between halo mass and formation time discussed in the previous section, using the full range of halo masses (black diamonds) improves the correlation coefficient in most cases. The SFR of the central galaxy can be seen to be only weakly correlated with halo formation time, regardless of the choice of halo formation time used. $t_x$ shows a moderately strong correlation with each of the formation times, behaving similarly to mass-weighted age. According to the model, sSFR, \MsMh and mass-weighted age afford the best chance of detecting differences in halo formation time from galaxy properties.


\subsection{Summary}

Using the L-Galaxies semi-analytic model, in this section we investigated: (i) how different definitions of halo formation time propagate into galaxy observables; and (ii) how different galaxy observables compare as potential proxies for halo formation time. 

The scatter between halo mass and $f_{\rm core}$ is substantially smaller than the scatter between halo mass and $f_{1/2}$ or $f_{\rm vMax}$. This means that observationally one might be able to measure changes in $f_{\rm core}$ without exquisite halo mass observational estimates. On the other hand, $f_{\rm core}$ only produces significant changes in galaxies' properties for haloes with $f_{\rm core} > 8$ Gyrs. I.e., galaxy properties cannot predict  $f_{\rm core}$ if $f_{\rm core} \lesssim 8$ Gyr.

Observationally, we are restricted to measuring galaxy properties. We found that sSFR, \MsMh and mass-weighted age are typically better predictors of halo formation time than others considered. However, our analysis demonstrates that translating measured differences in these quantities into an absolute difference in formation time is difficult, and that different observables are more or less sensitive to different definitions of formation time, with an ever present dependence on halo mass. However, all definitions of formation time leave the same average quantitative trends on galaxy observable - e.g., a measurement of a larger \MsMh in a sample of galaxies controlled for halo mass, will always indicate an older halo according to the L-Galaxies model. 


\section{Simulating observations}

Whereas in principle some galaxy properties can provide a proxy for halo formation time - as demonstrated in the previous section - such quantities can be notoriously difficult to extract from data. All of the information is encapsulated in full star-formation histories, but compressed or integrated quantities, such as mass-weighted age or total stellar mass can be more robust to degeneracies and limitations of the modelling. In this Section we quantify how well a set of galaxy properties can be recovered from GAMA-like spectra, which we simulate using the L-Galaxies model and subsequently analyse using the full-spectral fitting code VESPA (the GAMA survey is summarised in Section~\ref{sec:GAMA}). An important aspect of this exercise is that the stellar population synthesis (SPS) models used to generate and analyse the simulated spectra are always made to match. The results in this section are robust to the choice of SPS models provided the above statement remains true. We therefore only present results using a single set of SPS models. However, it is well established that the choice of SPS models impacts on the interpretation of data (e.g.\citealt{Tojeiro2011}), and in Section~\ref{sec:application} we will assert the robustness of our results to the choice of SPS modelling. 

\subsection{Making simulated spectra} \label{sec:simulated_spectra}

In brief, we construct simulated spectra by convolving the model star-formation and chemical enrichment histories with a set of stellar population models, attenuating the light due to dust absorption, and adding simulated noise. As in the previous section, we work on a snapshot with $z\approx 0.15$, as it best approximates the median redshift of the GAMA sample we will use in the next section. At this redshift, the model star-formation histories are given in 13 bins, approximately logarithmically spaced in lookback time. The algorithm for binning the SFH is described in \cite{ShamshiriEtAl15} (see their Fig.1). Whereas this resolution smoothes over much of the natural short-length stochasticity of the SFHs, it retains enough complexity to reproduce observed magnitudes at $z=0$ \citep{ShamshiriEtAl15}. 

The model provides stellar mass formed in each bin of lookback time in the disc and bulge separately. The rest-frame luminosity of each component is computed by:

\begin{equation}\label{eq:Lgal_L}
L_\lambda = L_{\rm bulge} + f_{\rm dust, D}L_{\rm disc}
\end{equation}
where $f_{\rm dust}$ encaspulates the effects of dust attenuation and is defined in the next section. According to the L-Galaxies model, only the disc's light is attenuated by dust and bulges are dust-free. The luminosity of each component is computed as:

\begin{equation}
L_{\rm obs} = \sum_i m_i L_{\rm SSP}(t_i, Z_i).
\end{equation}
where $m_i$ is the total mass formed within the time bin $i$, $L_{\rm SSP}(t_i, Z_i)$ is the predicted luminosity, given by the Single Stellar Population (SSP) models, of stellar populations of age $t_i$ and metallicity $Z_i$. $L_{\rm SSP}(t_i,Z_i)$ is taken here to be a $\delta-$function episode of star-formation, at the mean age of the bin $i$. 

\begin{figure}
\includegraphics[scale=0.37]{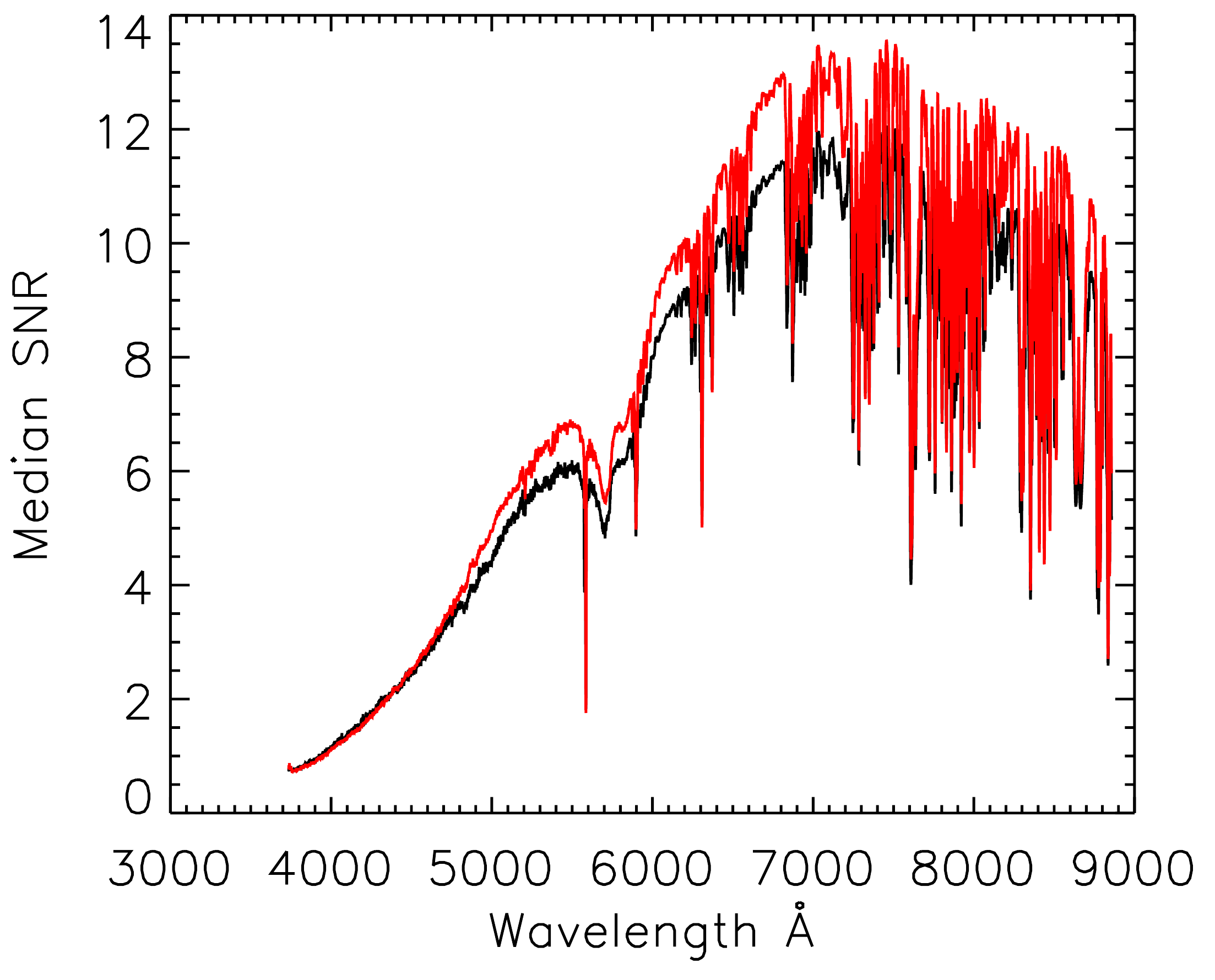}
\caption{Median S/N of approximately 10,000 GAMA galaxies as a function of observed wavelength. The S/N drops dramatically towards the blue, and the effect of skylines can be clearly seen redwards of the 5578AA skyline. The break in the continuum is caused by the 5700AA dichroic. The black line shows the median S/N for a random subsample of the full GAMA populations; the red line shows the median S/N for the BCGs used in the next section. Unless specifically stated otherwise, the model shown in the red line is used to create the mock galaxies in this section.}
\label{fig:SNR_GAMA_median}
\end{figure}

Gaussian photon noise emulating the effective signal-to-noise (S/N) ratio of GAMA galaxies is added to the spectrum in the observed frame. Fig.~\ref{fig:SNR_GAMA_median} shows the effective median SNR per pixel, as a function of observed wavelength for a representative sample of GAMA galaxies (in black) and the central BCG sample of galaxies used in Section~\ref{sec:application} (in red). 


Finally, the wavelength range is set by only taking the region between 3800\AA\ and 8500\AA\ in the observed range, to approximately match the wavelength range of the AAT spectrographs. 
 
\begin{figure*}
\includegraphics[scale=0.4]{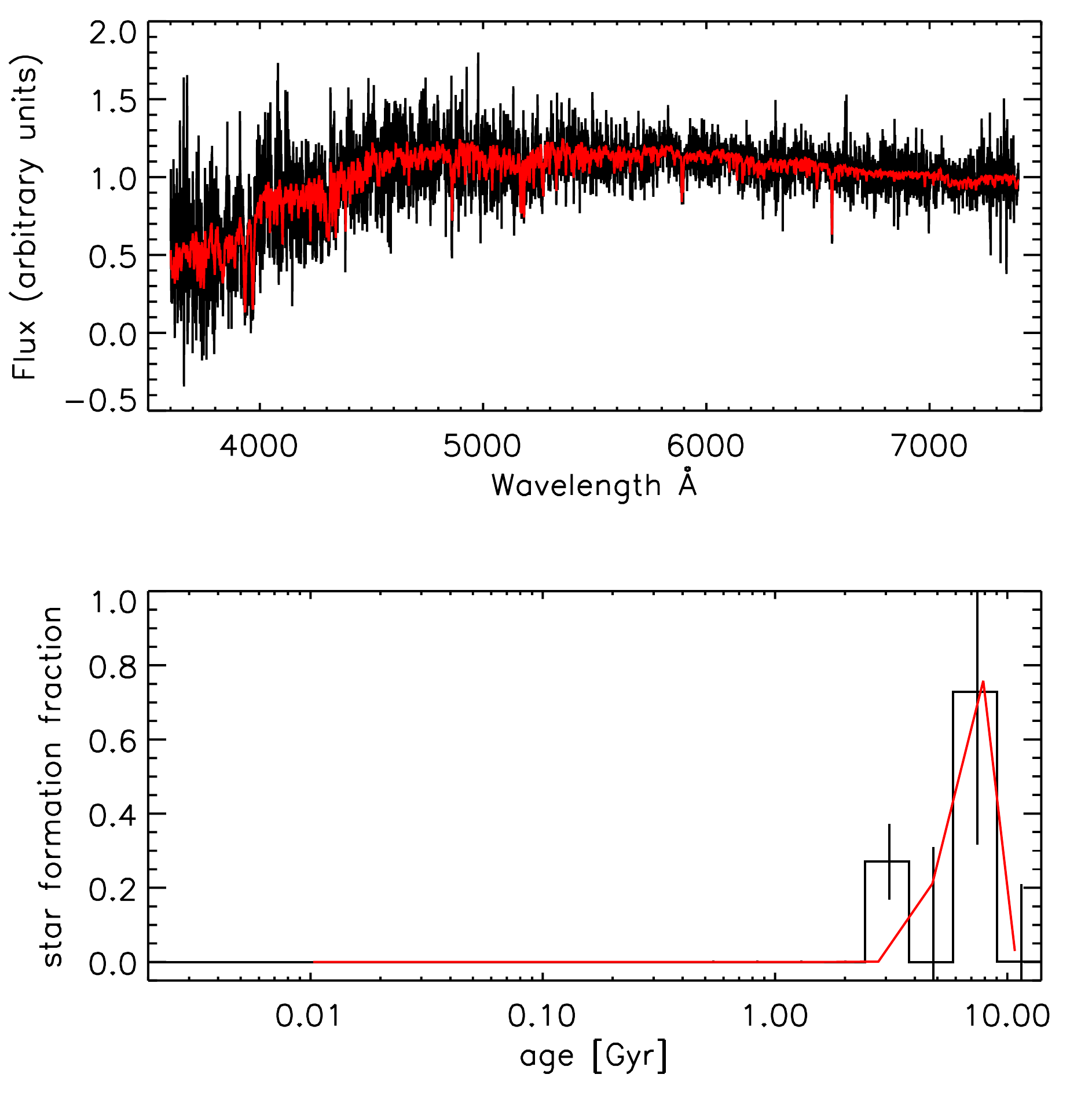}
\includegraphics[scale=0.4]{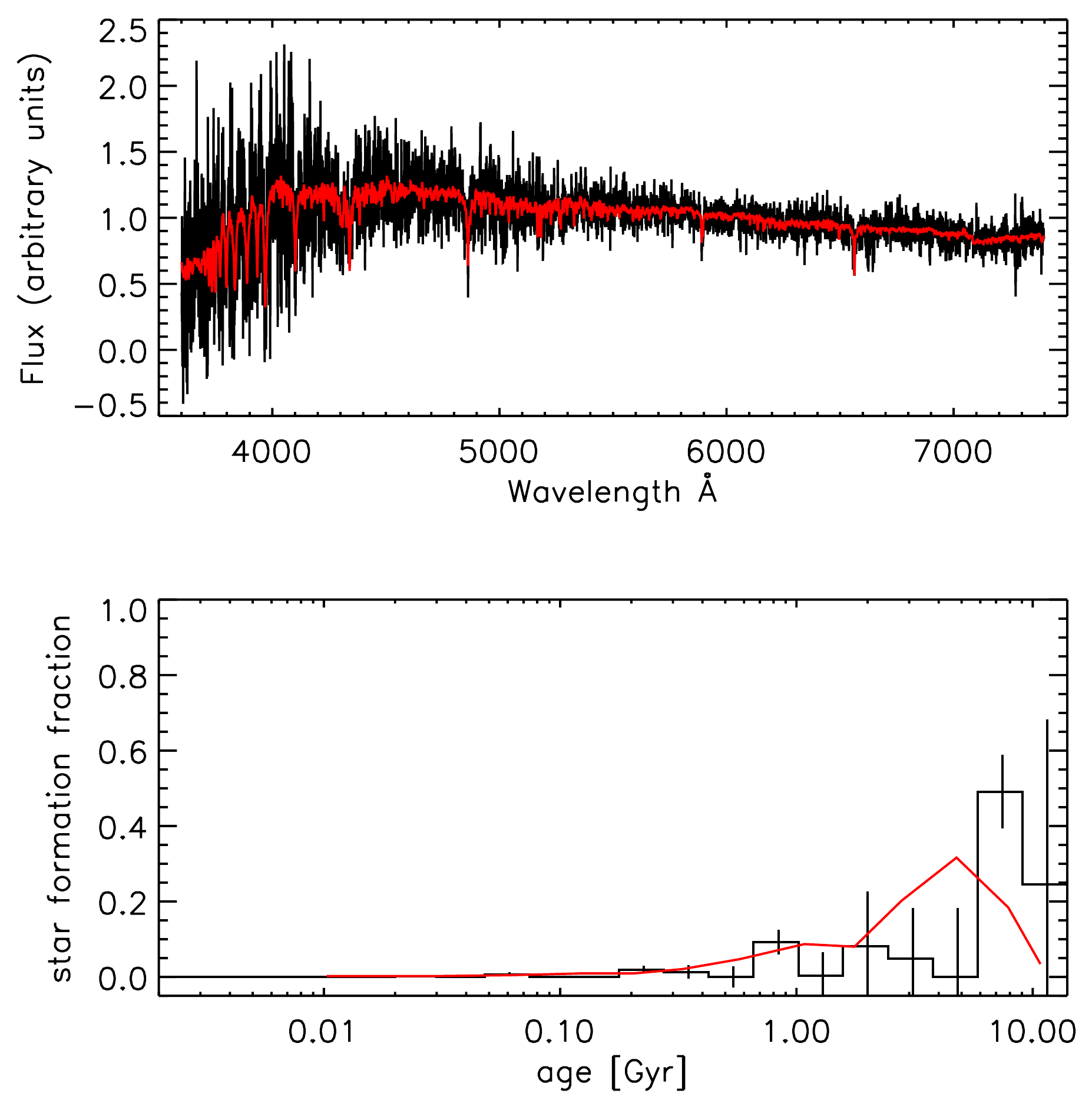}
\caption{Two galaxies from the L-Galaxies semi-analytic model. The top panel shows the noise-free spectrum in red and a noisy realisation in black, computed according to the details in Section~\ref{sec:simulated_spectra} and with the S/N template shown in Fig.~\ref{fig:SNR_GAMA_median}. The bottom panel shows the SFH in lookbacktime, as given by the model (in red) and as recovered by VESPA (in black). Further details on the VESPA reconstruction are given in Section~\ref{sec:vespa_on_lgal}. }
\label{fig:examples_BCGnodust}
\end{figure*}

We compute a total star-formation history for the galaxy by adding the star-formation histories of the disc and bulge, and computing a mass-weighted metallicity in each bin of lookback time\footnote{In this instance, the mass-weighted metallicity is computed as $Z_i = (Z_{\rm bulge,i} M_{\rm bulge,i} + Z_{\rm disc,i} M_{\rm disc}) / (M_{\rm bulge,i} + M_{\rm disc,i})$ in each time bin $i$. This should not be confused with a mass-weighted metallicity averaged over the age of the galaxy, used in Section~\ref{sec:vespa_on_lgal}, and defined as ${\rm MWZ} = \sum_i Z_i m_i / \sum_i m_i$}. It is these quantities that we will attempt to recover using VESPA. Fig.~\ref{fig:examples_BCGnodust} shows examples of simulated spectra, with and without noise, for two model galaxies. For each case, we also show the recovered star-formation history obtained with VESPA, which we describe in Section~\ref{sec:vespa_on_lgal}.

\subsubsection{Dust attenuation}\label{sec:dust}

To understand the effects of dust and modelling on the recovered solutions, we construct three distinct sets of spectra that only differ from one another by how the flux is attenuated by dust. The objective of this exercise is to reveal how different assumptions about the dust modelling - or any other modulation of the continuum - affect the ability to recover physical parameters of interest, rather than applying sophisticated dust models. For simplicity, we do not model dust emission (which is unimportant at optical wavelengths) and we assume that stars of all ages see their light attenuated in the same way. Briefly, the constructed sets are:

\begin{enumerate}
\item a set with no dust attenuation applied;
\item a set with a simple mixed-slab model applied to stars of all ages \citep{CharlotFall00}; 
\item a set with a dust attenuation as modelled by L-Galaxies.
\end{enumerate}
We now described the last two sets in more detail.

In (ii), we implement the same dust modelling that is assumed by VESPA. Light from stars of all ages is attenuated by a mixed-slab of absorbers and emitters for optical depths less than unity and a uniform slab of absorbers for larger optical depth values. The mixed-slab attenuates light according to:

\begin{equation}
f_{\rm dust}(\tau_\lambda) = \frac{1}{2\tau_\lambda}[1+(1-\tau_\lambda)\exp(-\tau_\lambda) - \tau^2_\lambda E_1(\tau_\lambda)],
\end{equation}
whereas the uniform-slab of absorbers has the simple form of:

\begin{equation}
f_{\rm dust}(\tau_\lambda) = \exp(-\tau_\lambda)
\end{equation}

In both cases, the wavelength dependence of the optical depth is given by $\tau_\lambda = \tau_{V}(\lambda / 5500 \AA)^{-0.7}$. $\tau_V$ is randomly chosen from a Gaussian distribution with mean of unity and a standard deviation set to 0.5, and the attenuation is applied to both morphological components, i.e., equation (\ref{eq:Lgal_L}) becomes $L_\lambda = f_{\rm dust} (L_{\rm bulge} + L_{\rm disc})$.

In (iii), we follow the prescription given in \cite{Henriques2015} (Section 1.14 of the supplementary material), with the exception that we do not consider separately the extinction of young stars. In brief, the amount of dust attenuation depends on the cold gas mass and the cold gas disk scale-length (with scaling factors to account for metallicity and redshift dependence), and the extinction law is given by \cite{Mathis83}. As detailed in equation (\ref{eq:Lgal_L}), dust attenuation is only applied to stars in the disk. 

Set (iii) provides the best attempt at simulating a complex galaxy, whereas sets (i) and (ii) allow us to assess the overall effect of including dust attenuation and of making different assumptions about the dust model or dust geometry than the one used to construct the spectra.

%

\subsection{Recovering star-formation histories with VESPA}\label{sec:vespa_on_lgal}

VESPA is a non-parametric full-spectral fitting code, which solves for the star-formation history of a galaxy via a regularised matrix inversion. VESPA divides the age of the Universe into 16 bins, logarithmically spaced between 0.002 and 16 Gyrs. It returns the amount of mass formed in each bin along with its metallicity, and up to two dust attenuation values - one that applies to stars of all ages, and one corresponding to a birth cloud component, that applies only to stars younger than 30 Myrs. Full details can be found in \cite{Tojeiro2007,Tojeiro2009}.

\subsection{Robustness of recovered results}


\begin{figure*}
\includegraphics[scale=0.52]{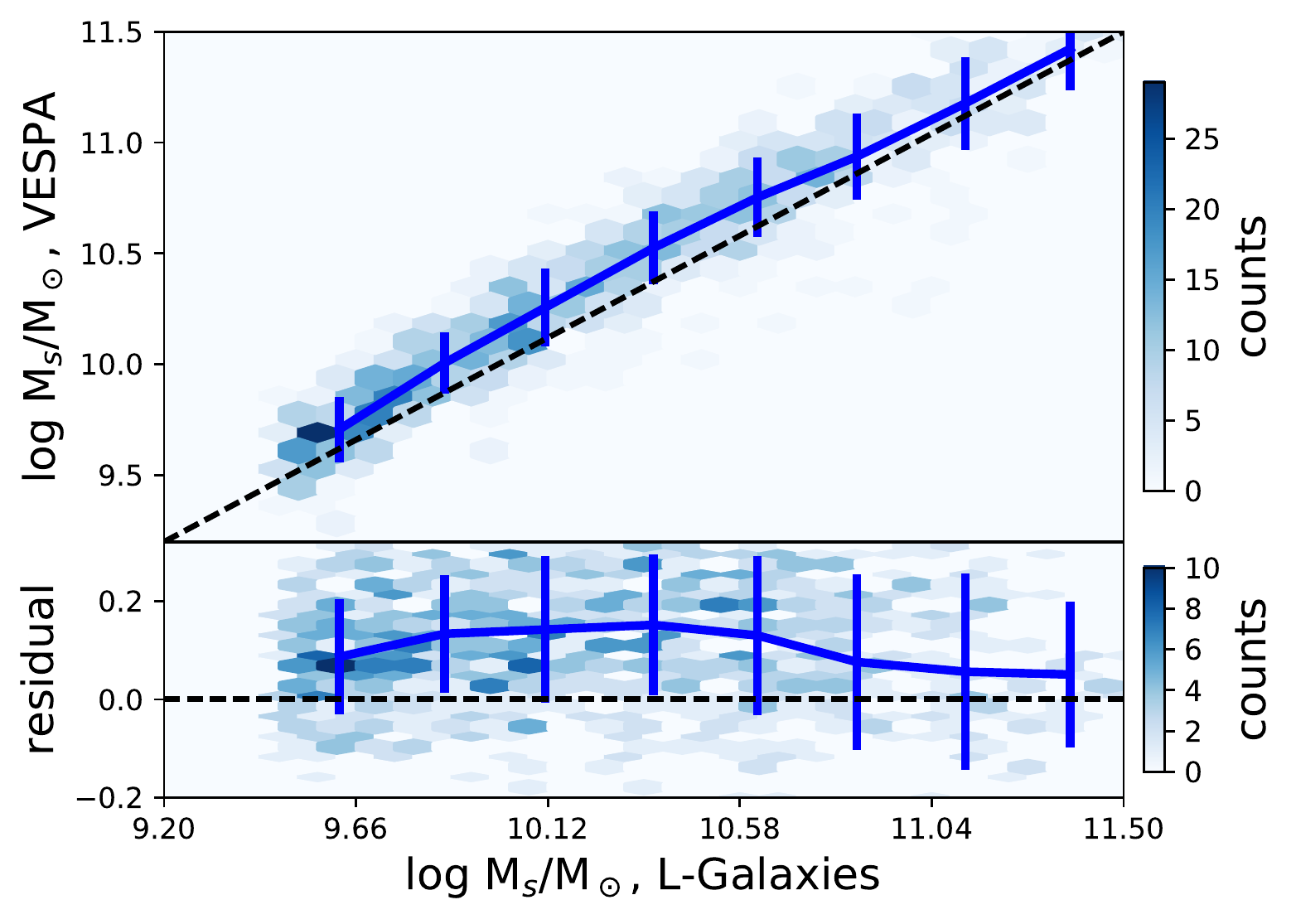}
\includegraphics[scale=0.52]{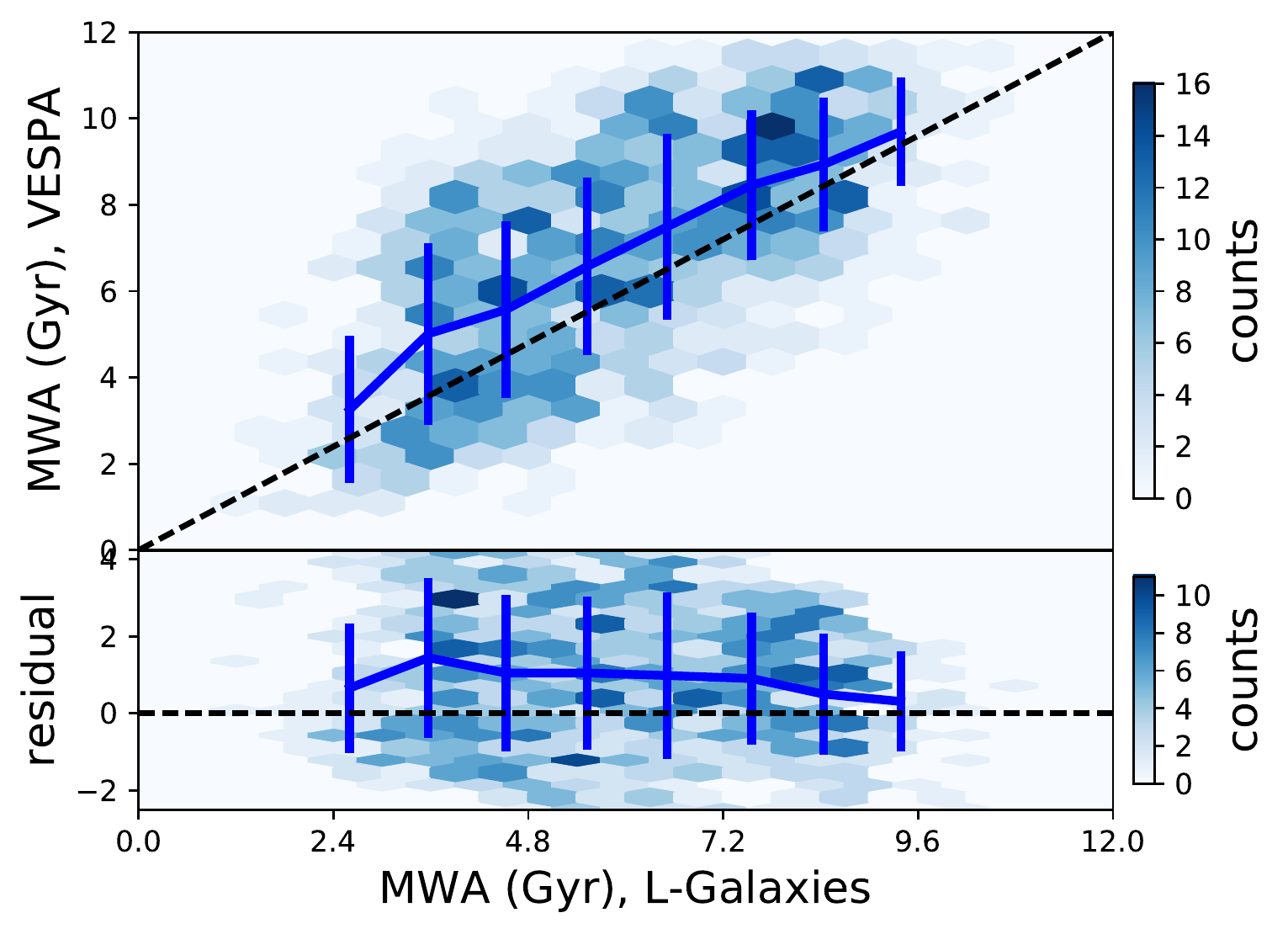}
\includegraphics[scale=0.52]{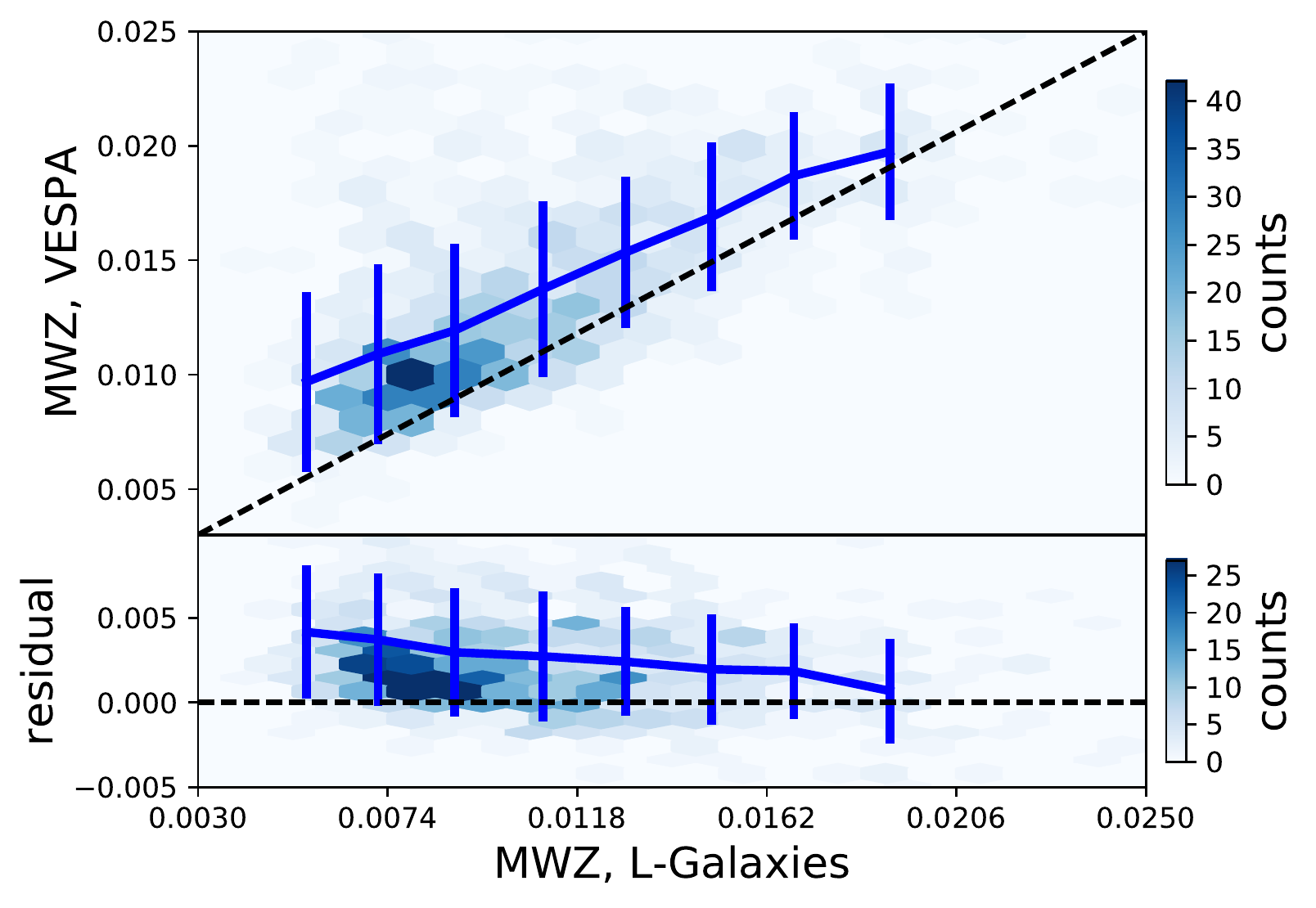}
\includegraphics[scale=0.52]{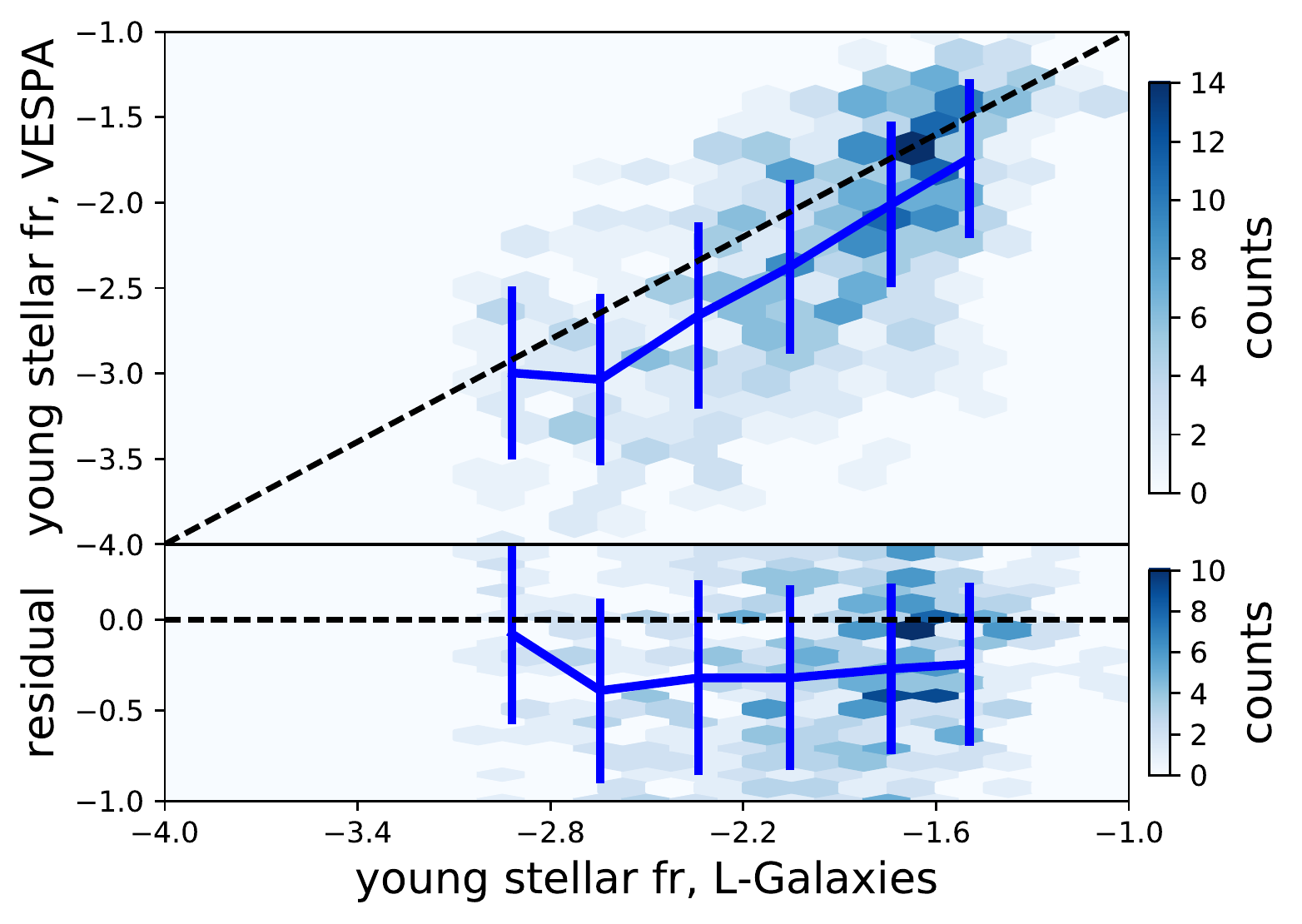}
\caption{Comparison of recovered and input values of the VESPA analysis of L-Galaxiesaxies-GAMA mock spectra: stellar mass (top left), mass-weighted age (top right), mass-weighted metallicity (bottom left) and fraction of young stars (bottom right). In all panels in the hexagonal density histogram shows number counts on a linear scale, and the blue line shows the median in each bin. The error bars show the 67\% and 33\% quantiles, and the dashed lines show the one-to-one relationship or zero residual, as appropriate.}
\label{fig:mocks_dust}
\end{figure*}

To understand the limitations of VESPA in analysing GAMA-like data, we recovered a star-formation history from each of our simulated spectra, from which we computed: mass-weighted age (defined as $\rm{MWA} = \sum_i t_i m_i / \sum_i m_i$), mass-weighted metallicity (defined as $\rm{MWZ} = \sum_i Z_i m_i / \sum_i m_i$), fraction of young stars (defined as fraction of stellar mass in stars  younger than 250 Myrs) and total stellar mass $M_s$. In this section we will focus on the set of mock spectra attenuated by the L-Galaxies dust model (case (iii) in Section~\ref{sec:dust}), since that set represents our best attempt at modelling realistic galaxies. We will refer to results using mock spectra without any dust attenuation or with a simpler dust model applied (cases (i) and (ii) in Section~\ref{sec:dust}), but present the plots from those runs in the discussion presented in Appendix \ref{sec:appendix_dust}.

For each parameter, we will be particularly interested in biases that are dependent on that parameter. Our goal is to interpret observed {\it differences in the mean} of these physical parameters in data subsets. These tests give us a handle on potential systematics due to the spectral analysis and a minimum error for the mean. Constant offsets are therefore inconsequential, but trends on the residuals are of concern. We select a sample of 1000 galaxies from the L-Galaxies simulation, with $\log_{10}(M_*) > 9.5$, and we construct simulated spectra for each galaxy according to the methodology detailed in Section~\ref{sec:simulated_spectra}. Fig.~\ref{fig:mocks_dust} shows the recovered parameters. 

Stellar mass is unsurprisingly very well recovered, showing a nearly constant offset of 0.1 dex. This offset is driven entirely by a mismatch in the dust modelling and/or geometry. This can be seen by inspecting the the equivalent panels in Figs.~\ref{fig:mocks_nodust} and ~\ref{fig:mocks_dustsimple}, where dust is either not included, or where the dust model assumed by VESPA is the same used to attenuate the mock galaxies. In these cases, there is only a negligible bias ($<0.01$ dex) in the case where the dust model is known exactly, and no bias in the case of no dust attenuation.

The mass-weighted age is recovered with a large scatter (around 1.7 Gyrs) and a nearly constant offset of 1 Gyr. If we consider a subsample of 200 mock galaxies (which is below the typical sample size used in Section~\ref{sec:application}) with a given mass-weighted age, the expected error on the mean of the recovered mass-weighted age for that sample is therefore less than 0.15 Gyrs. When analysing real data, we expect the achieved error on the mean to be larger than this estimate: the subsamples from which we will compute a mean weighted-age (defined in bins of halo mass and GE) will have an intrinsic distribution of ages of some typical width. The observed scatter of mass-weighted age in each subsample will be the addition in quadrature of the intrinsic width in mass-weighted ages with the scatter introduced by noise and other aspects of the spectral analysis - we are only accounting for the latter in this section (and only as a lower limit). Nonetheless, this exercise on mock spectra gives us an upper limit on the precision we might achieve using VESPA. Interestingly, with perfect knowledge of the dust model and geometry, this upper limit on precision would change only modestly by about 10\% (see Figs.~\ref{fig:mocks_nodust} and ~\ref{fig:mocks_dustsimple}). This is because imperfections on large-scale modulations on the spectra are preferentially picked up by the dust modeling, leaving only relatively small changes one the mean recovered star-formation histories. The mean bias, however, would be reduced by roughly a factor of two.

In Section 2 we showed that $t_{50}$ is at least as good an estimator as mass-weighted age. However, $t_{50}$ is poorly recovered from the mocks due to the lack of time-resolution in the VESPA age grid: many galaxies form over 50\% of their mass in the first bin. We found that $t_{85}$ is much better recovered. The results for $t_{85}$ follow the results of mass-weighted age very closely and are not shown here.

Mass-weighted metallicity shows a slight metallicity-dependent offset, with larger residuals at lower metallicities. The mass-weighted metallicity is always over-estimated by VESPA, even in the simplest case of no dust being added (see Fig.~\ref{fig:mocks_nodust}) - although in that case the tilt in the residual disappears. It is unclear why this is the case, but it does cast doubt on any metallicity measurements we obtain with VESPA on this dataset. We will show mass-weighted metallicities in the next section, given that as we will see they fall in line with expectations, but will not consider them in our final conclusions.

Similarly, the fraction of mass in stars younger than 250 Myrs is always under-estimated by VESPA, with an offset that is constant on all cases. Of all the galaxy properties we consider, this is the estimator with the largest fractional scatter. This is expected, given the lack of blue coverage in the GAMA spectrograph, the low S/N in the blue, and the lack of any modeling of emission lines. However, the results from the mocks indicate that it should remain an unbiased estimator when looking at differences in the mean. 

\subsection{Summary}

In this section we described the construction of spectral mocks of model galaxies, using the output of the L-Galaxies semi-analytic model. These galaxies have complex SFHs, and are a significant step beyond the simple mocks often used to assess the reliability of fitting codes. A thorough exploration of how several issues affect the recovery of physical parameters using similar mocks will be presented in a separate paper. Here we mainly focused on a set of mocks that roughly mimic GAMA spectra in terms of S/N and spectral range.

We found that a lack of knowledge of the real dust law imparts biases on the recovered physical  parameters. In principle, one can use these mocks to correct for biases in the measured parameters. However, with the exception of mass-weighted metallicity, we found these biases to be constant offsets of no consequence provided we focus on differences on mean quantities, which we will do in the next section. For simplicity, we therefore do not apply any corrections to our measurements. 

\section{ Application to GAMA}\label{sec:application}

\subsection{Sample definition and mass estimates}\label{sec:GAMA}

We use data from the GAMA survey \citep{Driver2009,Driver2011,Hopkins2013,Liske2015}, a spectroscopic survey of over 230,000 SDSS-selected galaxies over 230 deg$^2$ of sky. Galaxies were targeted to a petrosian magnitude limit of $r_{\rm pet} \lesssim 19.8$, with an impressive 98\% spectroscopic completeness, and yielding a median redshift $z\approx 0.2$. It is the unique combination of cosmologically useful volume with a high number-density of tracers and spectroscopic information that make GAMA a powerful dataset in order to explore the connection between geometric environment and halo formation time that we wish to do here.
 
\subsubsection{Group catalogue and halo masses}\label{sec:halo_masses}
\cite{Robotham2011} used a friends-of-friends algorithm to create a GAMA group catalogue (version G3Cv06 is used), from which we select all grouped central galaxies in $0.04 < z < 0.263$; this redshift cut is necessary to ensure the robustness of the geometric environment classifications. We take a central galaxy as being the brightest in the group (BCG), but note that our results do not change significantly if instead we choose one of the alternative ways to identify the central galaxy within a group offered by the catalogue (see \citealt{Robotham2011} for details). Without any cut on group multiplicity this sample consists of 13047 galaxies (we study the effect of group multiplicity in our results in Section~\ref{sec:results}). Of these, 4547 galaxies had been observed by SDSS and were not re-observed by GAMA. Therefore around one third of our galaxies have SDSS spectra. 

\cite{HanEtAl15} then used a maximum likelihood lensing analysis to investigate the scaling of halo mass with group properties. They consider power-law combinations of six physical observables to find the best group mass estimator when matching to the weak lensing masses. 
We use M$_h$ = M$_p$ (L$_B$/L$_0)^\alpha$, where L$_B$ is the total group luminosity, corrected by a factor B in order to obtain an unbiased median luminosity in the $r-$band (as required in \citealt{Robotham2011}).  $(\log_{10}($M$_p), $L$_0, \alpha) = (13.40 \pm 0.12, 2 \times 10^{11}h^{2}$L$_\odot, 1.09 \pm 0.22$) are the fitted parameters to weak lensing masses \citep{HanEtAl15}, with a correlation factor between M$_p$ and $\alpha$ of $-0.13$. In Section~\ref{sec:results_env} we will explicitly test the effect of the error in this calibration on our main results.

\subsubsection{Stellar masses}\label{sec:stellar_masses}

We consider two estimates of stellar masses: photometric masses, from \cite{Taylor2011}, and spectroscopic masses from our VESPA analysis of GAMA data. The photometric masses are estimated from broadband optical photometry, using libraries of parametric star-formation histories computed using the BC03 models. \cite{Taylor2011} use SExtractor AUTO aperture photometry on SDSS imaging, and scale the stellar masses to the $r-$band sersic magnitude in order to account for mass beyond the AUTO aperture. We scale the spectroscopic stellar masses to the same sersic magnitudes. This puts GAMA and SDSS spectra on the same magnitude scale, and allows spectroscopic estimates to be directly comparable to photometric estimates.

\begin{figure}
\includegraphics[scale=0.55]{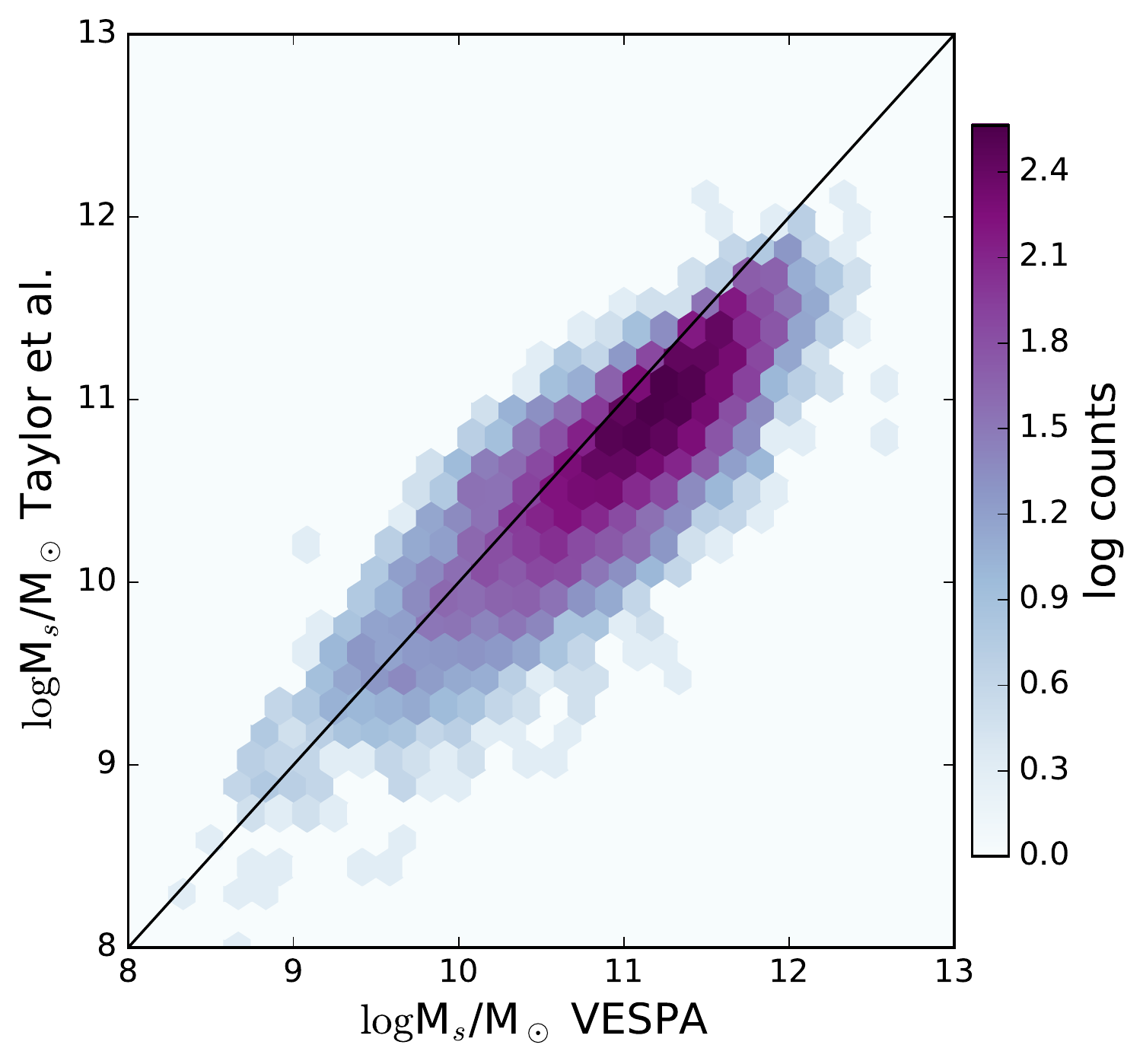}
\caption{A comparison of stellar mass estimates. The VESPA masses were computed using GAMA and SDSS spectra and full spectral fitting, and the Taylor et al. masses were computed using SDSS photometry and a library of star-formation histories. Both stellar mass estimates were computed using BC03 models.}
\label{fig:Mstar_comp}
\end{figure}

Fig.~\ref{fig:Mstar_comp} shows a comparison between the two estimates of stellar mass; the spectroscopic masses shown were computed with the BC03 models for consistency (and include GAMA and SDSS spectra). The two estimates show good agreement, with a small median offset of 0.2 dex. The large scatter between the two is primarily driven by the large uncertainties in the estimates of stellar mass from individual spectra, which in turn derive primarily from the low S/N. We analysed all spectra with two sets of stellar population synthesis codes: BC03 and FSPS (see Section~\ref{sec:implementation}). A comparison of the photometric masses with FSPS spectroscopic stellar masses is not shown - it presents a similar scatter but a larger median offset of 0.55 due to the differing SPS models. 

\begin{figure}
\includegraphics[scale=0.55]{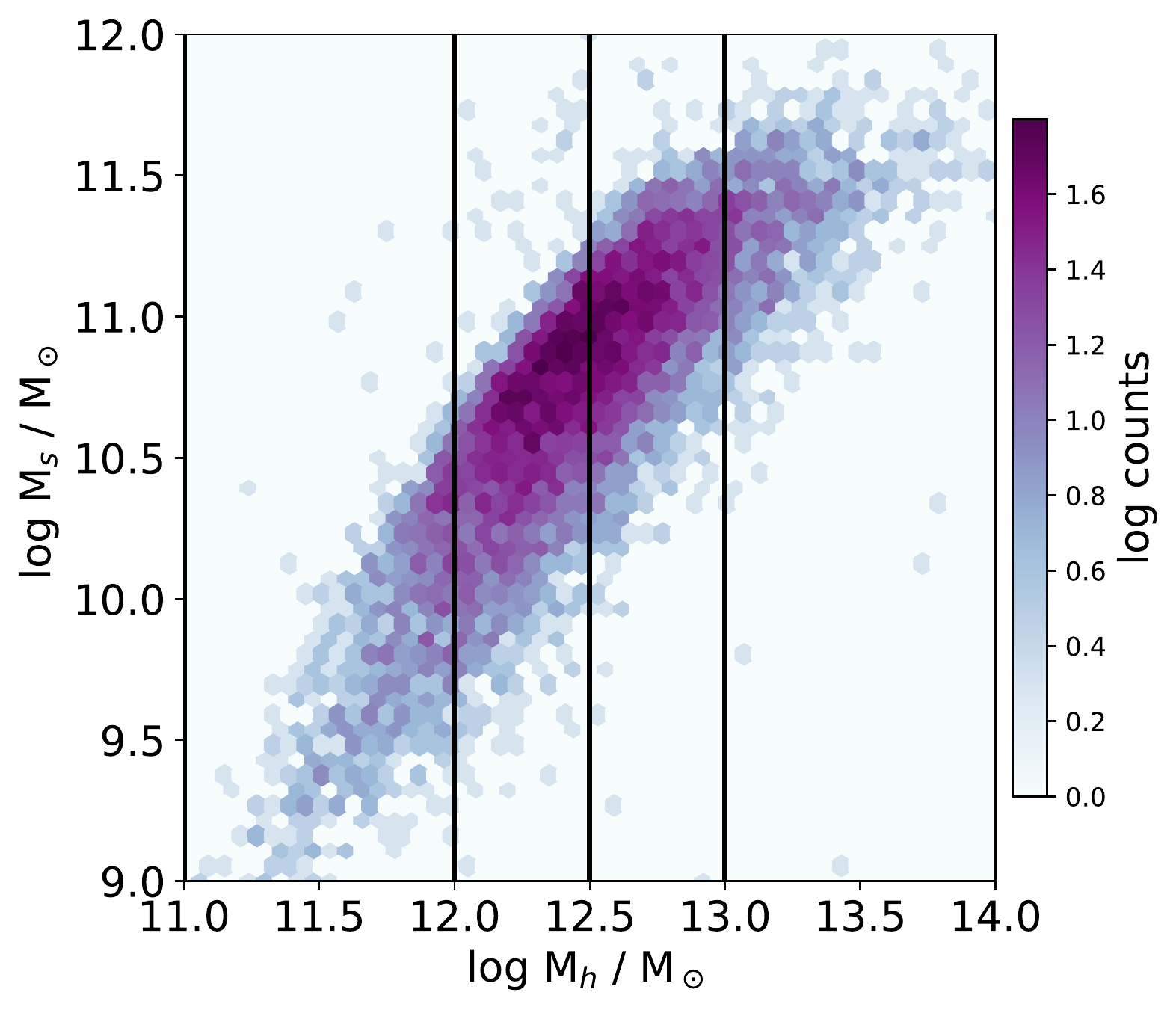}
\caption{The stellar mass to halo mass relation of the galaxies used in this study. The hexagonal density histogram shows number counts on a log scale. The edges of the four chosen bins in halo mass are represented by the vertical black lines.}
\label{fig:MsMh_data}
\end{figure}

Fig.~\ref{fig:MsMh_data} shows the halo mass - stellar mass relation of the central galaxies used in this paper; we show stellar masses computed from the photometry. We divide the sample into four bins of halo mass,  represented in Fig.~\ref{fig:MsMh_data} by the vertical black lines. The bins were chosen to keep a reasonable minimum number in each halo mass and GE bin, and  Table~\ref{tab:sample} shows the sample size each of these bins (GE classification is detailed in the following section). 

\begin{table}
\begin{tabular}{|l|c|c|c|c|c|c|}
\hline
 			& Bin 1 & Bin 2 & Bin 3 & Bin 4  & Total \\ \hline \hline
Voids 		& 257	& 703	& 388	& 25 	& 1373 \\
Sheets 		& 429	& 1700	& 1337	& 199	& 3665 \\
Filaments 	& 550	& 1904	& 1789	& 566	& 4809 \\
Knots 		& 217	& 577	& 522	& 306	& 1622 \\
Total		& 1453	& 4884	& 4036	& 1096	& 11469 \\ \hline
\end{tabular}
\caption{Sample size, as a function of geometric environment and bin of halo mass. As described in Section~\ref{sec:GAMA}, these numbers correspond to BCG-central galaxies in $0.04 < z < 0.263$, in groups with multiplicity greater than or equal to 2. This table does not include 68 galaxies for which the spectral analysis failed due to poor data, or galaxies beyond the lowest and highest halo mass bin boundary ($10^{11}$ and $10^{14}$ respectively).}
\label{tab:sample}
\end{table}

\subsection{Geometric environment classifications}

We use the geometric environment classifications of \cite{Eardley2015}, who compute an estimate of the tidal tensor from a smoothed galaxy density field in order to determine the dimensionality of collapse of any given region. This is done by computing the number of eigenvalues above a given threshold: a region is classified as a void if all eigenvalues are below the threshold (no collapse), a sheet if one eigenvalue is above the threshold (collapse in one direction), a filament if two eigenvalues are above threshold (collapse in two dimensions) and a knot if all eigenvalues are above the threshold (collapse in all dimensions). \cite{Eardley2015} compute geometric environment classifications using two combinations of density field smoothing length $\sigma$ and threshold value. Here we show results obtained by smoothing the density field with a 4 $h^{-1}$ Mpc smoothing scale, and a threshold value for the eigenvalues of 0.4, but note that using the alternative (10 $h^{-1}$ Mpc, 0.1) combination yields the same conclusions.

\begin{figure}
\includegraphics[scale=0.22]{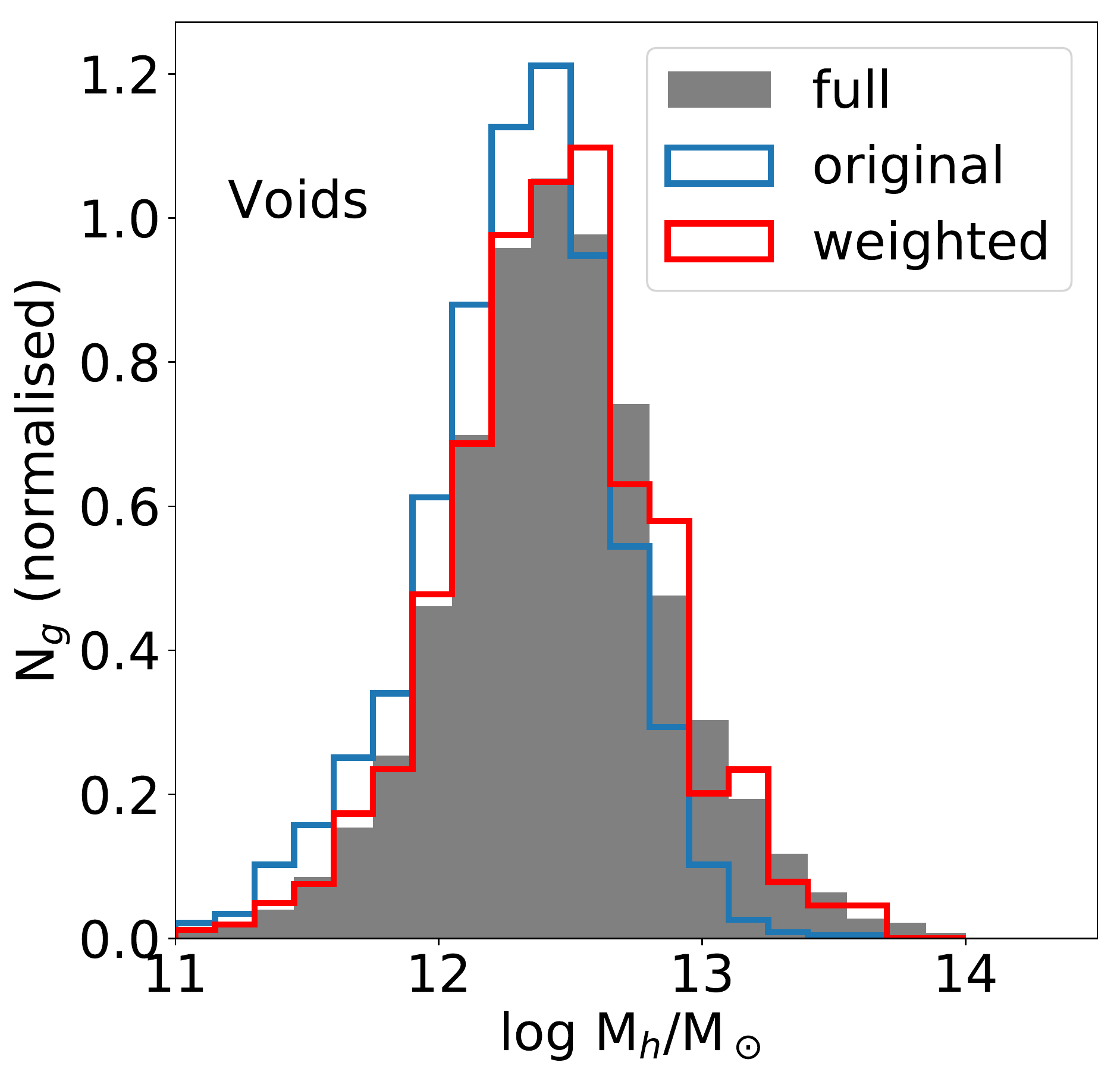}
\includegraphics[scale=0.22]{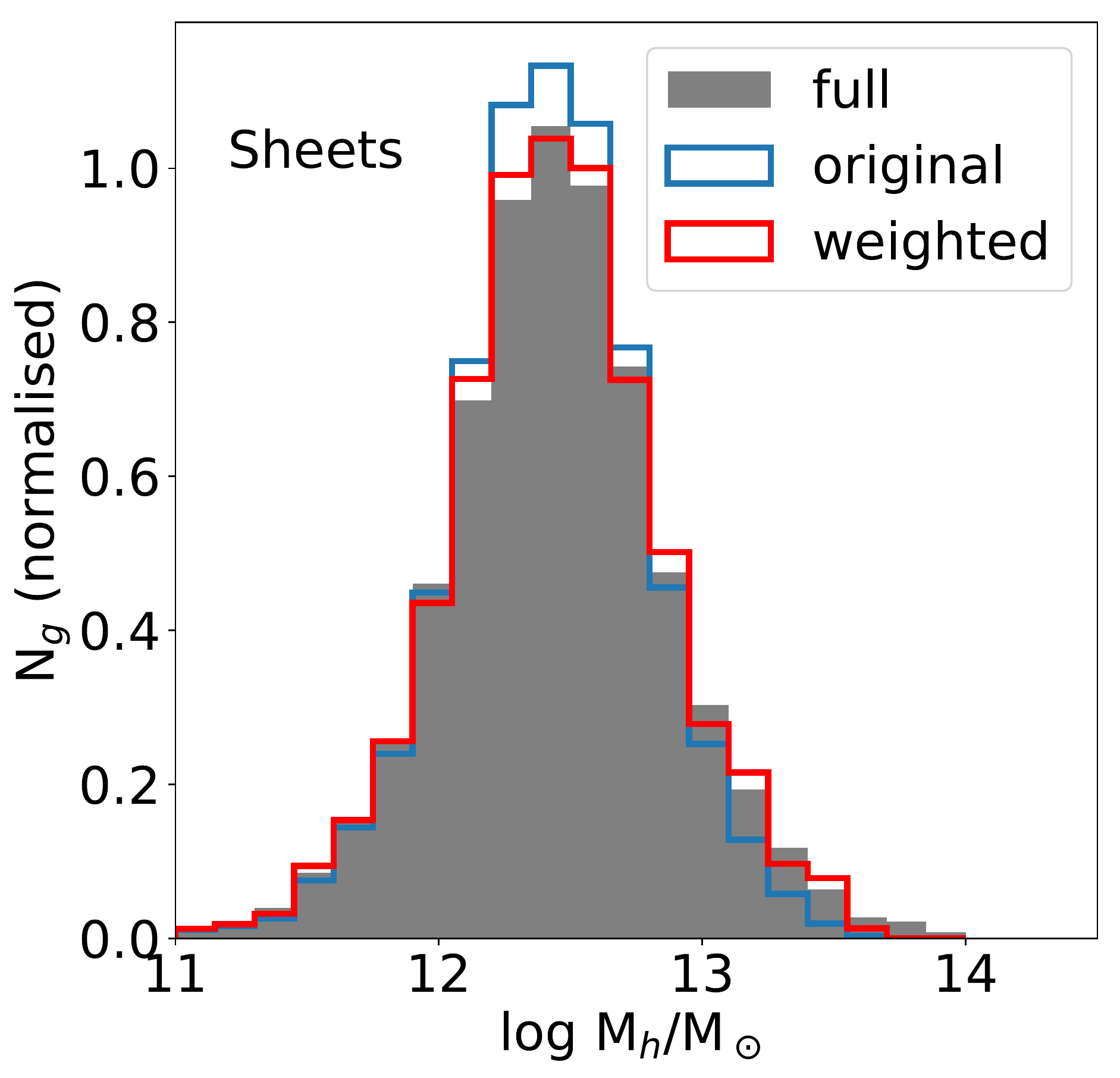}
\includegraphics[scale=0.22]{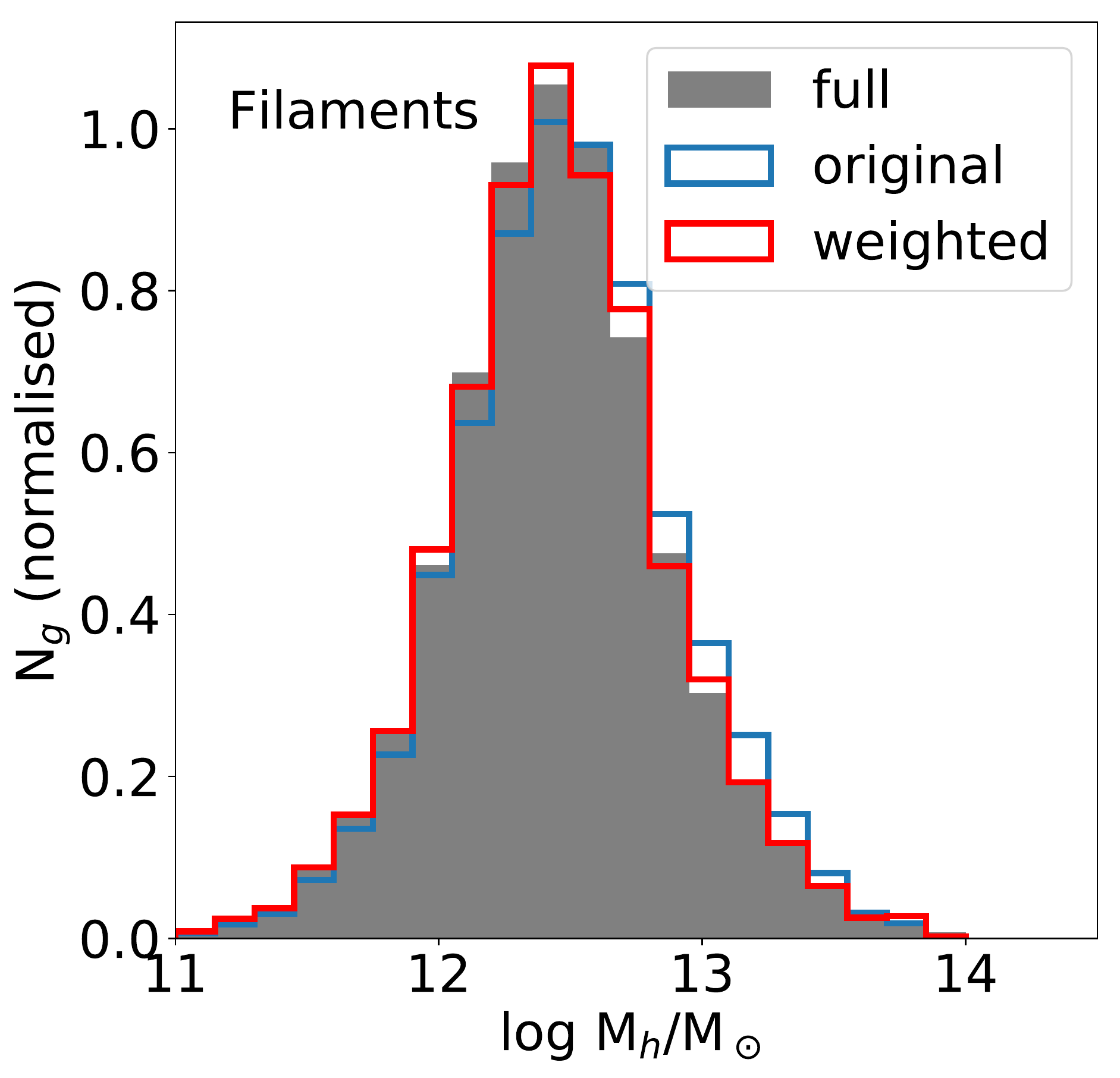}
\includegraphics[scale=0.22]{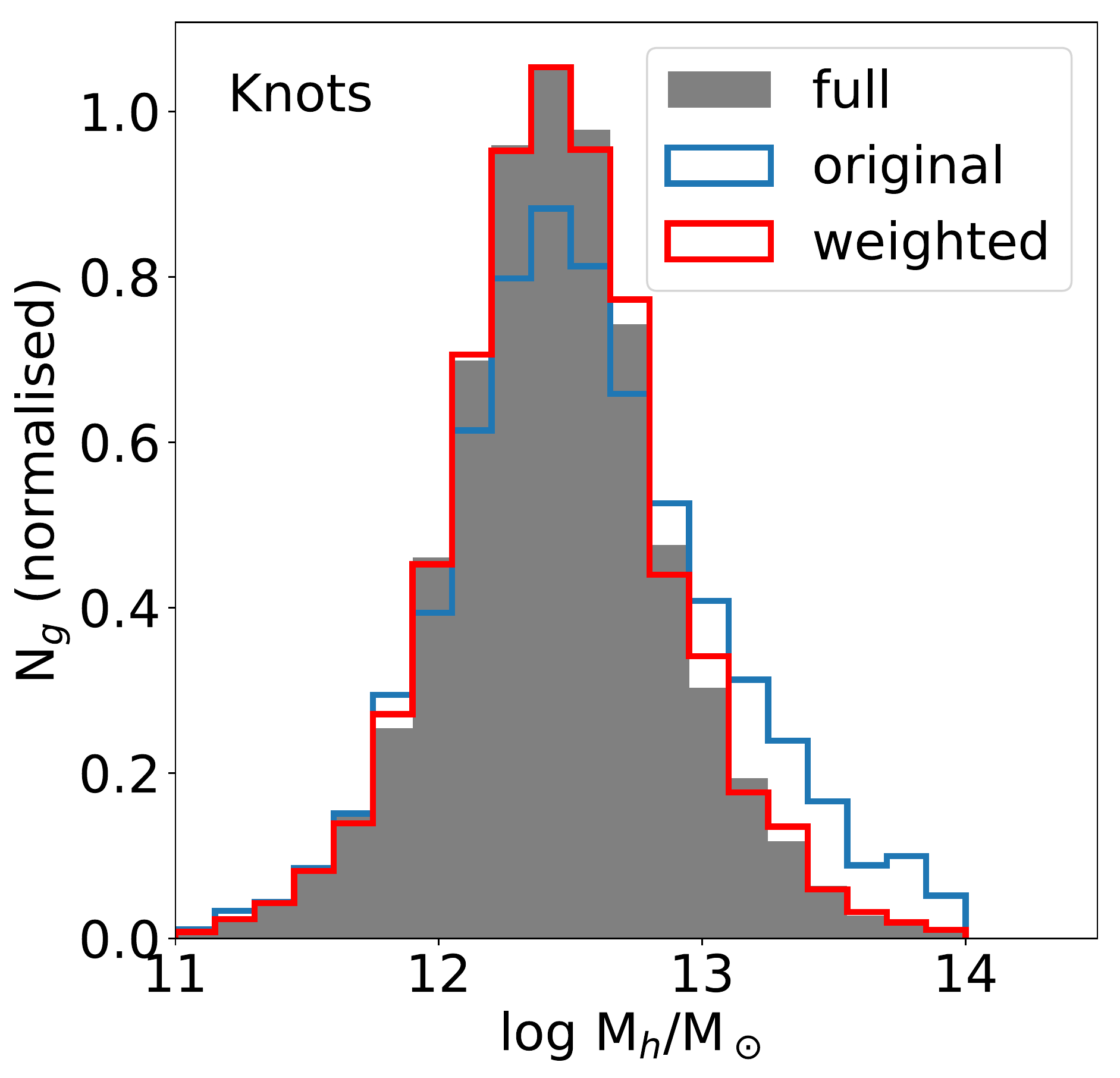}
\caption{Weighting the galaxies in each GE such that their weighted halo mass distributions are matched across all environments. In each panel (one for each GE, as indicated), we show the distribution of halo masses in the full sample (grey), original halo mass distribution for each GE (blue), and the weighted halo mass distribution (red). This scheme relatively downweights high-mass haloes in denser GE environments with respect to lower mass haloes, with the reverse happening in under-dense GE environments.}
\label{fig:weights}
\end{figure}

We expect local density - and therefore halo mass - to be correlated with geometric environment (e.g. \citealt{AlonsoEtAl14,Eardley2015}), and for the distribution of halo masses to be skewed towards high masses in knots and towards low masses in voids. Due to the small size of our sample, we need to work on large bins of halo mass, where we expect a residual effect from this dependence. We therefore compute weights for each galaxy, based on their GE classification, such that the weighted halo mass distributions in each environment match the full sample halo mass distribution. Fig.~\ref{fig:weights} shows the initial and weighted halo mass distributions for the 4 GE environments considered: as expected, we downweight high-mass haloes in denser GE environments relatively to lower mass haloes in the same GE; the reverse happens in under-dense GE environments. The amplitude of this effect is small compared to the theoretical expectation (see \citealt{AlonsoEtAl14}). This is explained by the effective low halo mass cut-off due to limitations in group-finding and by the small statistics at M$_h > 10^{14}$M$_\odot$, where the effect is most pronounced. Our weighting scheme is poor in terms of shot noise: e.g., the properties of galaxies in poorly populated regions of parameter space (such as galaxies living in high-mass haloes in voids) are significantly up-weighted. However, we will see later that with the exception of high-mass haloes in voids, the effect of these weights on mean properties is small.

\begin{figure}
\includegraphics[scale=0.44]{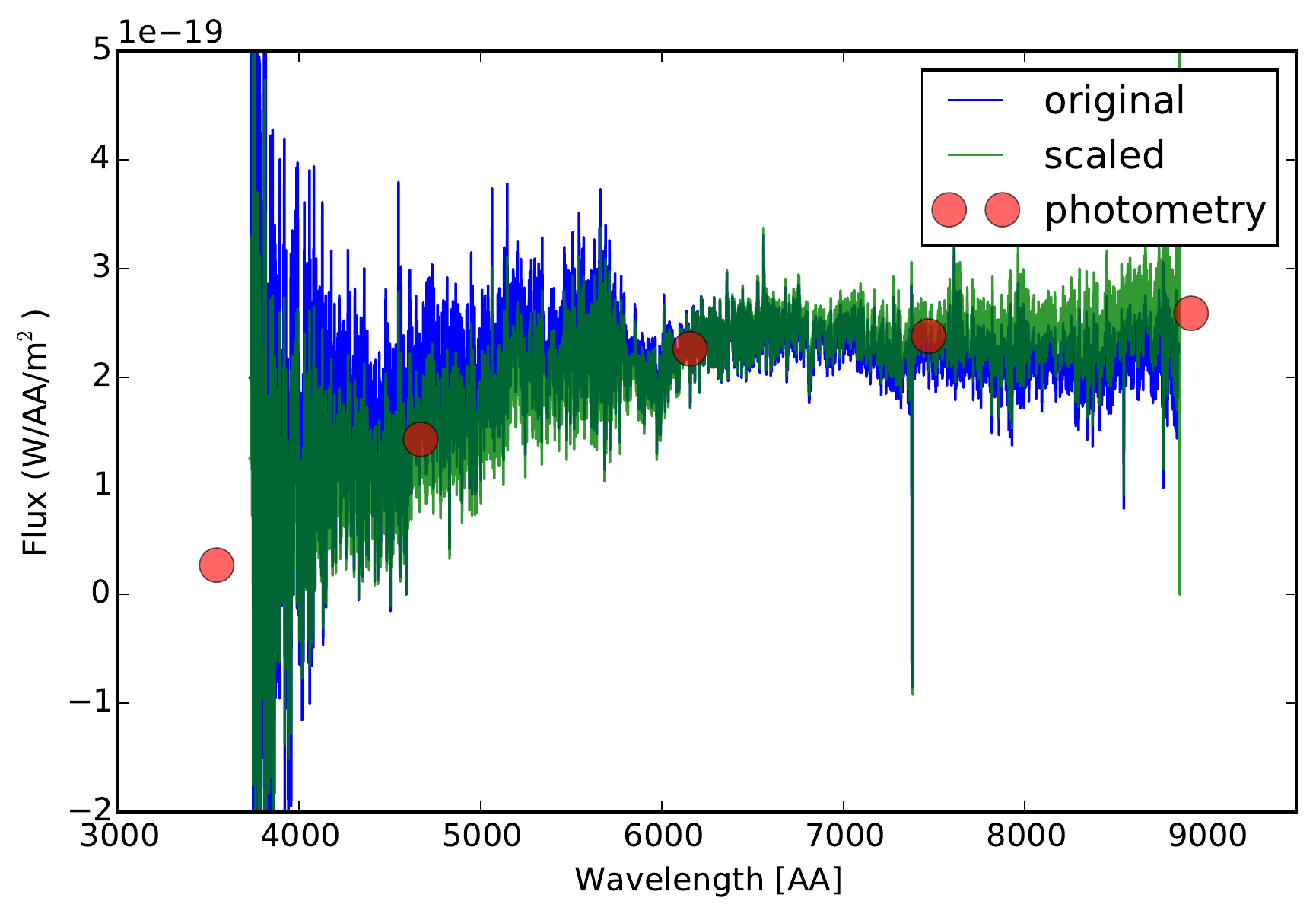}
\includegraphics[scale=0.44]{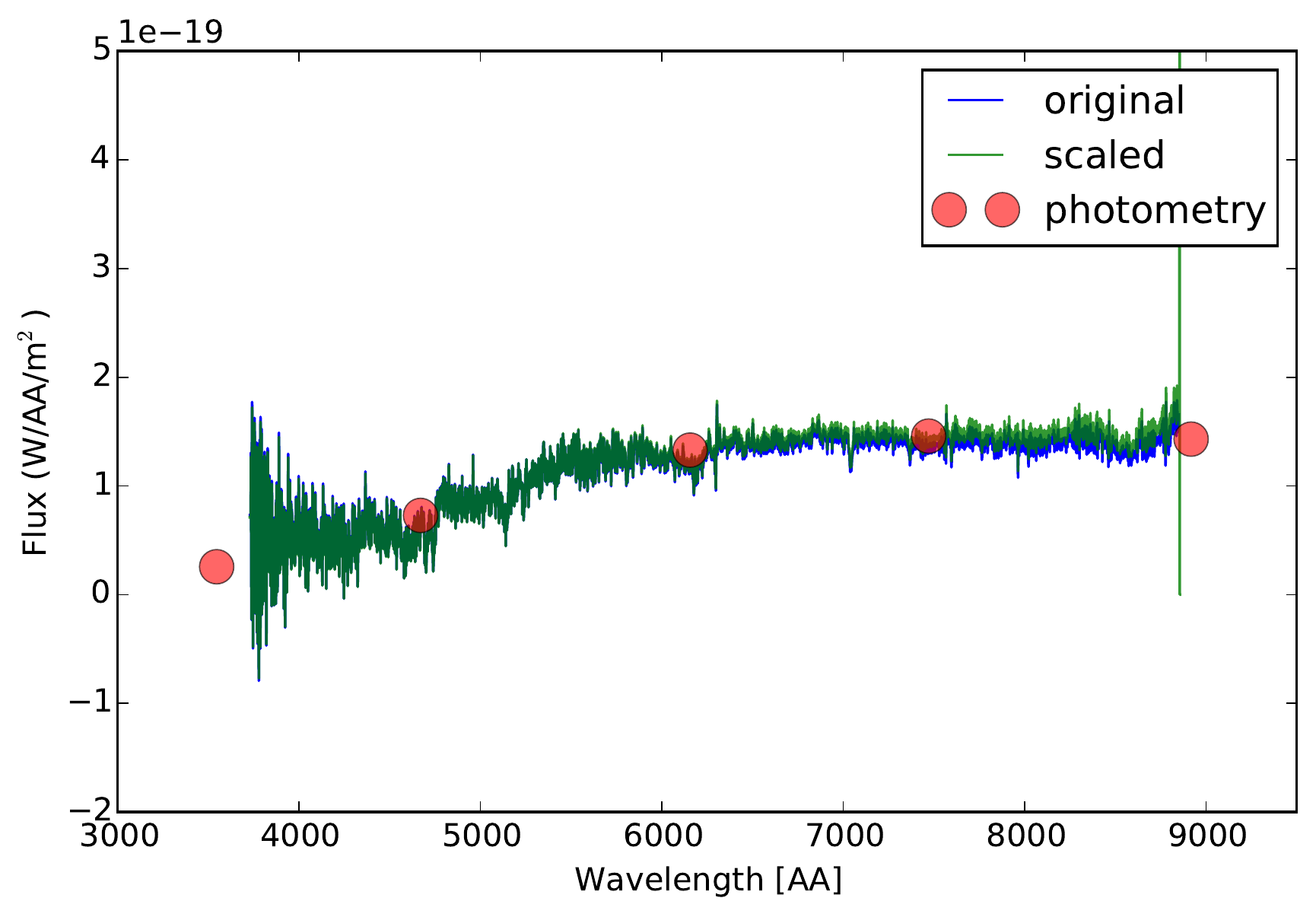}
\includegraphics[scale=0.44]{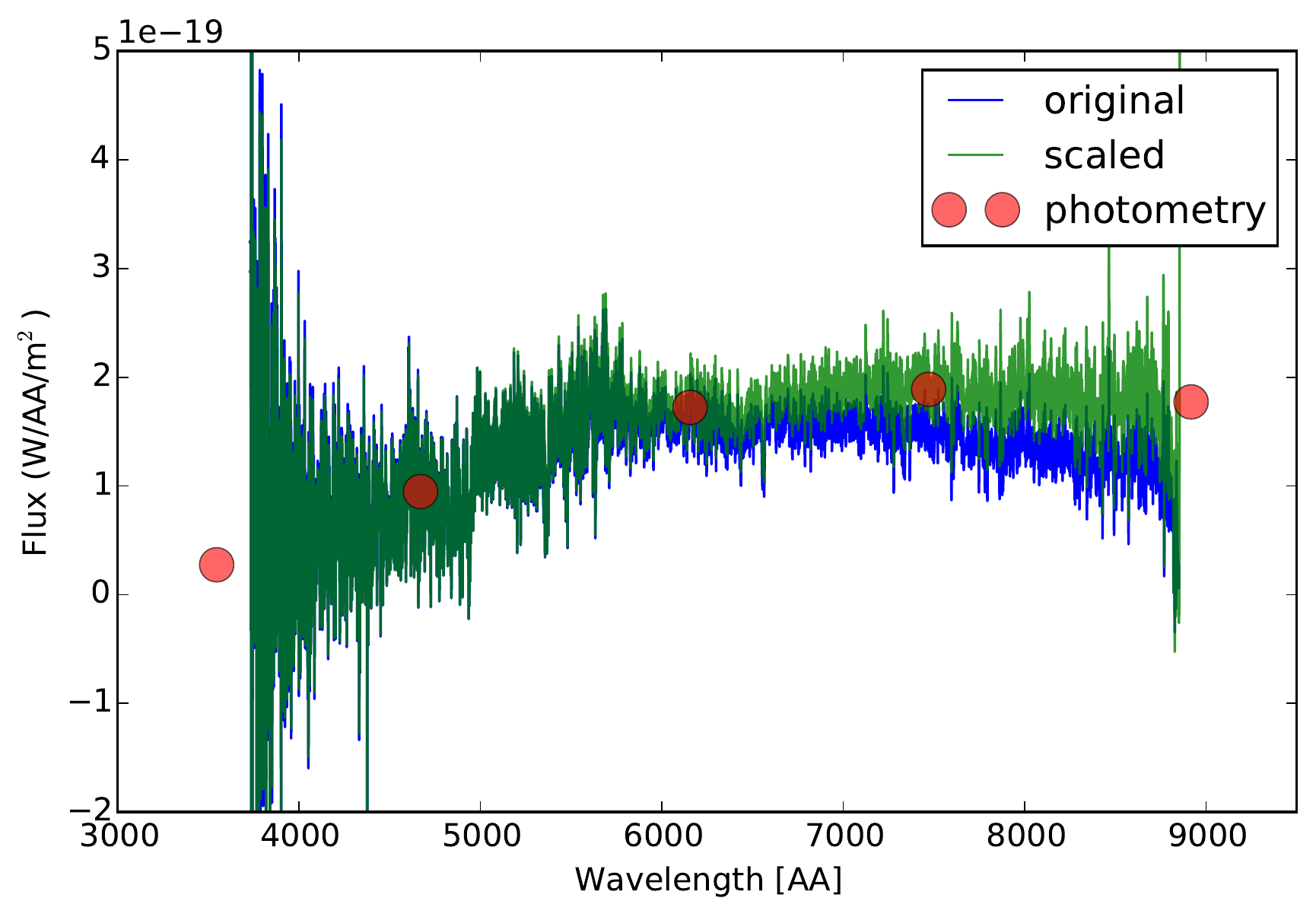}
\caption{Scaling GAMA spectra to SDSS $g-r-i$ bands, according to the methodology detailed in Section~\ref{sec:scaled_spectra}. In each panel the blue line shows the original spectrum, the green line the scaled spectrum, and the red circles show the SDSS aperture-matched Petrosian magnitudes. Note that although we show $u-g-r-i-z$ photometry, only $g-r-i$ were used in our scaling procedure.}
\label{fig:ScaleSpectra}
\end{figure}

\begin{figure*}
\includegraphics[scale=0.38]{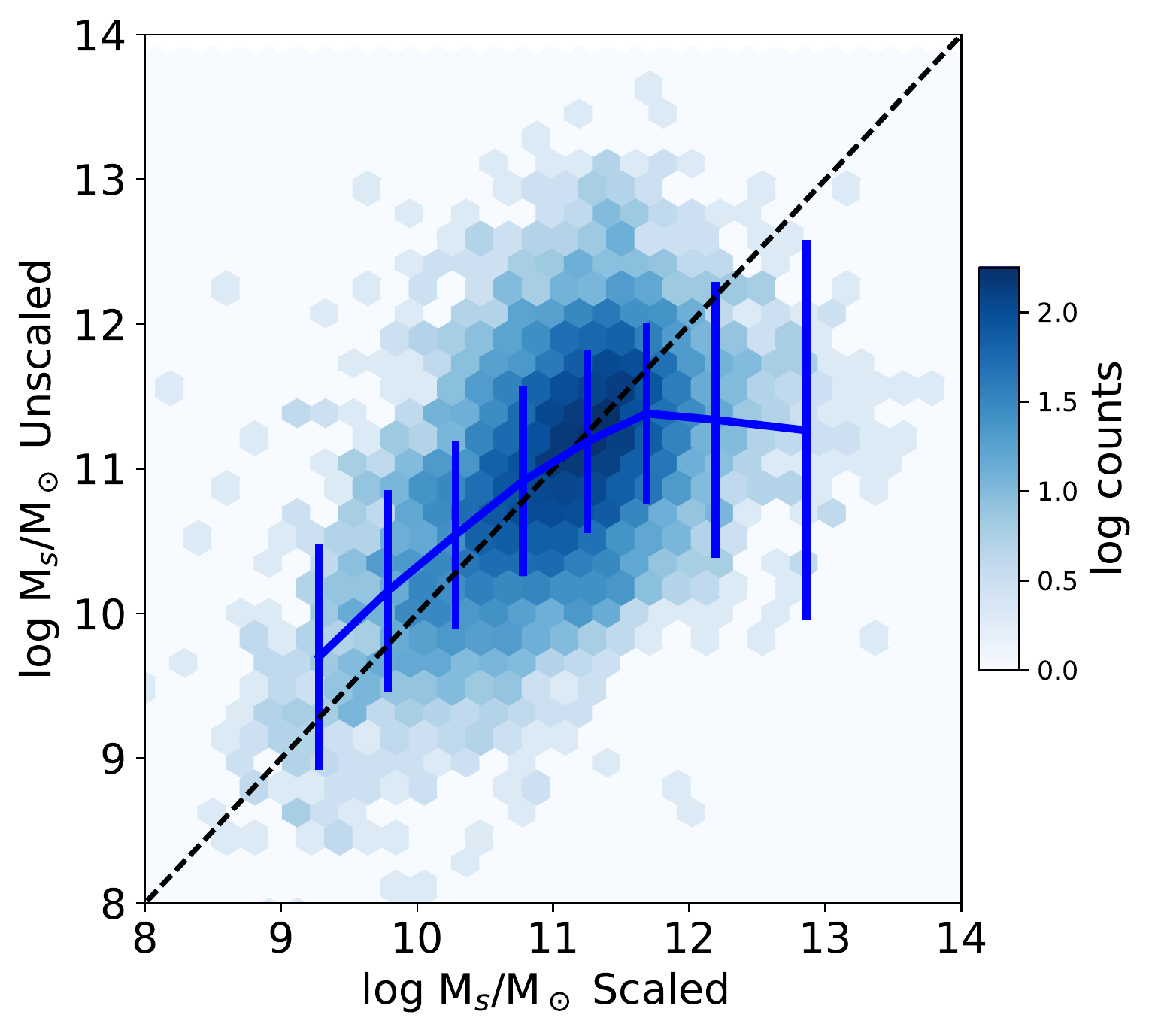}
\includegraphics[scale=0.38]{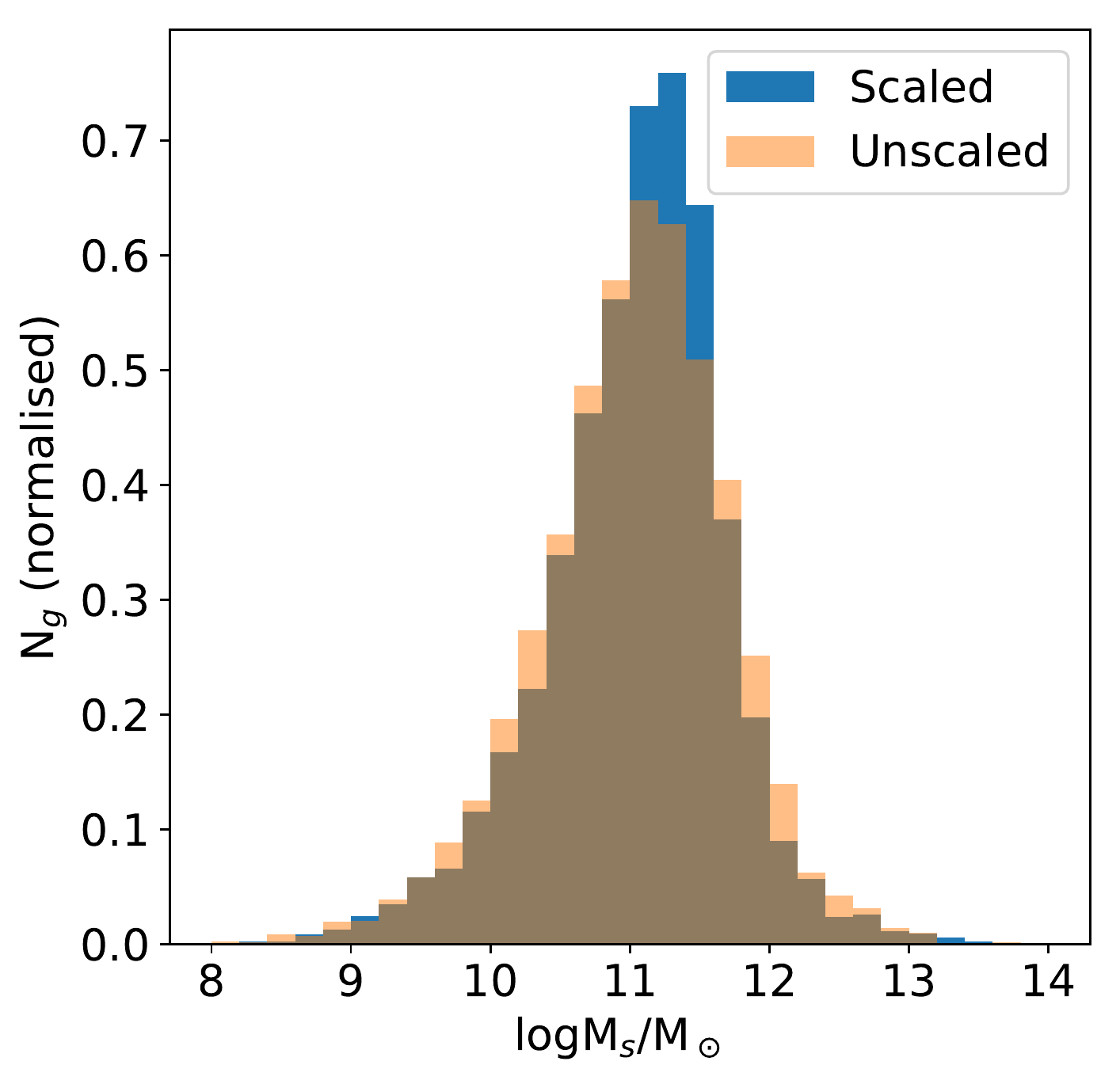}
\includegraphics[scale=0.38]{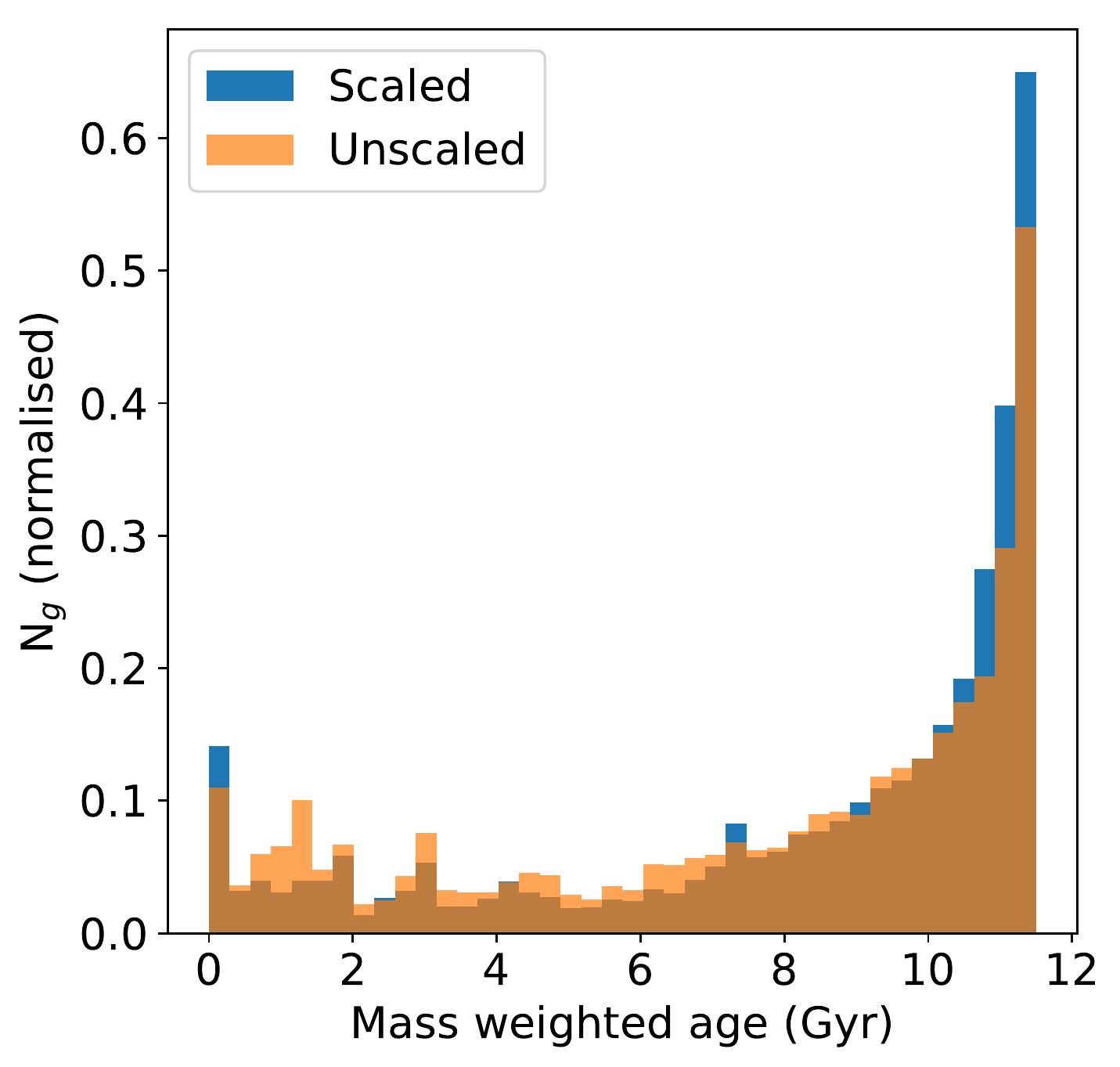}
\caption{The effect of scaling the GAMA spectra according to the methodology detailed in Section~\ref{sec:scaled_spectra}. {\it Left panel}: the direct comparison on stellar masses, on a galaxy-by-galaxy basis, as a two-dimensional histogram that shows number counts on a logarithmic scale. The solid blue line shows the mean trend and the error bars represent the standard deviation in each bin of stellar mass. There is a large scatter between the two estimates of stellar mass, typically around 0.65 dex.  {\it Middle panel}: the distribution of stellar masses recovered using scaled and unscaled spectra as labelled. Here it is clear that the unscaled case has a larger scatter, but the mean of the distribution differs only by 0.02 dex. {\it Right panel:} the distribution of mass-weighted ages computed from scaled and unscaled spectra, as labelled. Unscaled spectra yields on average younger galaxies, by approximately 0.6 Gyrs. Although not explicitly shown here, the scatter on a galaxy-by-galaxy basis is very large, of the order of 2 Gyrs. Whereas the effects of the spectrophotometric scaling are important on any individual galaxy, the total effect on the mean population is significantly more gentle.}
\label{fig:ScaleComp}
\end{figure*}

\subsection{VESPA analysis}\label{sec:vespa_on_gama}

\subsubsection{Scaled spectra}\label{sec:scaled_spectra}

The spectrophotometric calibration in GAMA spectra is only accurate to 10-20\% \citep{Hopkins2013}. Such large modulations to the continuum are potentially problematic for full-spectral fitting techniques, as they can introduce biases in the recovered parameters. Note that we are not concerned about an overall normalisation (this is set by the photometric scale chosen to match that adopted for the photometric masses - see section~\ref{sec:stellar_masses}), but rather with changes to the spectral continuum introduced by the observations and pipeline. From an analysis point of view, such a concern may be tackled by either removing the continuum altogether (e.g. Swann et al., in preparation) or by allowing some nuisance modulation of the large-scale modes in the spectral fitting. Some of the latter is unavoidably done by VESPA via the dust fitting; so errors in the continuum primarily result in biased dust parameters. However, VESPA does not presently allow for a purpose-built nuisance large-scale modulation to correct for potential errors in spectrophotometric calibrations.

In order to assess the magnitude of the resulting effect, and partially correct for it, we re-scale the GAMA spectra to SDSS photometry within the $g-r-i$ photometric bands. The GAMA spectra cover these three bands, and it is desirable that the integrated spectra across the wavelengths of each individual band matches the photometric flux. This was done using extinction-corrected aperture-matched Petrosian magnitudes from SDSS. We interpolated the SDSS filter transmission functions to the wavelengths of the spectra to provide weightings for the integration, allowing for an estimate of the flux that would contribute to the photometry of each band. Comparison with a ‘standard spectrum’ of constant flux whose expected magnitude can be calculated allows for an estimate of the g, r and i magnitudes of each spectrum. The difference between these ‘spectro-magnitudes’ and the SDSS magnitudes provides an estimate for the flux calibration error at the effective wavelength of each filter. We implemented a linear interpolation in two regimes, across the blue and red side of the $r-$band effective wavelength individually, to provide an estimated magnitude difference for each wavelength of the spectrum, $\Delta M(\lambda)$. At each wavelength, we scaled the flux by the appropriate linearly-interpolated value, $k(\lambda)$, once the magnitude-quantity had been converted to a flux-quantity using $k(\lambda) = F_{\rm scaled}(\lambda)/F_{\rm original}(\lambda) = 10^{0.4}\times \Delta M(\lambda)$. We repeated this process three times for each spectrum, where each iteration used the scaled fluxes to calculate a new spectro-magnitude. Fig.~\ref{fig:ScaleSpectra} shows three specific examples of the application of this technique. 

Fig.~\ref{fig:ScaleComp} shows a summary of the effects of scaling the GAMA spectra as detailed above. The effect of errors in spectrophotometric calibrations - and therefore of our scaling procedure - are large on individual spectra and recovered parameters, but are largely stochastic, leaving a much smaller signal on mean parameters. The results presented in the rest of this section hold regardless of whether we use scaled or un-scaled spectra, and we will focus on results based on scaled spectra. 

\subsubsection{Implementation}\label{sec:implementation}

We analyse all of the 13047 BCG centrals with VESPA, which provides a star-formation history in terms of the stellar mass formed as a function of lookback time. From these recovered SFHs we compute a total stellar mass, a mass-weighted age, a mass-weighted metallicity, the fraction of mass in stars younger than 250 Myrs, and the time at which 85\% of the stars currently in the galaxy had formed. 

We use two sets of stellar population models to analyse our sample: the Flexible Stellar Population Synthesis (FSPS) models of \cite{Conroy2009} and the ubiquitous models of \cite{Bruzual2003} (BC03). In all cases we use a one-parameter dust model, where a single attenuation value is applied to stars of all ages. This avoids a strong degeneracy between the amount of mass at young ages and a birth-cloud dust component that is impossible to break with data of this quality and wavelenth range \citep{Tojeiro2009}. As we wish to focus on mean results over ensembles of galaxies at given halo mass and geometric environment, we always run VESPA to its full resolution, yielding SFHs in 16 bins logarithmically spaced in lookback time between 0.002 and 14 Gyrs. Although the results become dominated by noise in any given galaxy, tests on mocks have shown that this procedure is more robust for the mean quantities we are attempting to recover.

\subsubsection{Unphysical solutions}

Some of the solutions recovered using VESPA from GAMA spectra are unphysical. These mostly manifest in values of total stellar mass and mass fraction in young stars that are too large, with stellar-to-halo mass ratios and mass fraction in young stars that are close to unity. From a fitting perspective, it is easy to understand how such cases arise. In massive galaxies, dominated by old populations, VESPA can easily miss younger stars (especially if they are dust obscured), therefore artificially deflating the mass-to-light ratio of the galaxy and overestimating the stellar mass. Similarly, in the case of young/low-mass galaxies, VESPA is unable to detect any older populations beyond the light-dominating young populations, resulting in a large fraction of young stars - 100 per-cent in extreme cases. 

These extreme cases are a small component of our sample - around 3.7 per-cent have \MsMh$ > 0.2$ and 2.7 per-cent have a young-star mass fraction greater than 20 per-cent. As such, they have a limited effect on median quantities. In what follows, we remove the small number of galaxies with  \MsMh$ > 0.2$ and those with more than 20 per-cent of their mass in stars younger than 250 Myrs. This does not affect our results, which are based on median values of samples.

These outliers are not correlated with S/N and they do not appear in our simulations, so our hypothesis is that these catastrophic fits are driven by errors in the spectrophotometric calibrations that were not addressed by our simple re-scaling, or by shortcomings in our modeling, or a combination of both.

\subsection{Results}\label{sec:results}

\subsubsection{Dependence on halo mass}\label{sec:results_Mh}

\begin{figure*}
\includegraphics[scale=0.22]
                       {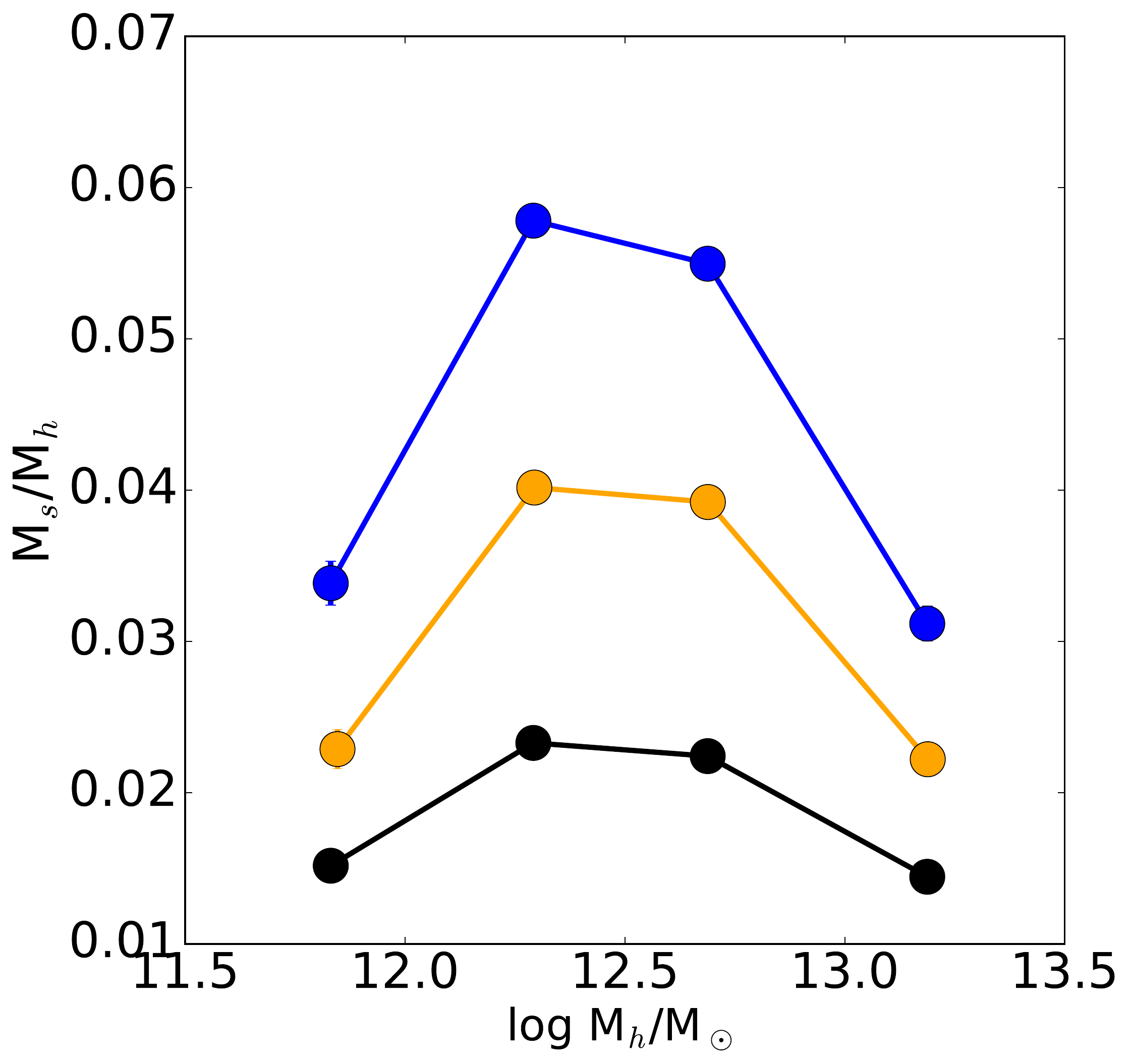}
\includegraphics[scale=0.22]
                        {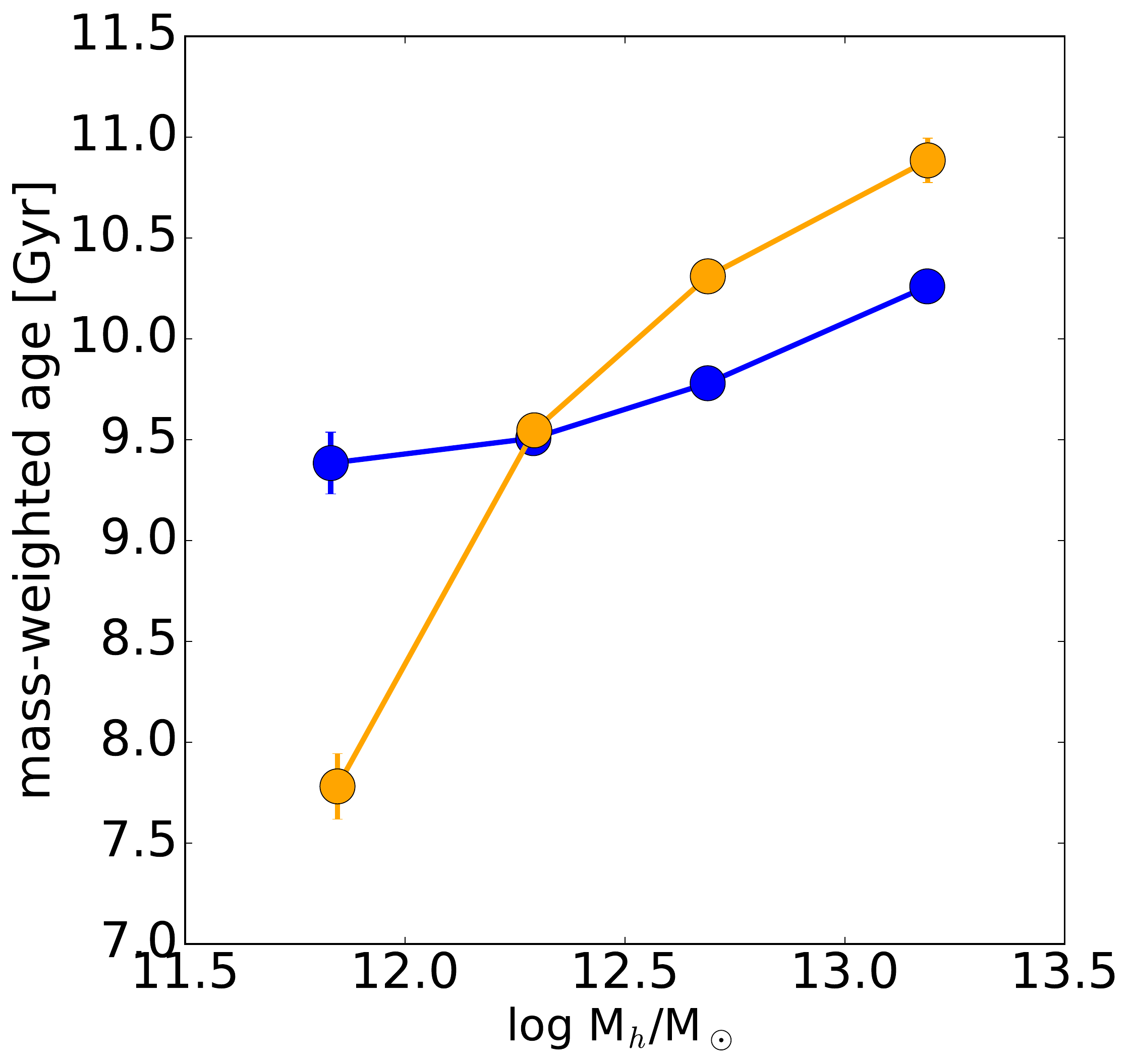}
\includegraphics[scale=0.22]
                        {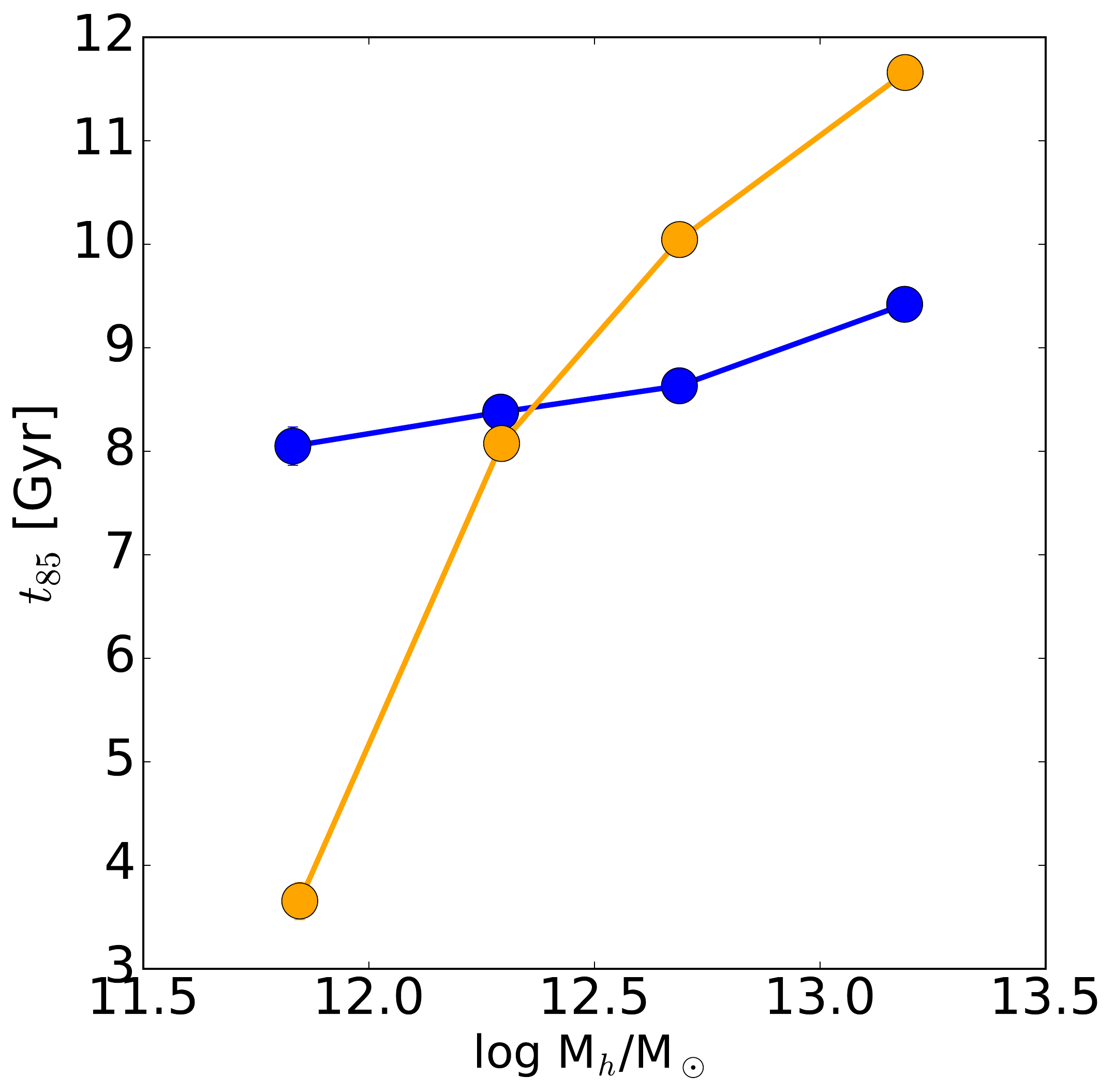}
\includegraphics[scale=0.22]
                        {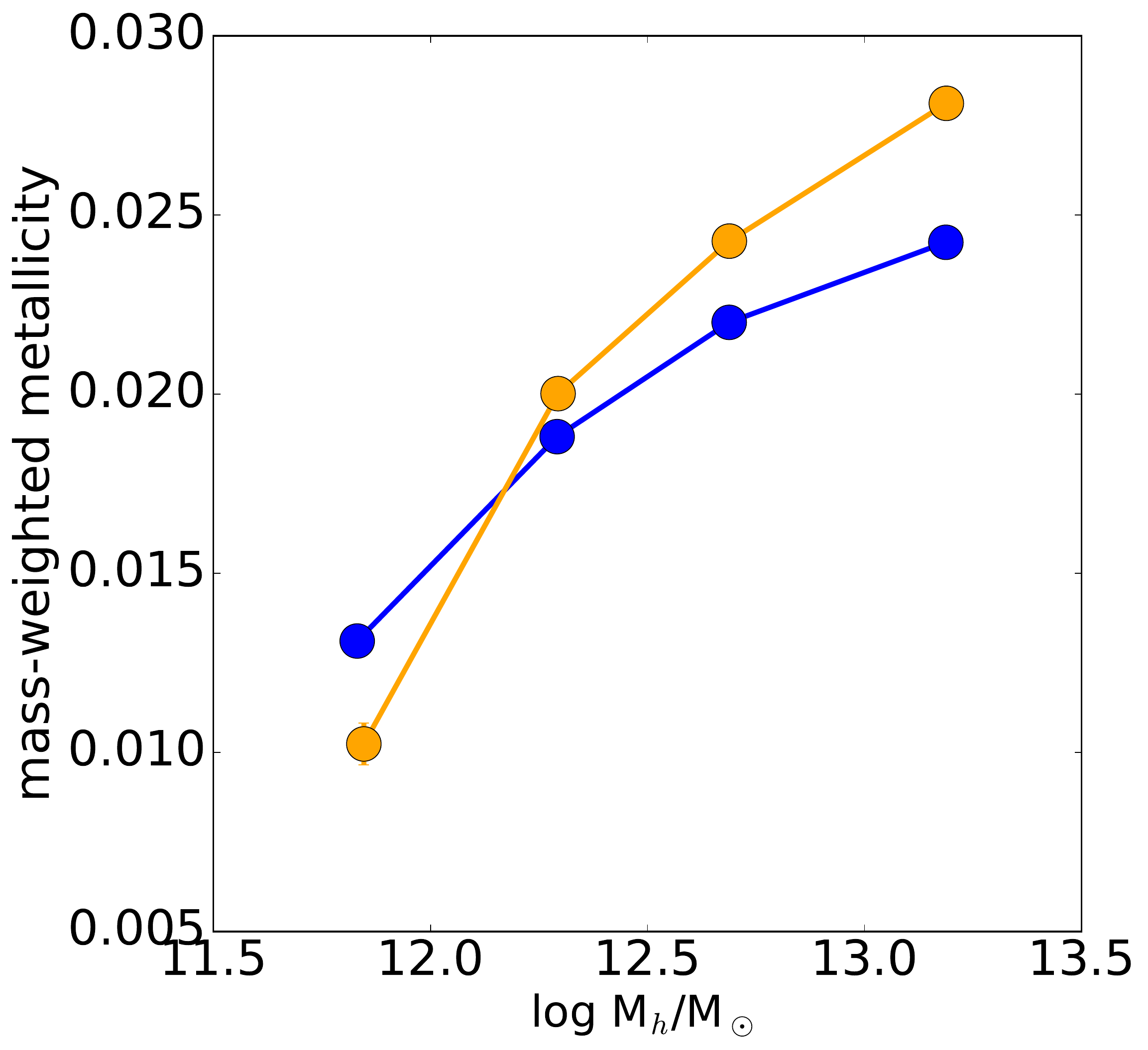}
\includegraphics[scale=0.22]
                        {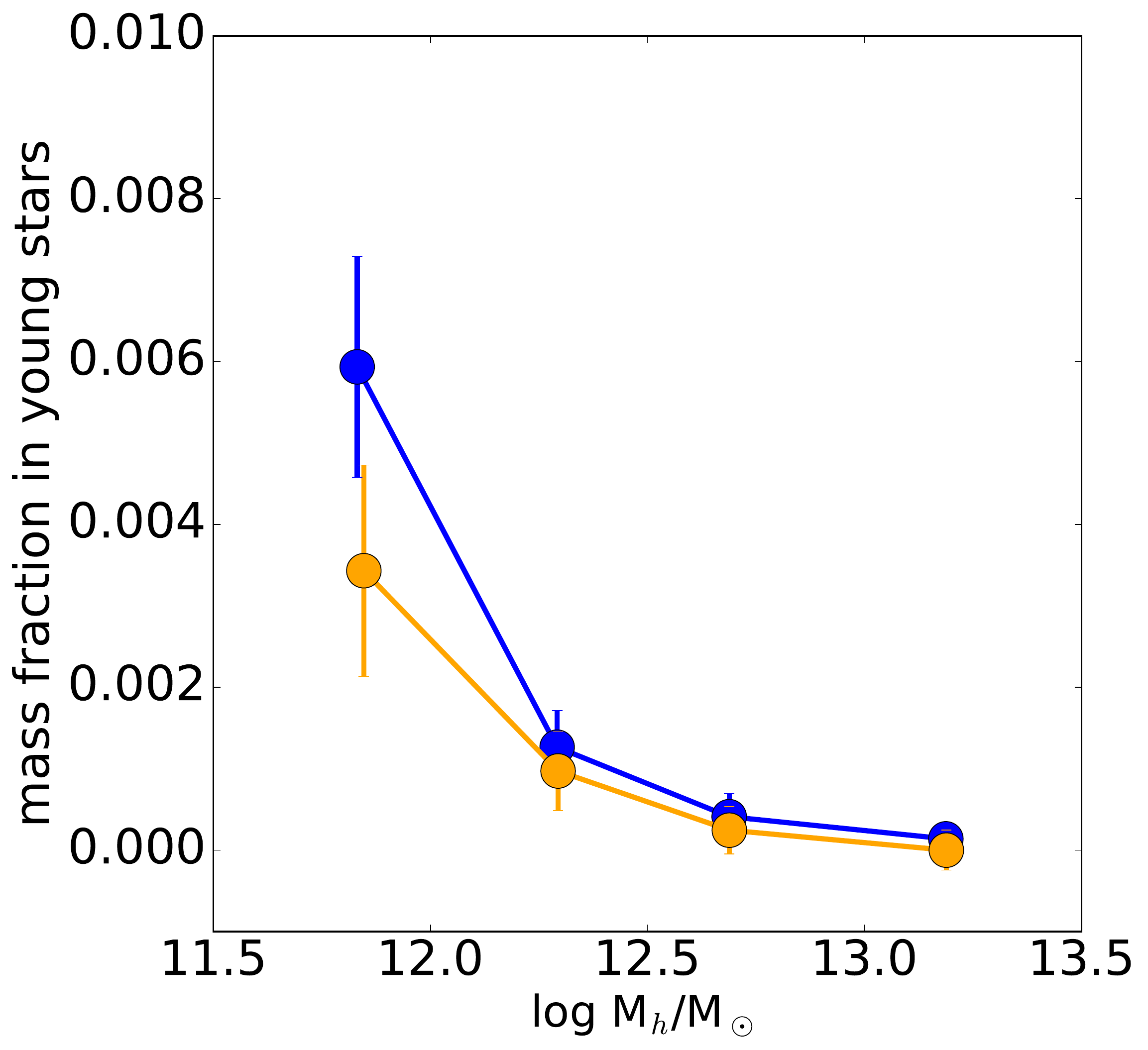}
\caption{Median M$_s/$M$_h$, mass-weighted age, time at which 85\% of the stars were formed, mass-weighted metallicity and fraction of young stars as a function of halo mass in GAMA BCG-centrals. The blue lines show results obtained using the FSPS stellar population models, and the yellow lines show the results obtained using the BC03 models. We show M$_s/$M$_h$ obtained using photometric masses in black. All error bars show the standard error on the median. The qualitative trends with halo mass are robust to the change in modeling - namely, both sets of models produce galaxies that are older and more metal rich with increasing halo mass. However, the slope of these relationships with halo mass changes substantially, depending on the stellar population modeling used. The offset in different estimates of stellar masses discussed in Section~\ref{sec:stellar_masses} is seen here as a change in amplitude of M$_s/$M$_h$.}
\label{fig:GAMA_res_Mh}
\end{figure*}

Fig.~\ref{fig:GAMA_res_Mh} shows the relationship between M$_s/$M$_h$, mass-weighted age, time at which 85\% of the stars were formed, mass-weighted metallicity and fraction of young stars as a function of halo mass. We show in each panel results obtained with FSPS and BC03 models, and we show in black the M$_s/$M$_h$ estimate using photometric masses. All measures of overall age show the expected trend with halo mass - galaxies in lower mass haloes formed their stars earlier, have a more extended star-formation history, and a larger mass fraction in stars younger than 250 Myrs than galaxies in higher mass haloes. One can interpret this as evidence for halo downsizing: galaxies in higher mass haloes formed most of their stars at higher redshift. Qualitatively, the trends are robust to the choice of SSP modeling, but quantitatively there are substantial differences. As we are ultimately concerned with relative changes in these quantities with geometric environment {\it at fixed halo mass}, such differences do not impact on our final conclusions. The increased scatter and offsets in stellar mass estimates discussed in Section~\ref{sec:stellar_masses} is seen here as an offset in amplitude of the M$_s/$M$_h$ vs M$_h$ relation, as well as a marked broadening of the stellar mass distribution at fixed halo mass, as seen here in the increased error bars. 

\begin{figure}
\includegraphics[scale=0.55]{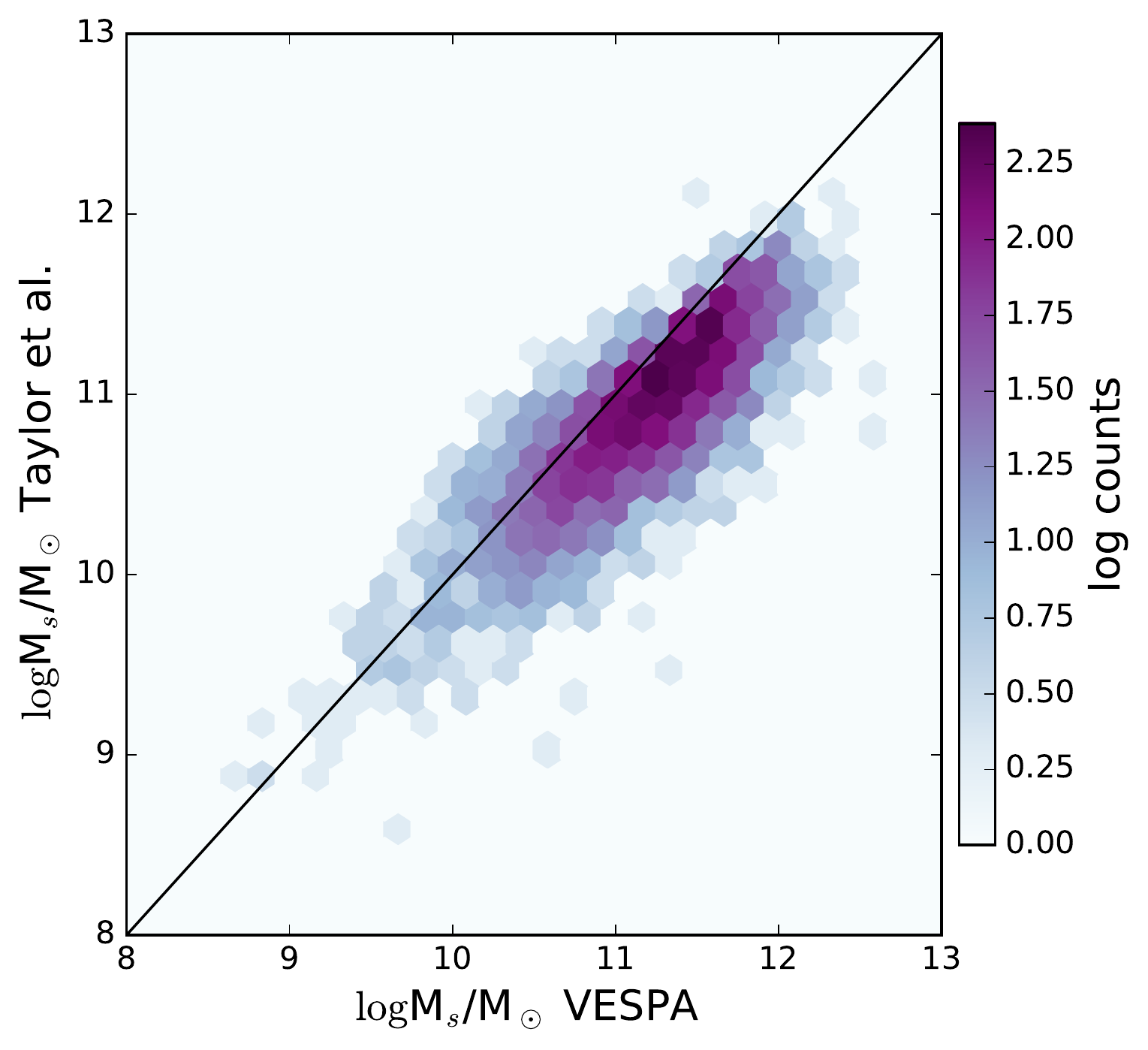}
\includegraphics[scale=0.55]{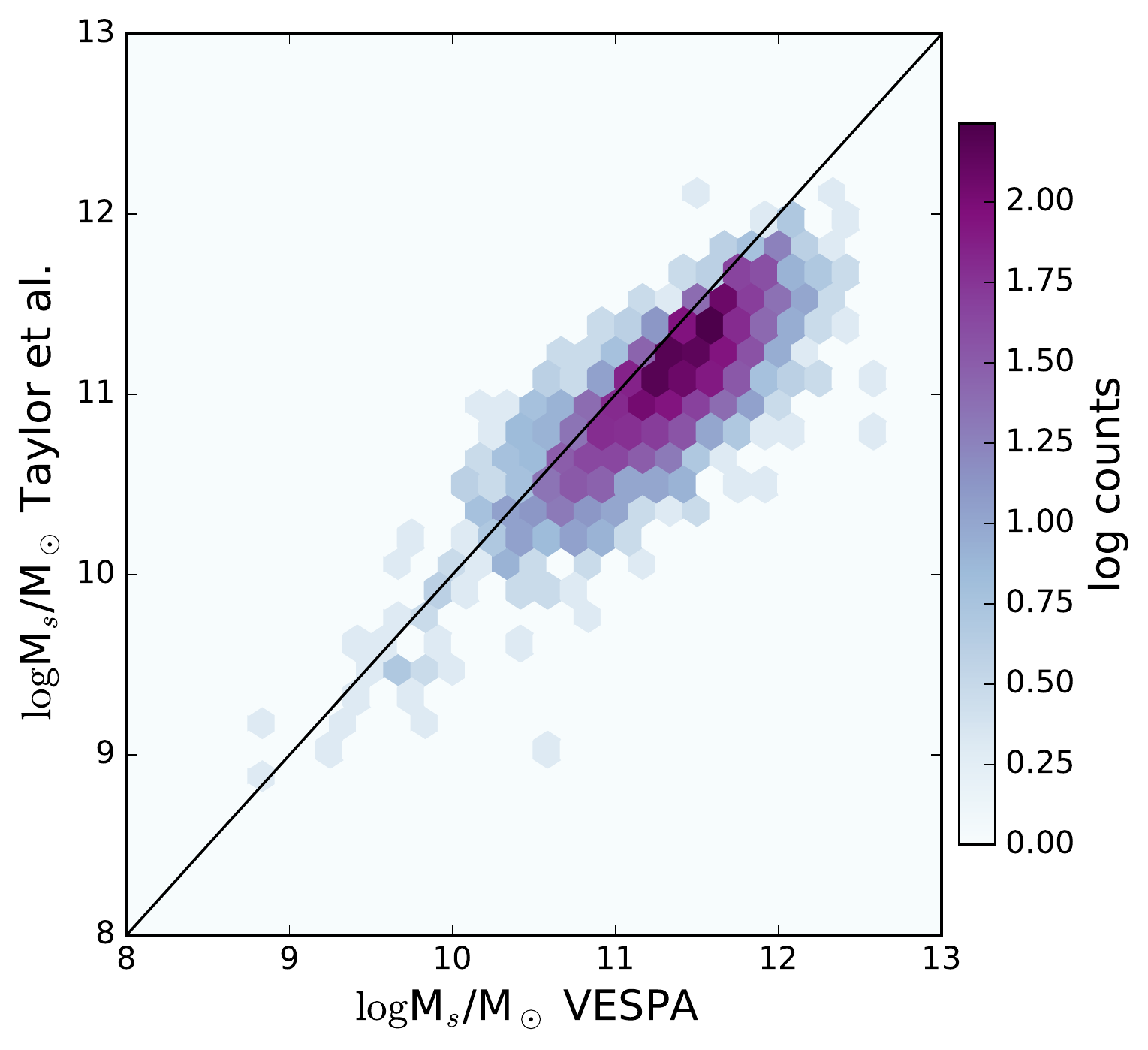}
\caption{Same as Fig.~\ref{fig:Mstar_comp}, but limiting the sample to central galaxies in groups with multiplicity equal to or greater than three (top panel) and four (bottom panel). The median offset between the two estimates remains roughly constant, but the mean mass is shifted to larger mass as we exclude the low mass haloes. The scatter between the two estimates is reduced at large stellar mass, as we loose some lower S/N objects. }
\label{fig:Mstar_comp_Nfof}
\end{figure}

\begin{figure*}
\includegraphics[scale=0.22]
                       {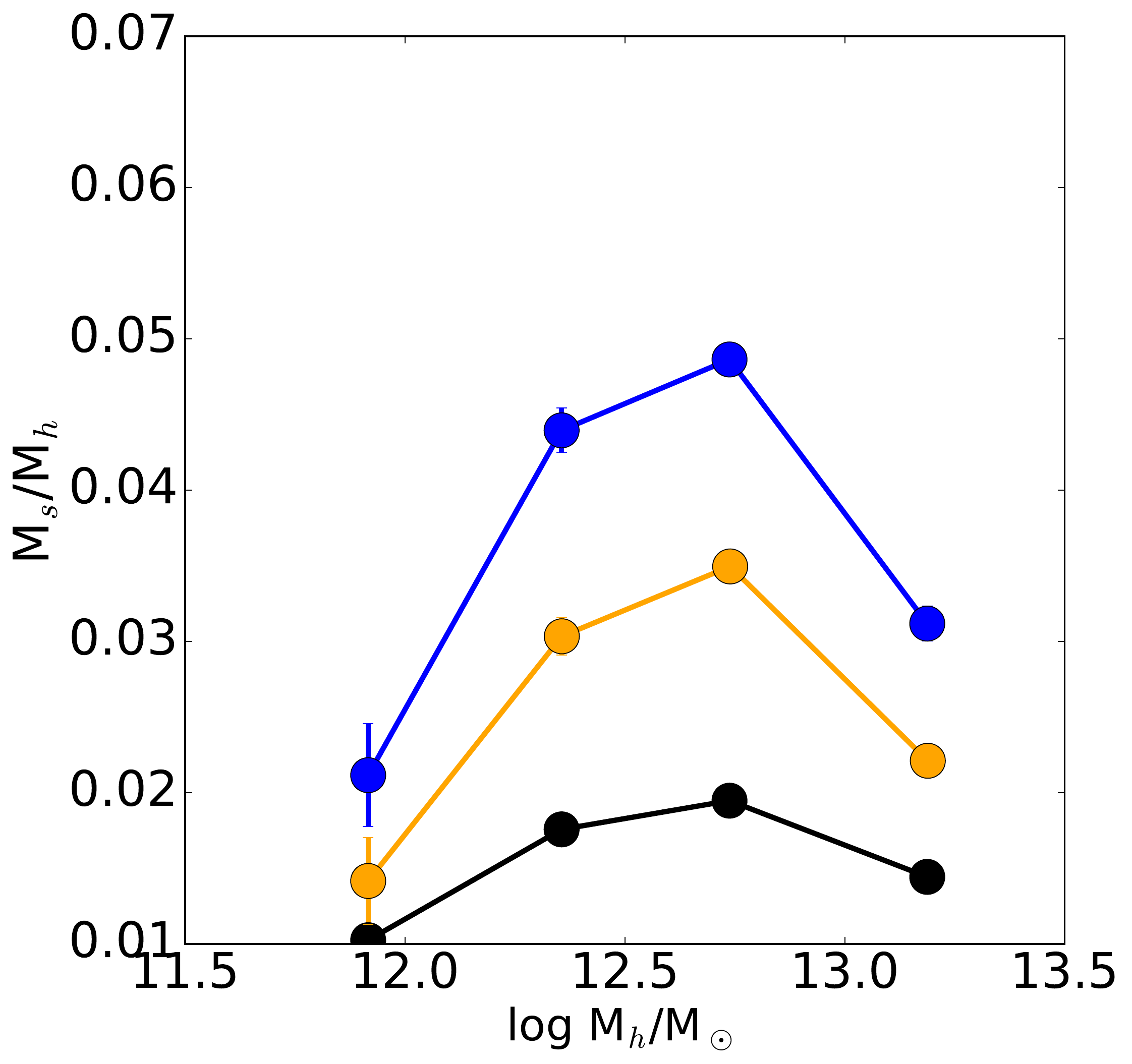}
\includegraphics[scale=0.22]
                        {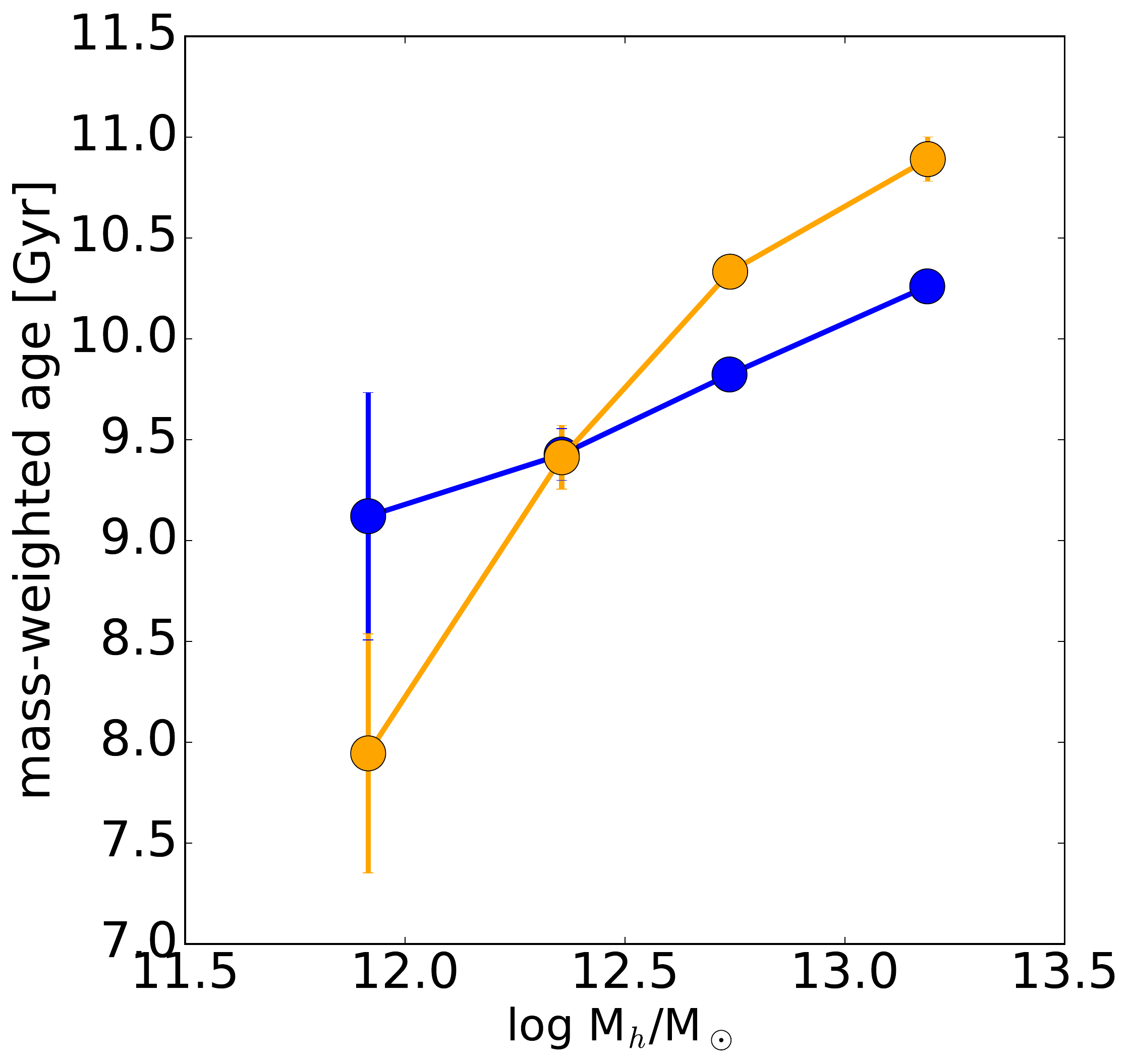}
\includegraphics[scale=0.22]
                        {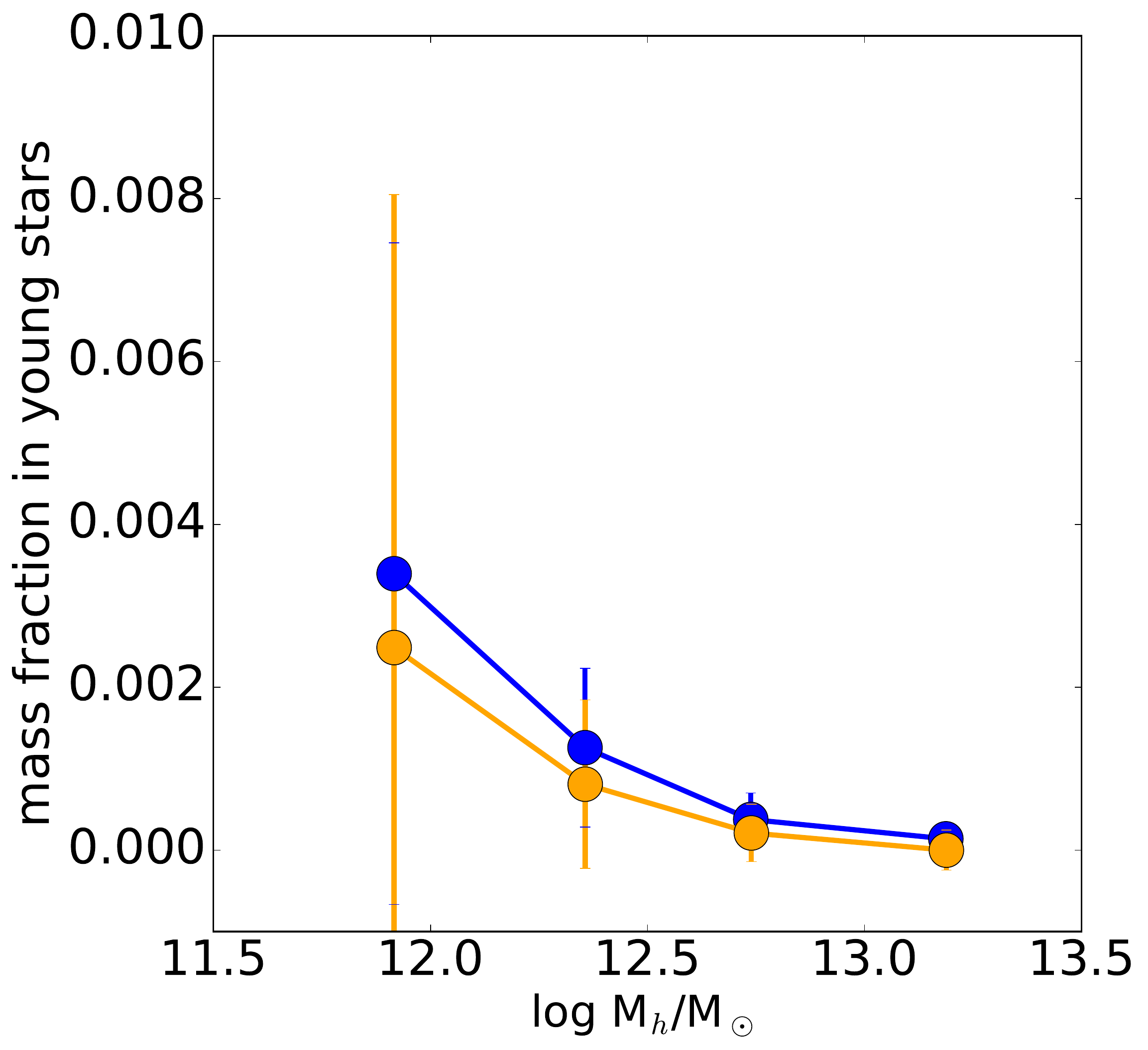}
\caption{As in Fig.~\ref{fig:GAMA_res_Mh}, but using groups with multiplicity greater than or equal to three.}
\label{fig:GAMA_res_Mh3}
\end{figure*}

\begin{figure*}
\includegraphics[scale=0.22]
                       {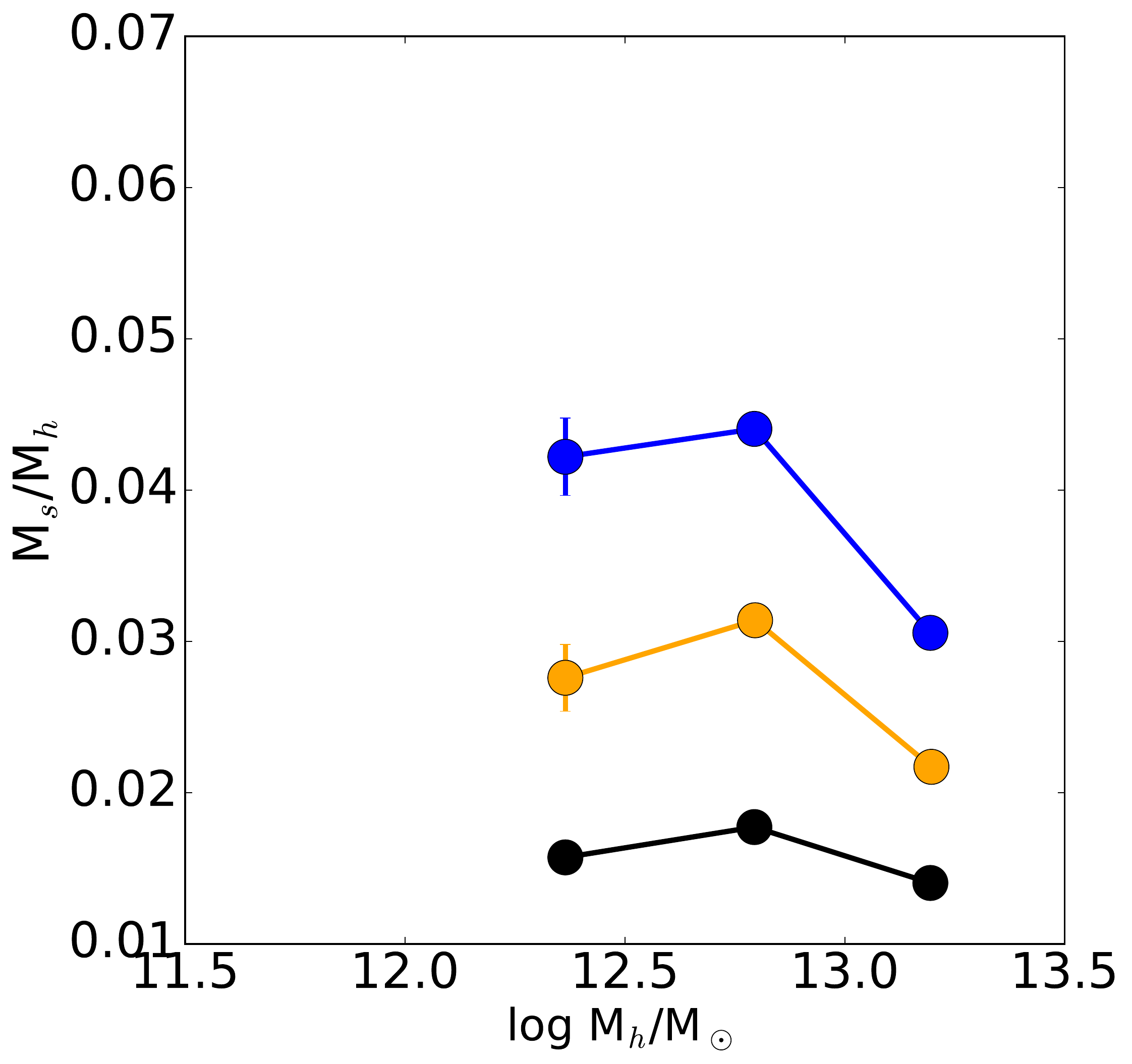}
\includegraphics[scale=0.22]
                        {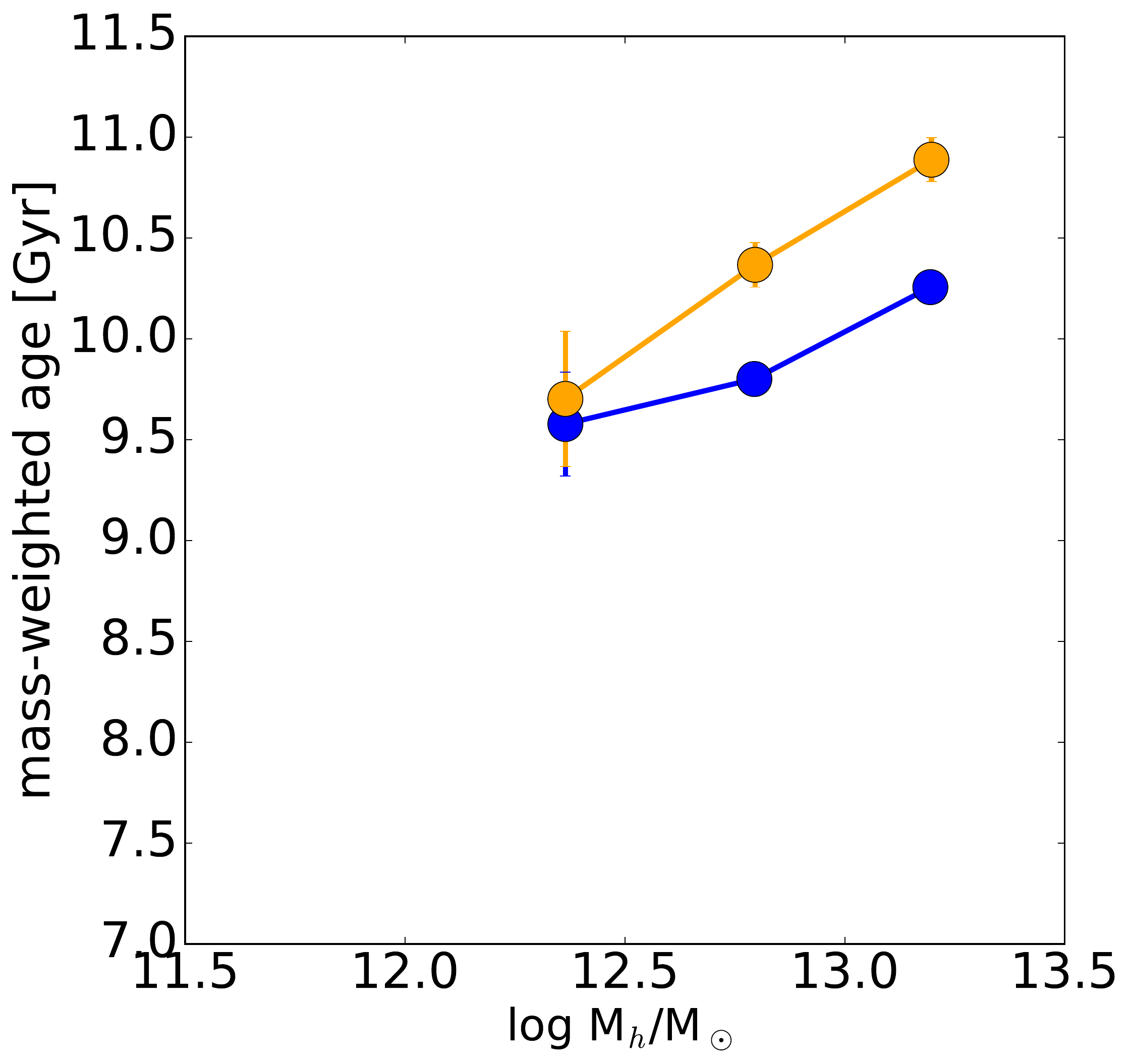}
\includegraphics[scale=0.22]
                        {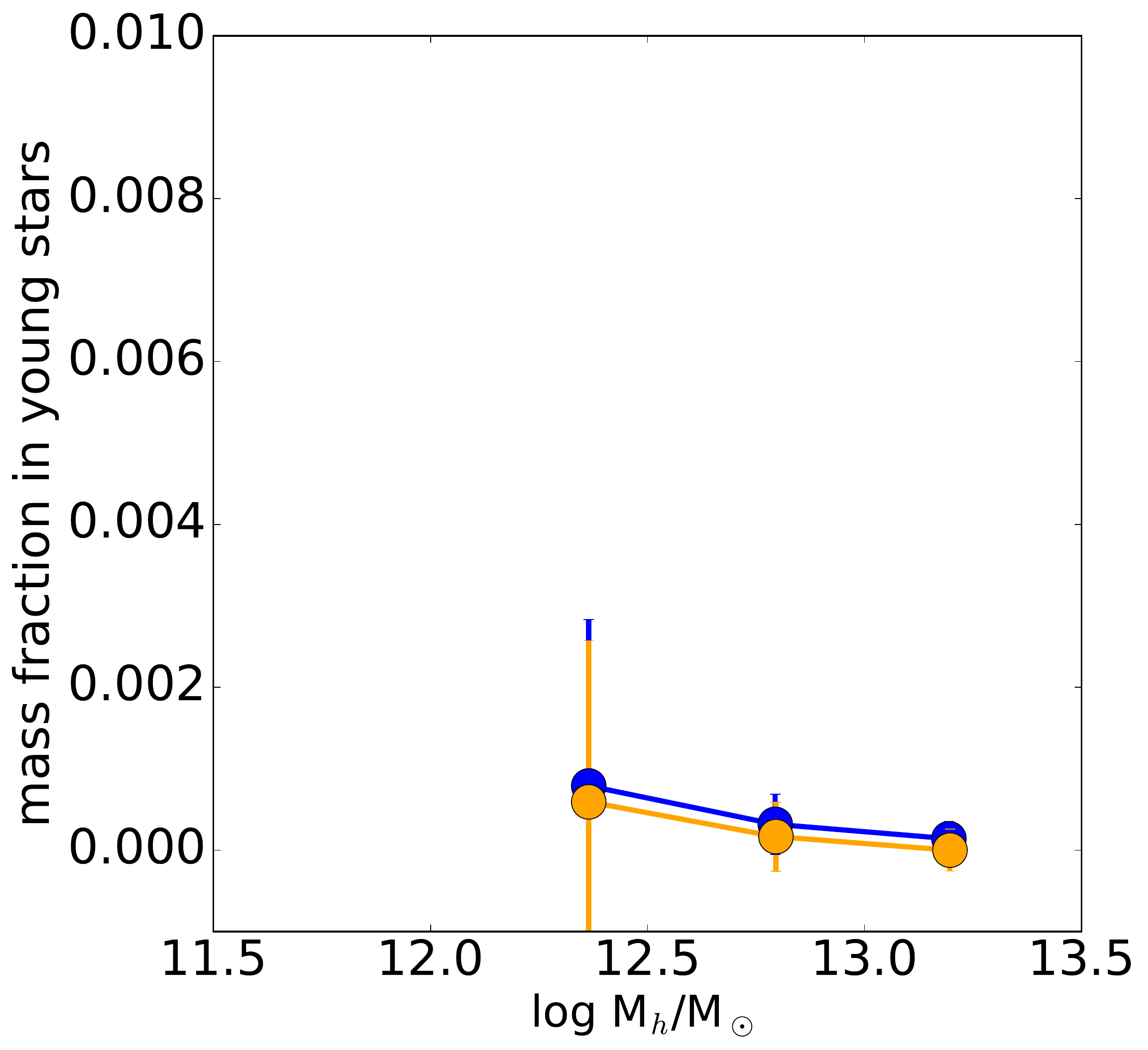}
\caption{As in Fig.~\ref{fig:GAMA_res_Mh}, but using groups with multiplicity greater than or equal to four. We only plot bins with at least 50 galaxies. }
\label{fig:GAMA_res_Mh4}
\end{figure*}

Fig.~\ref{fig:GAMA_res_Mh} includes all groups with multiplicity greater than or equal to two. Low multiplicty groups present two challenges: the halo masses are more uncertain, and nearly half of these groups are likely to be spurious superpositions \citep{Robotham2011}. To assess the impact of removing low multiplicity groups, we re-do our analysis by selecting groups with multiplicity greater than or equal to three and four. This returns a further biased sample of groups: we are preferentially removing low mass haloes, and therefore on average low mass galaxies (see Fig.~\ref{fig:Mstar_comp_Nfof}). The point of this exercise is to examine the effect of removing a likely contamination of spurious low mass haloes, which we show in Figures~\ref{fig:GAMA_res_Mh3} and Figures~\ref{fig:GAMA_res_Mh4}. We focus on M$_s/$M$_h$, mass-weighted age and young mass fraction - $t_{85}$ behaves similarly to mass-weighted age and mass-weighted metallicity is not considered in our interpretation due to the biases found in Section~\ref{sec:vespa_on_lgal}. Overall trends remain, and are completely consistent within the errors. In the case of M$_s/$M$_h$ from spectroscopic masses (blue and yellow lines), we note that the median values are lowered with higher multiplicity. This is entirely due to the reduction of the large scatter in stellar mass at the high-mass end due to low S/N spectra; this is seen also in Fig.~\ref{fig:Mstar_comp_Nfof}.
 This test does not allow us to explicitly test the effect of including low multiplicity groups in our lowest halo mass bin, but it allows us to state that the properties of the galaxies in these groups follows on average the properties of the galaxies in larger multiplicity groups, at a given halo mass. 
 

\subsubsection{Dependence on environment}\label{sec:results_env}

In this section we will focus on figures showing results obtained with FSPS models, as they give on average better fits to the data. We show the equivalent figures for BC03 in Appendix, and refer to them as required throughout this section. We describe how we assess significance of a detection and the effect of the lensing group luminosity to halo mass calibration in the next section.

\begin{figure*}
\includegraphics[scale=0.25]
                        {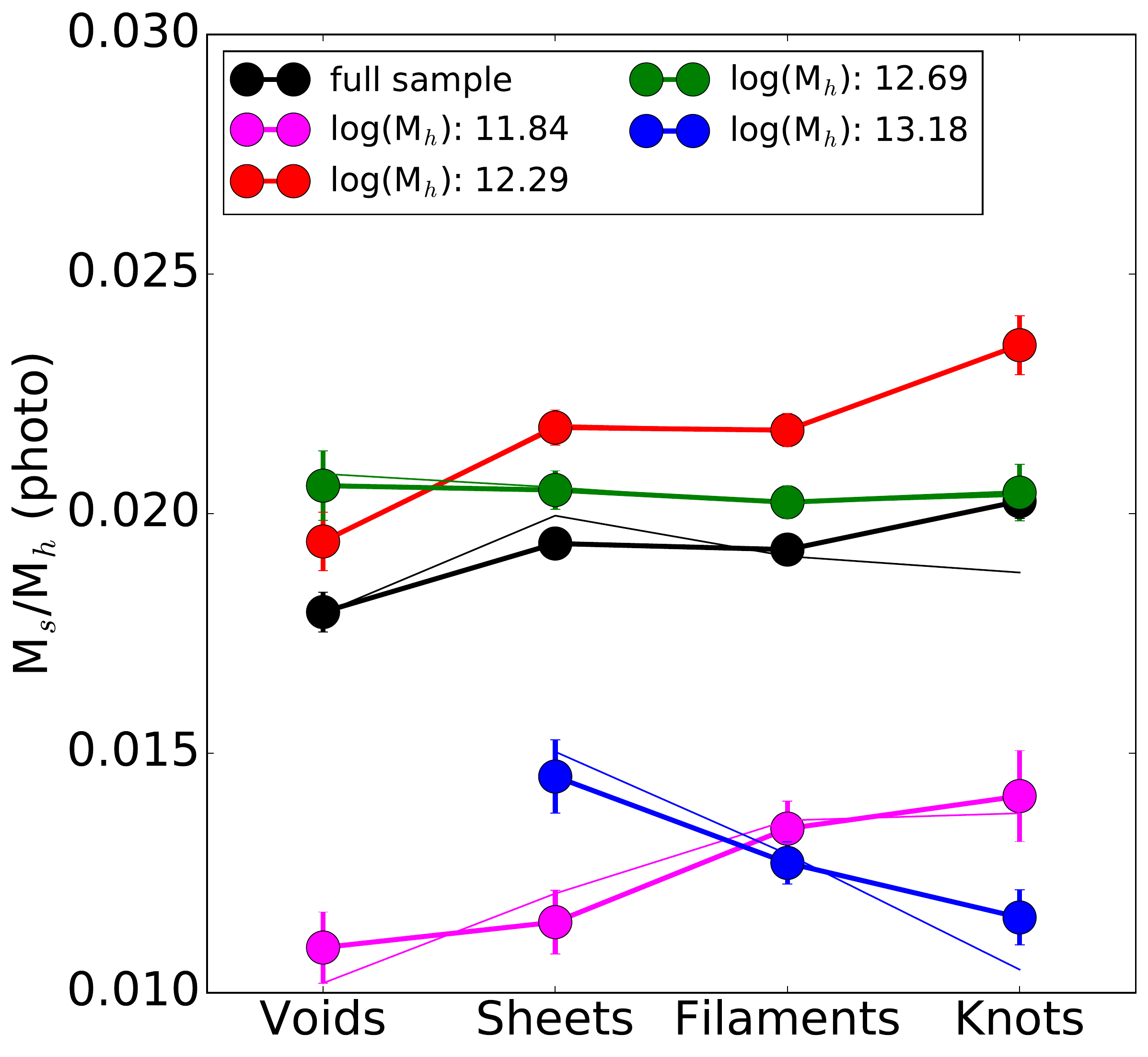}
\includegraphics[scale=0.25]
                        {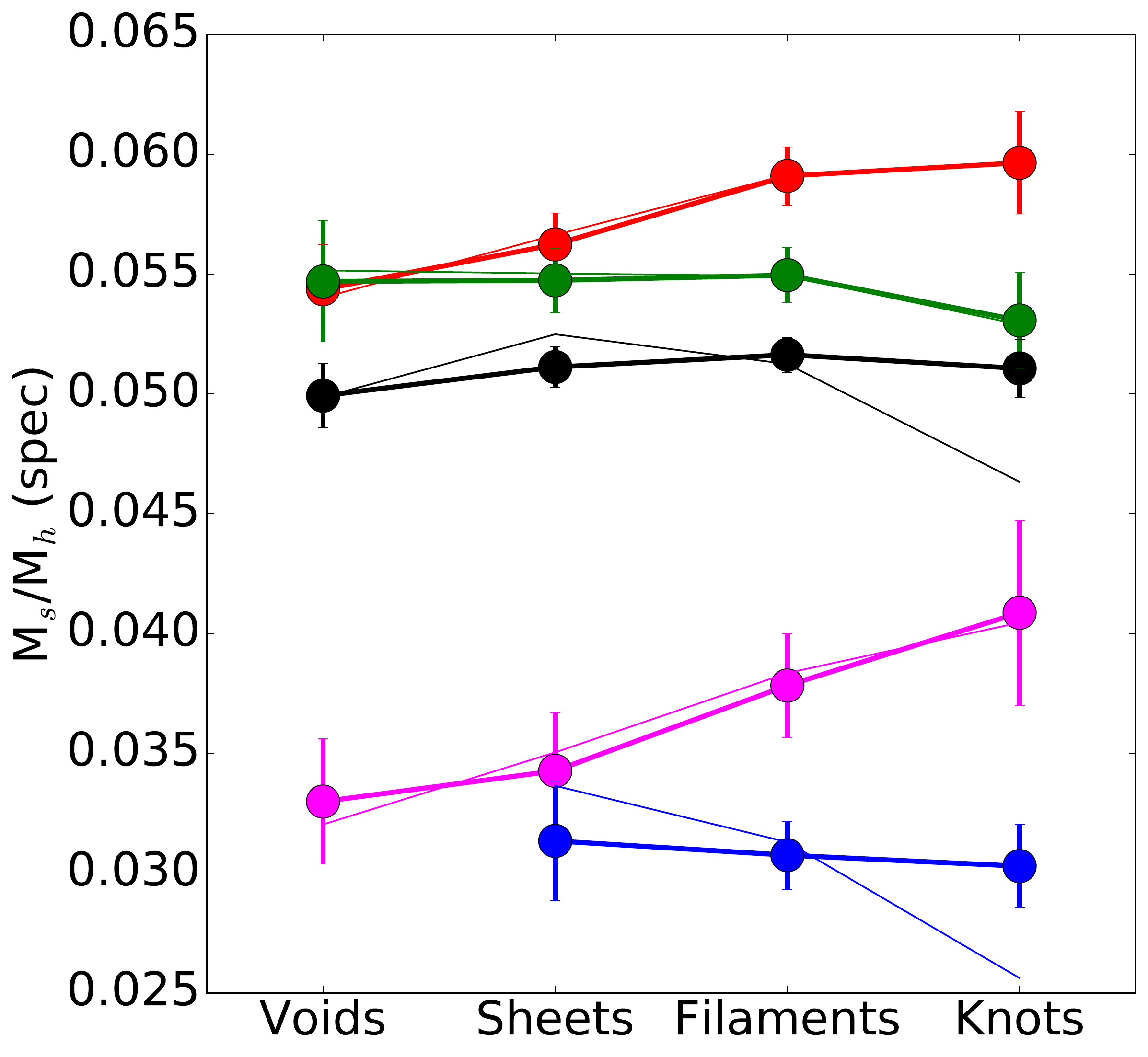}
\includegraphics[scale=0.25]
                        {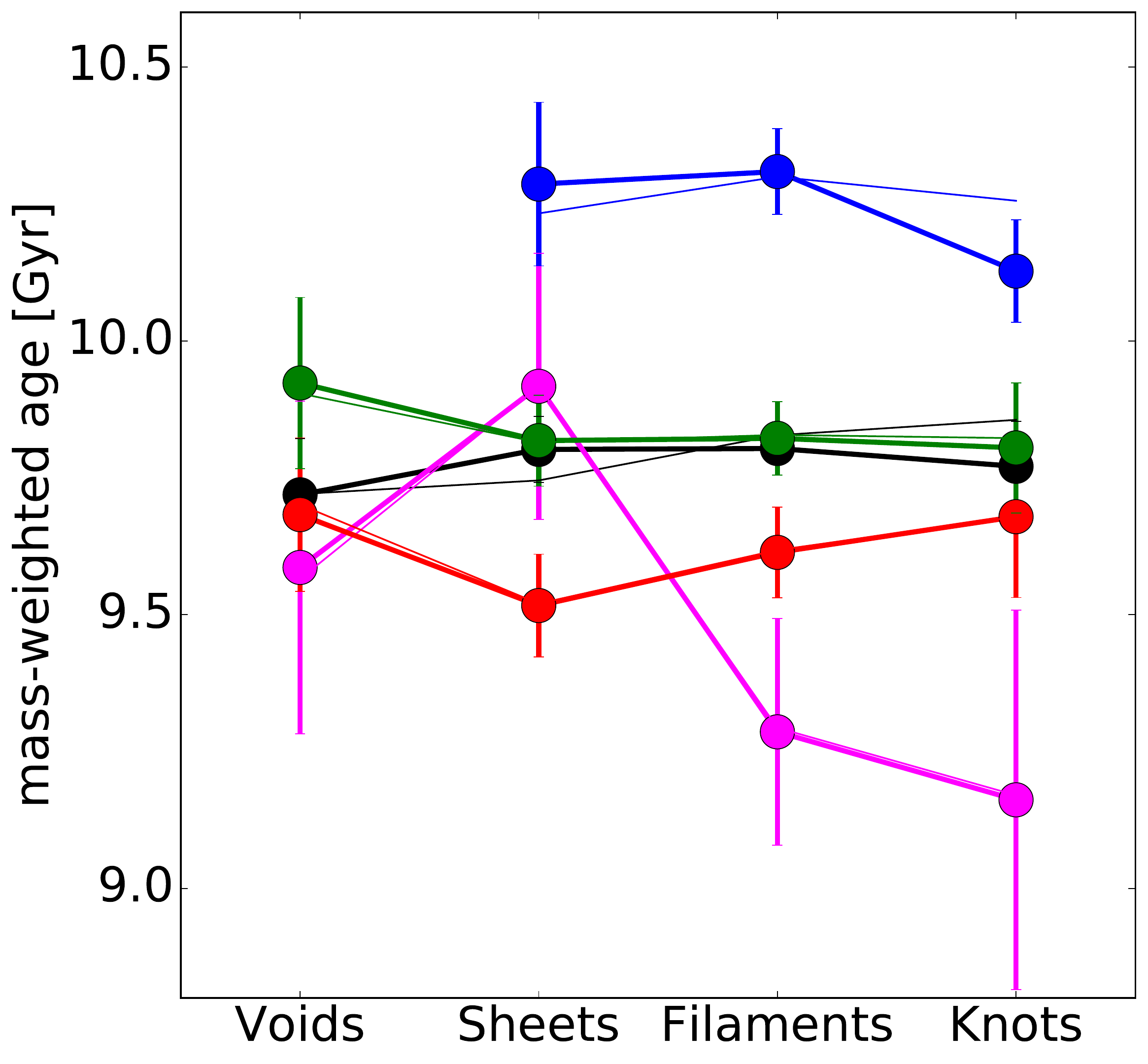}
\includegraphics[scale=0.25]
                        {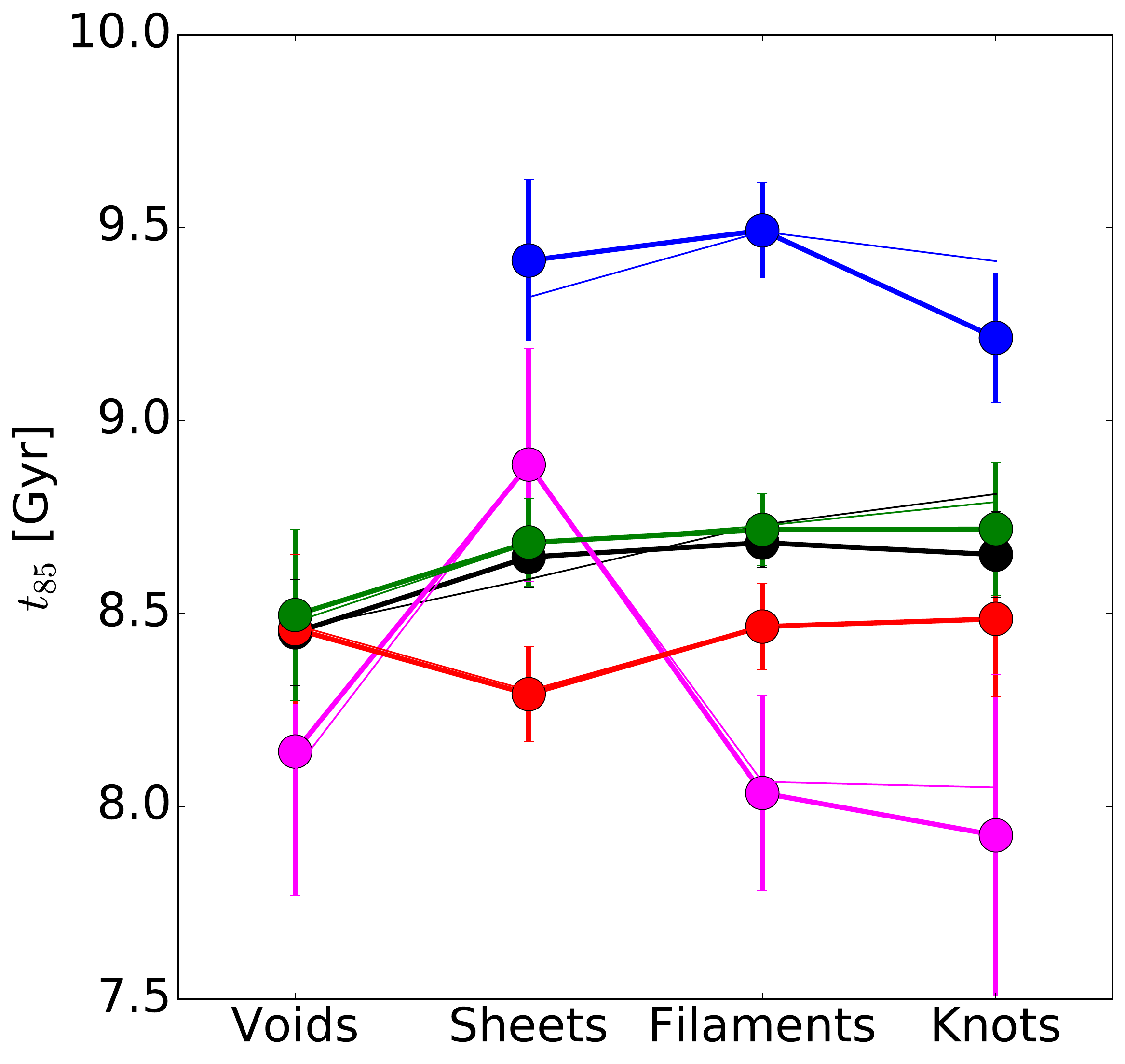}
\includegraphics[scale=0.25]
                        {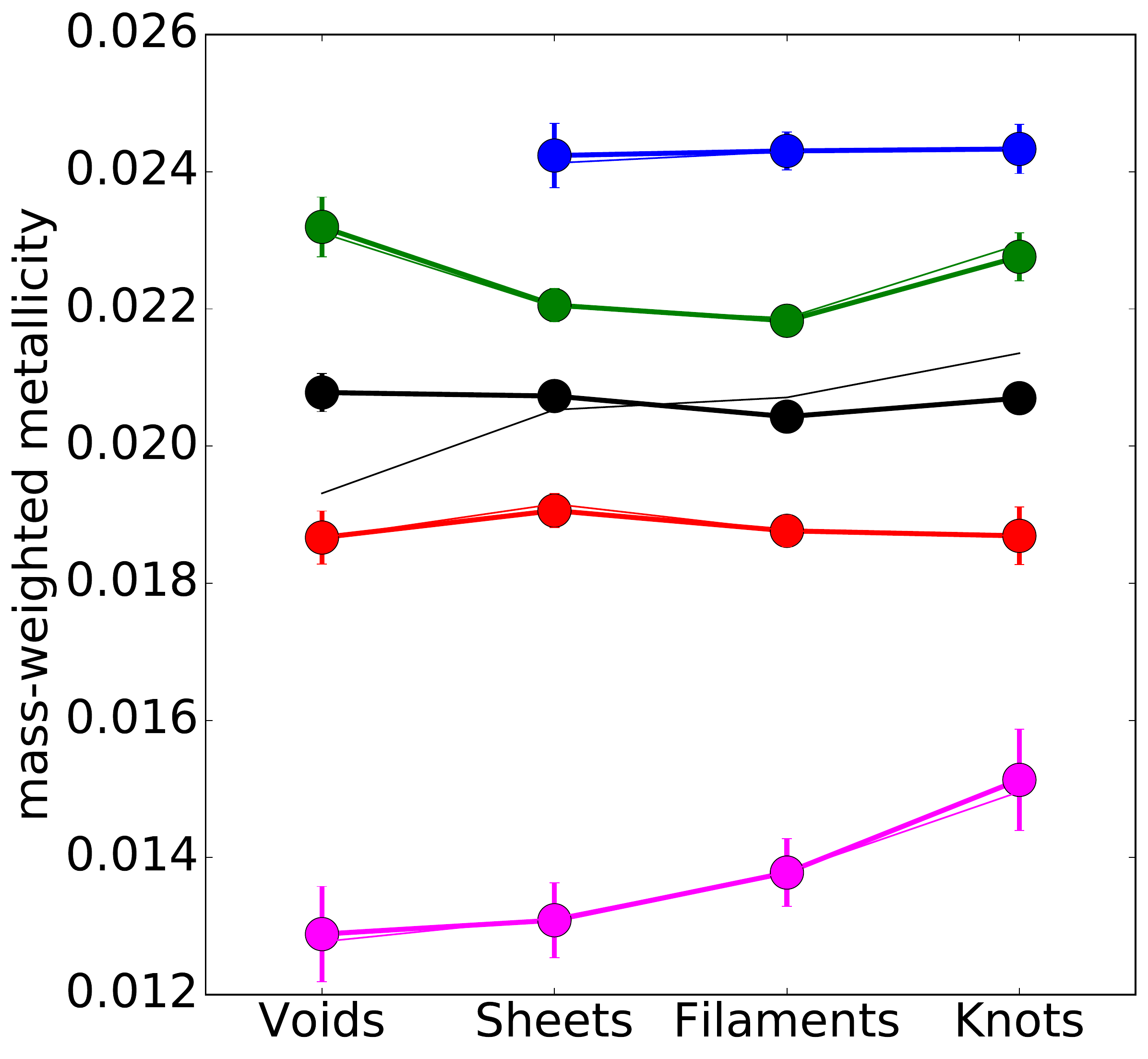}
\includegraphics[scale=0.25]
                        {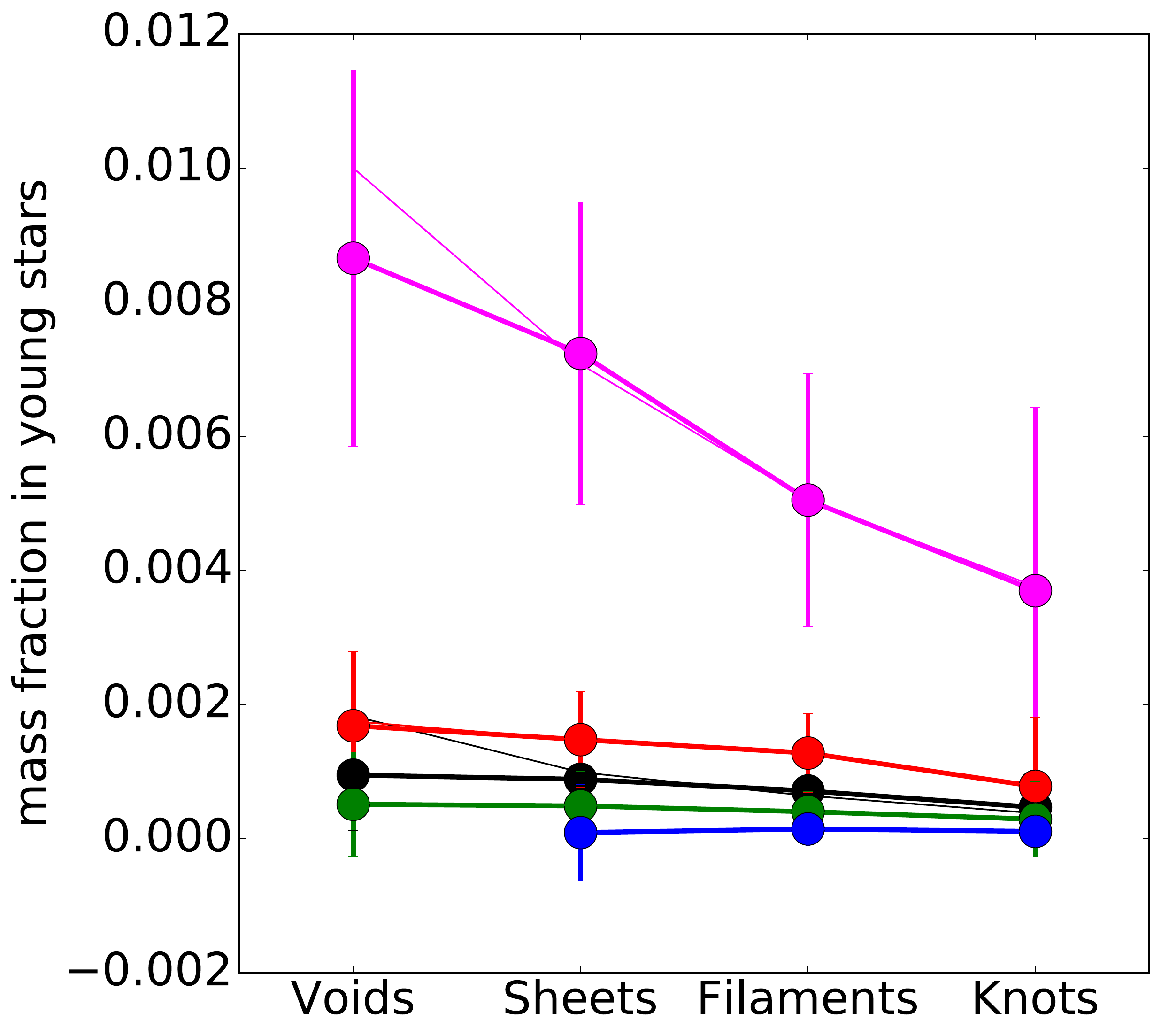}
\caption{M$_s$/M$_h$ , mass-weighted age, time at which 85\% of the stars were formed, mass-weighted metallicity and fraction of young stars as a function of geometric environment of GAMA BCG-centrals. The different coloured thick lines show the weighted-mean in bins of halo mass as defined in Fig.~\ref{fig:MsMh_data}; the black line shows the mean over the full sample. We show unweighted means using thin lines of the same colour. The error bars show the error on the mean. Note that the measurements of mass-weighted age, $t_{85}$ and young mass fraction are strongly correlated: all quantities are computed from the same SFH.}
\label{fig:GAMA_res}
\end{figure*}

Fig.~\ref{fig:GAMA_res} shows the \MsMh (estimates computed using photometric and spectroscopic stellar masses are shown in different panels), mass-weighted age, time at which 85\% of the stars were formed, mass-weighted metallicity and fraction of young stars measured in central galaxies as a function of geometric environment and halo mass. The lines show mean values and the error bars show the standard error on the median. The error bars are given only as a visual indication of the error on the median; significance is computed using a Kruskal-Wallis test, using the full set of points in each bin of halo mass and GE (see next section). We only plot bins with at least 50 galaxies.

We first note that the downsizing signal observed in Fig.~\ref{fig:GAMA_res_Mh} {\it is independent of geometric environment}. I.e., we see very little evidence of downsizing changing in any significant way as a function of geometric environment.

We see a clear trend of \MsMh increasing from voids to knots in low mass haloes (in the two lowest mass bins) and decreasing in highest mass bin, with a clear transition at $\log $ M$_h \approx 12.68$, where \MsMh is flat with GE. Although the median trend is visible using both photometric and spectroscopic stellar masses, the larger scatter in spectroscopic masses, driven by the low S/N of the spectra, removes the statistical significance of the trend of spectroscopic \MsMh with GE. The trend is very similar using BC03 models (see Fig.~\ref{fig:GAMA_res_BC03}). Interpreting such a measurement within the context of assembly bias and the L-Galaxies model, our analysis shows that low mass haloes that reside in knots are older that haloes of the same mass residing in voids. At high mass, the trend is reversed. Such a change in direction is in line with the predictions from simulations.

When looking at stellar ages, we find no significant trend of age with GE at fixed halo mass. In the lowest mass bin, we observe a difference in mass weighted age of approximately 0.5Gyr between galaxies residing in voids and knots. However, the error on the median is large and we note that this behaviour is not observed when using BC03 models (see Fig.~\ref{fig:GAMA_res_BC03}). There is no discernible trend of mass fraction in young stars as a function of environment at fixed halo mass; the errors are, however, very large so we are simply not able to put any constraints on this relationship. Mass-weighted metallicity shows no trend with GE at fixed halo mass, although we note that according to the spectral GAMA mocks, our estimate mass-weighted metallicity is likely to show a small bias as a function of metallicity.

\subsubsection{Significance and halo mass uncertainties}\label{sec:stats}
Our main goal is to assess whether galaxy properties are different in distinct GE at fixed halo mass. The halo mass measurements are derived from group luminosity estimates and calibrated to halo masses, and subsamples defined according to these calibrated halo mass measurements. To understand the effect of the error on the lensing calibration on our result, we Monte Carlo (MC) halo mass estimates by drawing correlated pairs of M$_p$ and $\alpha$ according to the uncertainties and correlation factor quoted in \cite{HanEtAl15} and Section~\ref{sec:halo_masses}. For each draw, we repeat our full analysis with the new halo mass estimates, changing only the bin boundaries by the difference in mean of halo masses; this keeps approximately the same number of galaxies in each of the four bins. We perform 100 MC draws. 

In each MC realisation, we estimate whether there is a significant change of galaxy property with GE, at fixed halo mass, by using a Kruskal-Wallis (KW) test. The KW test tests the null hypothesis that an arbitrary number of samples are drawn from the same distribution or, in other words, whether they share the same median. The KW test is non-parametric and makes no assumption on the shape of the underlying distributions. If the null hypothesis is proved incorrect (assessed by a p-value smaller than 0.05), then at least one of the samples has a median that is inconsistent with the median values of the other samples. The KW test makes no statement on whether there is a consistent trend with GE, or which sample(s) are the outliers. We use the KW test to identify which bins of halo mass show a consistent rejection of the null hypothesis and we further investigate how often they show a strictly monotonic trend with GE (ascending or descending). As expected from the visual inspection of Fig.~\ref{fig:GAMA_res}, the only consistent detection comes from \MsMh using photometric masses, which we summarise in Table~\ref{tab:KW}.  The observed trends of \MsMh with GE are consistent and significant in the lowest and highest mass bins and a trend is seen 69\% of the time in the second halo mass bin.  The null hypothesis is confirmed on our third mass bin ($\log $ M$_h \approx 12.7$), where no significant trend is seen in 94\% of the cases, as expected from Fig.~\ref{fig:GAMA_res}. Using the best-fit values for halo masses calibration (i.e., those quoted in \citealt{HanEtAl15} and Section~\ref{sec:halo_masses}), the trend with GE shown in Fig.~\ref{fig:GAMA_res} is significant and monotonic in all bins except Bin 3. We conclude from this analysis that the observed dependence of \MsMh with GE at low and high mass is robust to uncertainties on the halo mass calibration from weak lensing.

\begin{table}
\begin{tabular}{|l|c|c|c|c|c|}
\hline
 				& Bin 1 & Bin 2 & Bin 3 & Bin 4   \\ \hline \hline
N$_{\rm dect}$ 	& 	91	&	73	& 13	& 98 		\\
N$_{\rm trend}$ & 	92	& 	77	& 19	& 93		\\
N$_{\rm both}$ 	& 	85	& 	69	& 6	& 92		 \\ \hline
\end{tabular}
\caption{Summary of results from our analysis of the 100 MC runs detailed in Section~\ref{sec:stats}, shown separately for each bin in halo mass. Table shows number of significant detection of an outlier median in that mass bin (N$_{\rm dect}$),  number of times a monotonic trend was found as a function of environment N$_{\rm trend}$, and number of times that both conditions were met.}
\label{tab:KW}
\end{table}

\begin{figure*}

\includegraphics[scale=0.22]
                        {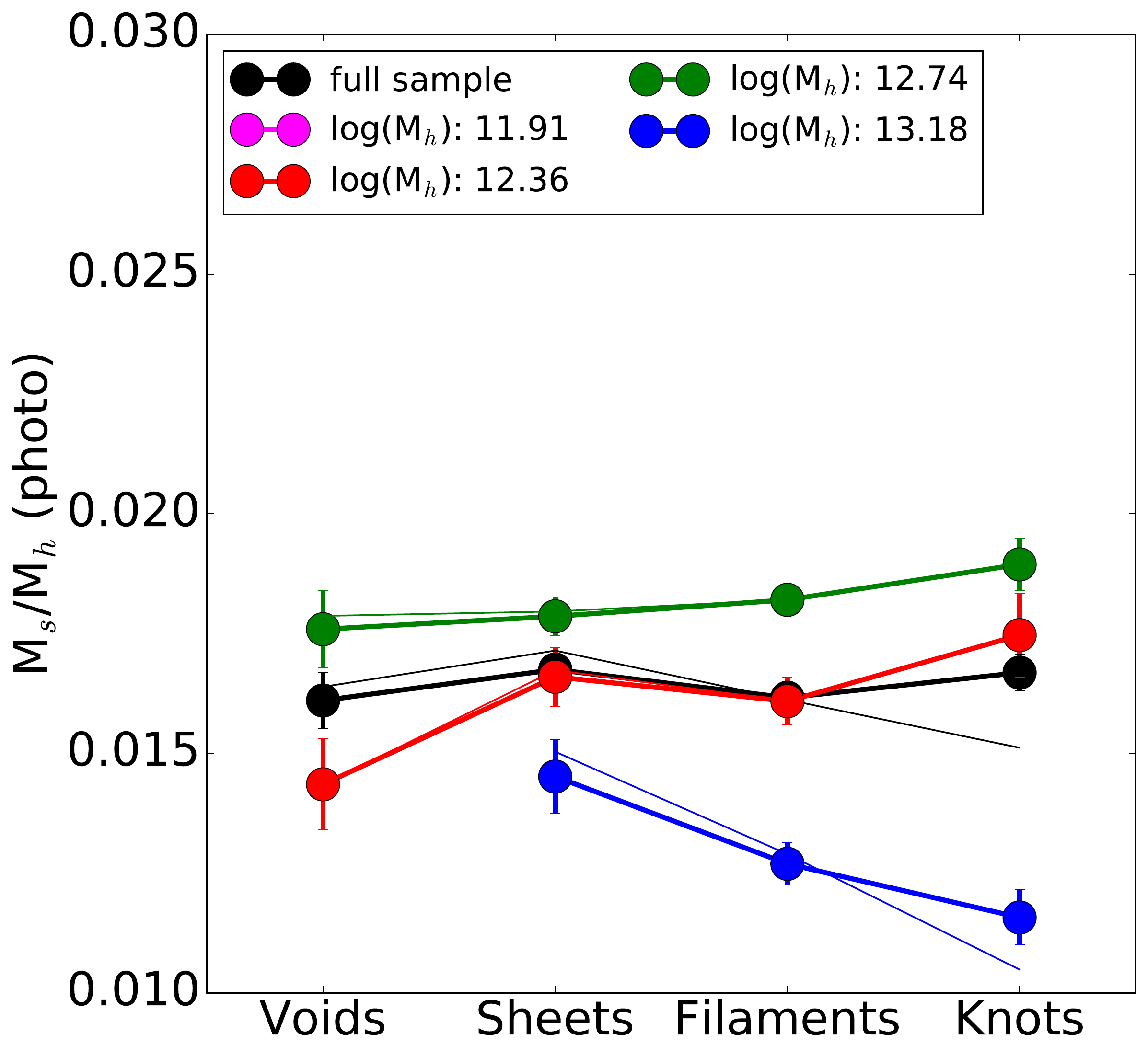}
\includegraphics[scale=0.22] 
                        {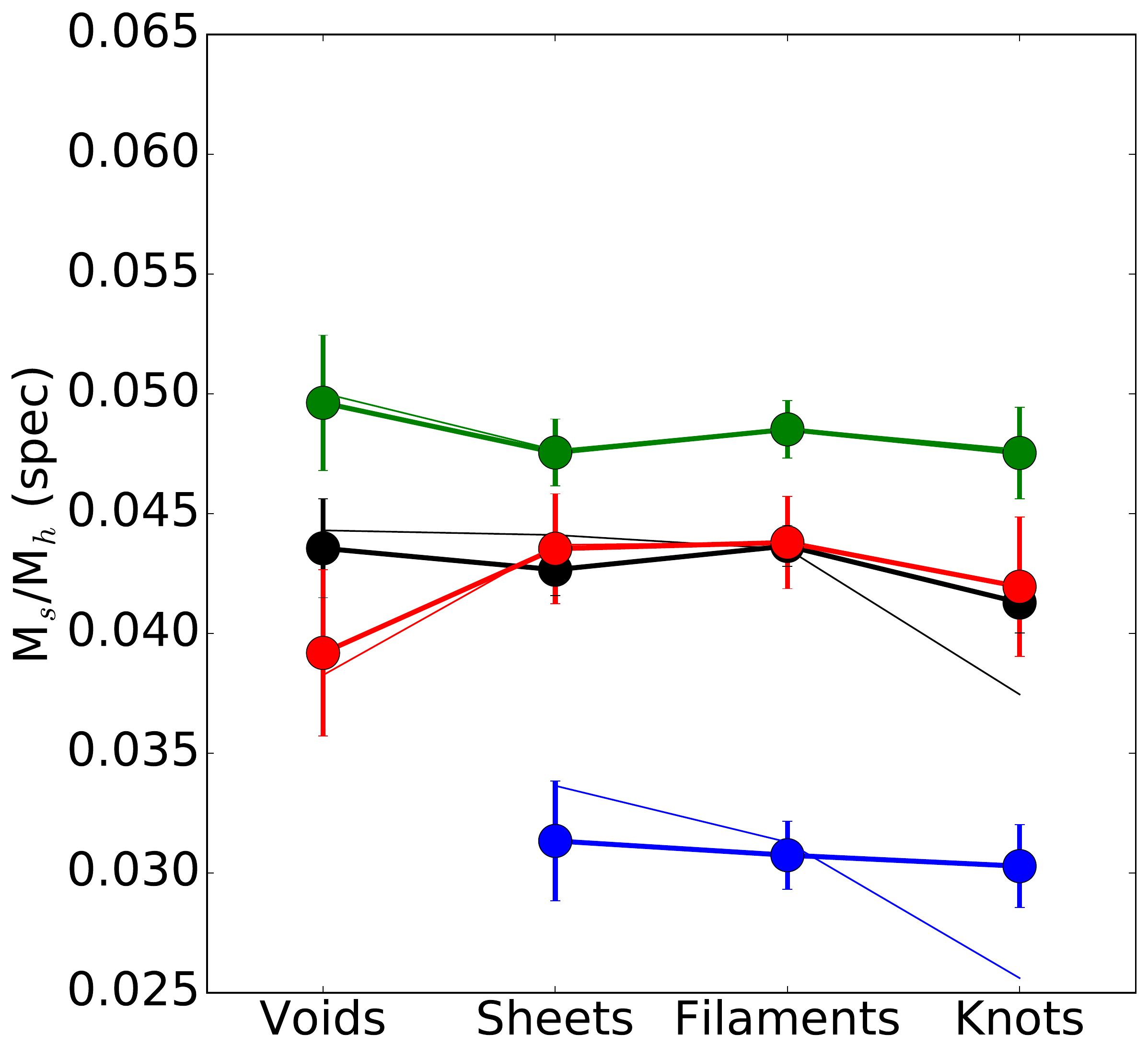}
\includegraphics[scale=0.22]
                        {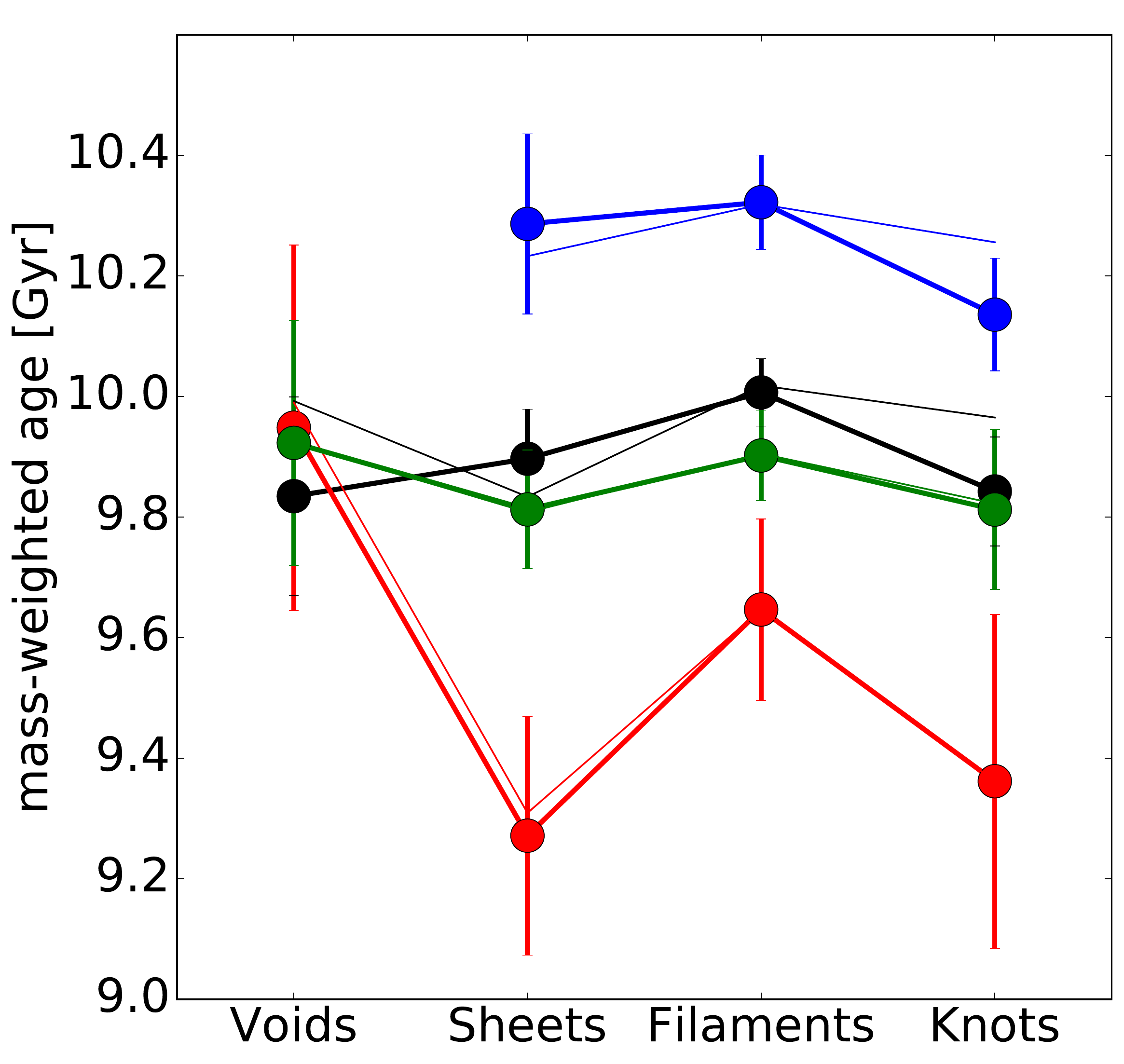}
\caption{M$_s$/M$_h$ using photometric and spectroscopic stellar mass estimates and mass-weighted age as a function of halo mass and GE, using only groups with multiplicity greater than or equal to three. Only bins with more than 50 galaxies are shown; this effectively removes our previously lowest halo mass bin, but we retail good numbers in our second lowest halo mass bin (M$_h \approx 10^{12.3}$). The trends at low and high halo mass remain unchanged: at low mass, M$_s$/M$_h$ increases from voids to knots; at high halo mass,  M$_s$/M$_h$ decreases. We continue to see a transition at intermediate halo masses, where no trend of M$_s$/M$_h$ is GE is observed.} 
\label{fig:GAMA_res_Nfof3}
\end{figure*}
\begin{figure*}
\includegraphics[scale=0.22]
                        {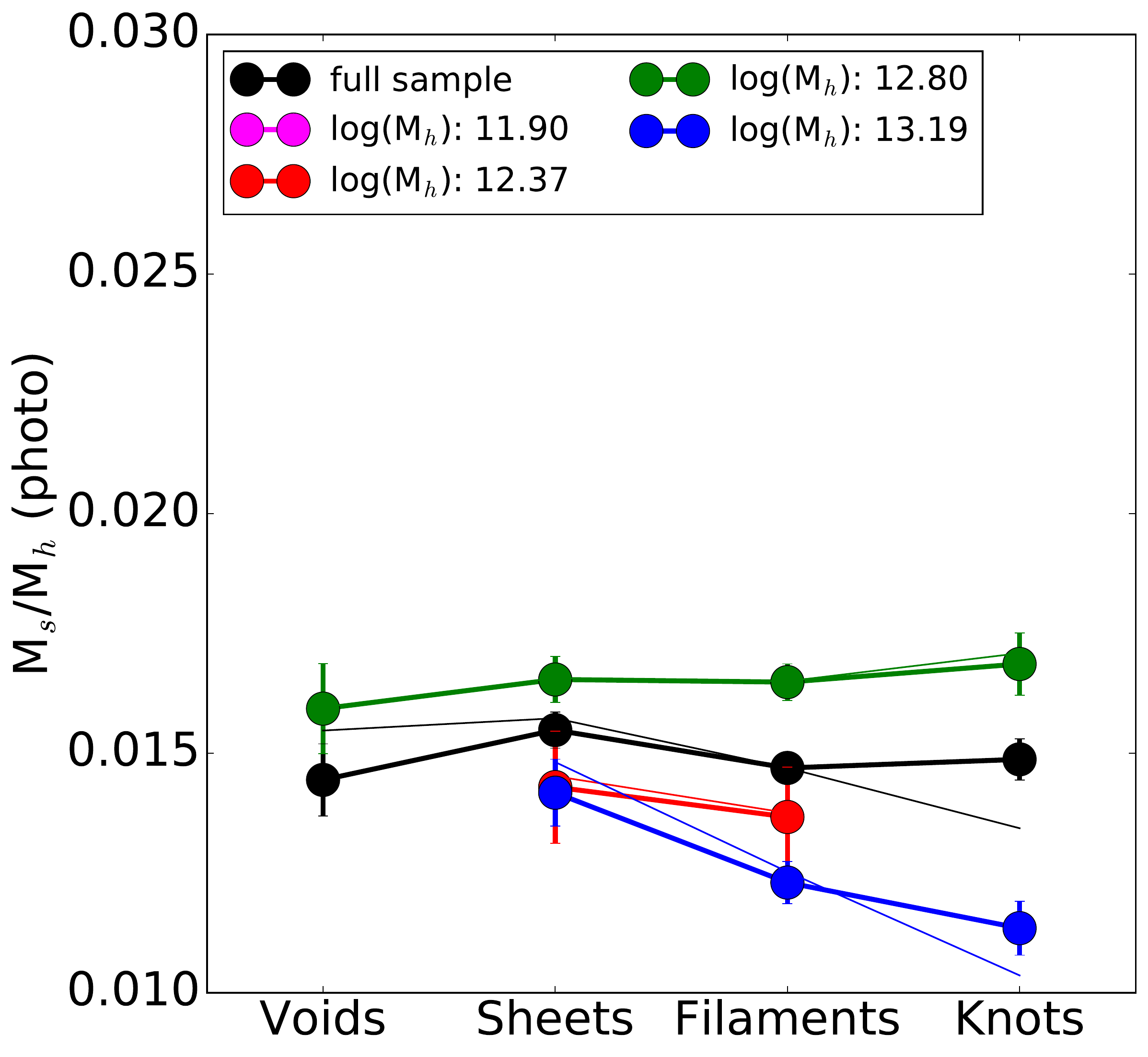}
\includegraphics[scale=0.22] 
                        {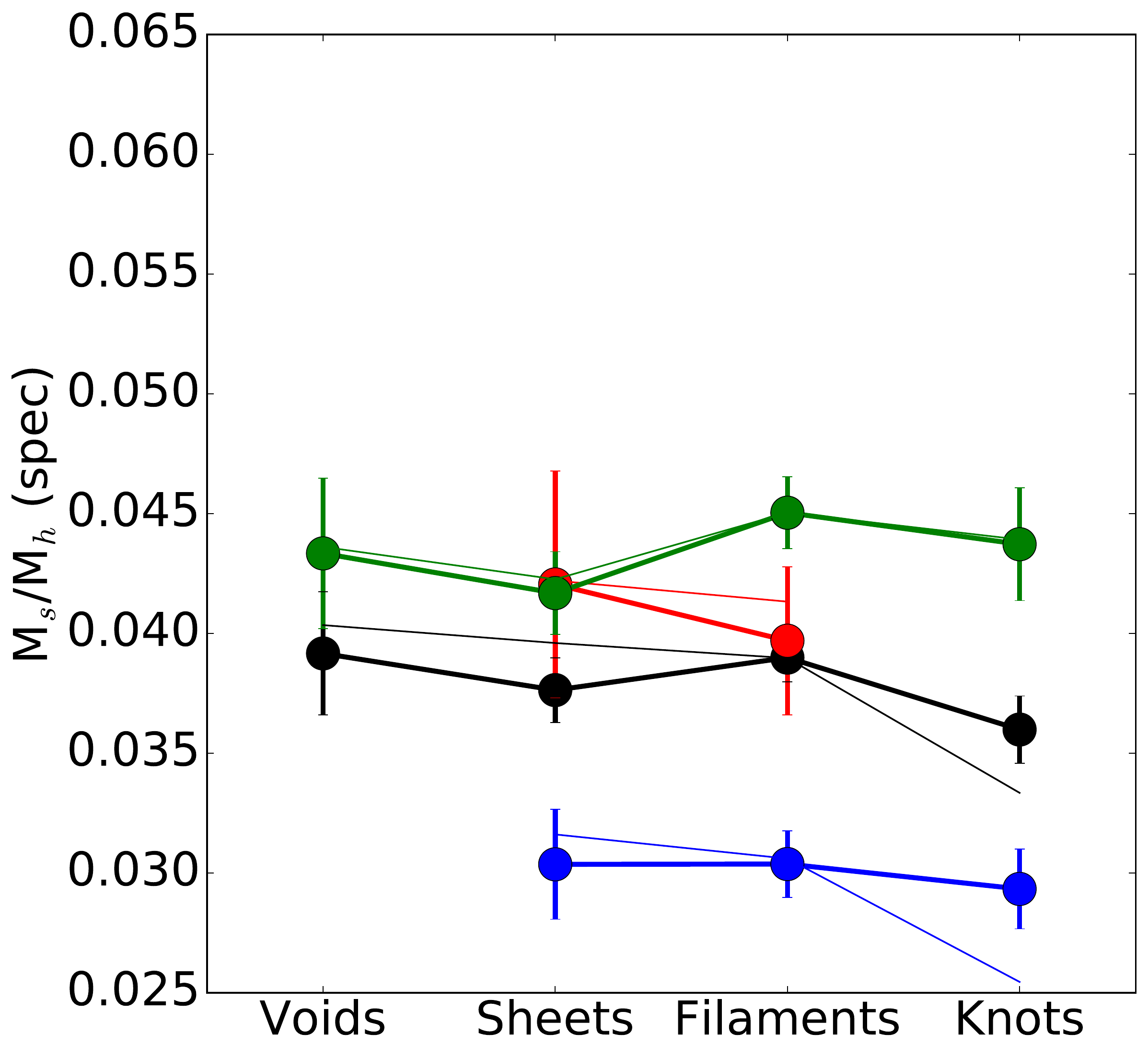}
\includegraphics[scale=0.22]
                        {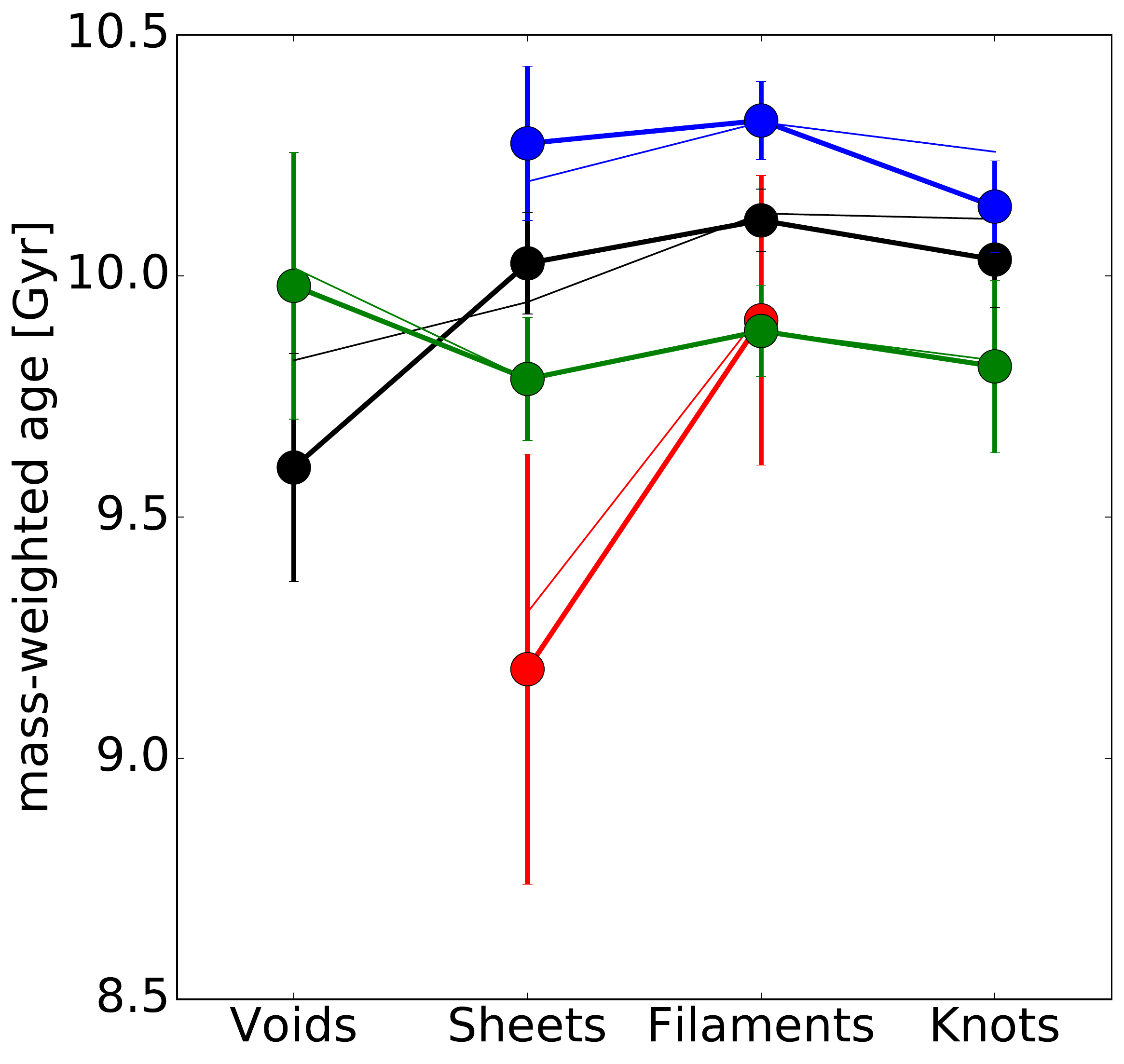}
\caption{M$_s$/M$_h$ using photometric and spectroscopic stellar mass estimates and mass-weighted age as a function of halo mass and GE, using only groups with multiplicity greater than or equal to four. Only bins with more than 50 galaxies are shown; this effectively removes our original two lowest halo mass bins. The trend at  high halo mass remain unchanged: M$_s$/M$_h$ decreases from sheets to knots. We continue to see a transition at intermediate halo masses, where no trend of M$_s$/M$_h$ is GE is observed.}
\label{fig:GAMA_res_Nfof4}
\end{figure*}

We show the effect of removing low multiplicity groups in Figs.~\ref{fig:GAMA_res_Nfof3} and \ref{fig:GAMA_res_Nfof4}, where we remove all groups with multiplicity equal to or lower than two and three, respectively. As in the previous section, we systematically loose the galaxies in the least massive haloes. The trend of increasing \MsMh(photo) towards knots on low mass haloes and decreasing towards not on high mass haloes is conserved where enough galaxies remain. The statistics in \MsMh(spec) and mass-weighed age, already insufficient when using all groups, are further degraded but the results remain entirely consistent with those observed in Fig.~\ref{fig:GAMA_res}.

We conclude from this section that GAMA data shows convincing evidence for a dependence of \MsMh as a function of GE at fixed halo mass, which we interpret in the next section in the context of halo assembly bias.

\subsection{Interpretation}

According to the study in Section~\ref{sec:L-Galaxies}, the formation time - or age - of a dark matter halo is correlated with the star-formation histories of the galaxies. However, we see no strong evidence of any direct measure of the shape of the SFH (i.e., mass-weighted age, $t_{85}$ or fraction in young stars) of a galaxy changing with GE, at fixed halo mass. The only significant trend comes from M$_s$/M$_h$. We interpreted this result as an indication that low-mass haloes (M$_h < 10^{12.3}$M$_\odot$ - equivalent to our two lowest mass bins, shown on the pink and red lines) are younger in voids than haloes of the same mass in knots; conversely, high-mass haloes (M$_h > 10^{13.2}$M$_\odot$ - corresponding to our largest mass bin, shown in the blue line) are older in voids and younger in knots. 

{\it Why do we not see a signal in our age estimators?} According to the expectations from the model, we should expect the age of galaxies in low mass haloes to increase from voids to knots. We can refer back to the investigations shown in Sections 2 and 3. We can compute the expected average difference in mass-weighted age, young mass fraction and \MsMh in galaxies that reside in haloes with $M_h < 10^{12} M_\odot$ with $f_{\rm 1/2}$ of roughly 9 and 10.5 Gyrs. That results in 0.6 Gyrs, $<1\%$ and 0.006 respectively (the latter is close to what we measure: a difference of around 0.005 between voids and knots at $M_h \lesssim 10^{12.3}$). According to the results in Section 3, and our obtained errors on the mean in Fig.~\ref{fig:GAMA_res}, a 1\% difference in young mass fraction is far beyond what we can measure from our data, and a change in 0.6 Gyrs on the mass-weighted age is just below or at the limit of what we might currently. Although this statement is obviously model-dependent, it provides some guidance on how to interpret our results. According to the L-Galaxies model, improving on our analysis requires a significant reduction of the scatter of mass-weighted ages at fixed halo mass and GE - this can be accomplished by a larger sample, better spectral data, or both.


{\it How do our results compare to expectations from dark-matter simulations?} Theory and simulations make different predictions according to the halo mass regime being considered. \cite{Hahn2009} specifically investigated tidal effects on the assembly rate of low mass haloes as a physical explanation of assembly bias. They find that growth of these haloes in the vicinity of high-mass haloes can be suppressed (see also \citealt{Borzyszkowski16}, and that this suppression at late times is indicative of an earlier formation time. This effectively increases the number density of small haloes in regions of a high density of large haloes, boosting their clustering amplitude. Or, in different words, low mass haloes that reside in areas of strong tidal fields are older - our results are therefore direct observational evidence supports this theory (see also \citealt{Hearin2015b}). At high-mass halos, we expect a dependence of bias with peak curvature (or concentration) and formation time (see \citealt{Dalal2008, Zentner2007}), from the consideration of the Gaussian statistics of rare high-density peaks. Several other authors have numerically established the change in trend in assembly bias from low to high mass haloes, which we observe here. The majority of such work is done in terms of large-scale bias, whereas here we focus on geometric environment. According to \cite{Fisher2016}, working on the Millennium simulation, at fixed halo mass there is a monotonic increase of halo bias with geometric environment from voids to knots. This puts our results in agreement with well-established predictions from simulations, namely that at low-mass younger haloes cluster less strongly and that at high-mass, younger halos cluster more strongly.

{\it Are our results in agreement with past work on the role of the cosmic web in galaxy properties?} Previous work is converging on a small role of GE on galaxy properties, once local density and stellar mass are taken into account (e.g. \citet{Eardley2015,BrouwerEtAl16,AlpaslanEtAl16}, see Section 1 for further detail). Of particular interest to us is the work of \cite{BrouwerEtAl16}, who explicitly look for changes in halo mass with GE, in the haloes of galaxies selected according to their stellar mass. They found no evidence that the halo mass changes with GE, once stellar mass and local density are taken into account. This is seemingly in contrast with our findings on the variation of \MsMh with GE in bins of halo mass - and potentially worrying for our conclusions given that their halo masses are computed from stacked weak-lensing profiles. However, we note that their stellar mass weighting is done over the full sample. This leaves room for a dependence of \MsMh on GE at fixed halo mass, especially as the trend we find is markedly weaker when we consider a wide range of halo masses, due to the change of the dependence on GE with halo mass (Fig.~\ref{fig:GAMA_res}). With the exception of \MsMh we do not find any strong dependence of further galaxy properties with GE. This agrees with much of the literature.%

\section{Summary and conclusions}

This paper presents a comprehensive joint analysis of simulations and data, with the aim to establish whether the dark matter haloes that reside in different geometric environments have assembled at different times. We were motivated by work in dark-matter N-body simulations that strongly suggests this should be the case, especially considering tidal effects in the growth of dark matter haloes. At the same time, work on data has failed to converge. Here we back up our work on the GAMA data (Section 4) with an exploration of the L-Galaxies semi-analytic model (Section 2), which we also use to create mock data with complex SFHs (Section 3). Our main conclusions are as follows:

\begin{enumerate}
\item The relation between halo formation time and galaxy properties is complex, especially once we consider different definitions of halo formation time. According to the L-Galaxies model, M$_s$/M$_h$, sSFR and mass-weighted age are the best proxies for halo formation time of the ones considered, and we find that sensitivity is almost always better at low halo masses. We find that instantaneous SFR is an extremely poor indicator of all halo formation times that we consider, at all halo masses.
\item Using realist SFHs to create spectral mocks with GAMA-like properties, we found that a lack of knowledge of the dust attenuation model, extinction curve and dust geometry imparts significant biases on the recovered physical parameters. With the exception of mass-weighted metallicity, these biases are constant offsets that may be neglected if one considers only differences on the mean of well-defined samples. 
\item At all halo masses, GAMA central galaxies residing in different geometric environments show no significant difference in mass-weighted age, mass-weighted metallicity, mass fraction of young stars or time at which 85\% of stellar mass formed. However, low mass haloes show a steadily increasing M$_s$/M$_h$ from voids to knots, with the trend being reversed at the high-mass end. Using the results from L-Galaxies and agreeing with much of the literature, we interpret this measurement as an indication that low-mass haloes living in knots are older than haloes of the same mass living in voids, and that high-mass haloes living in knots are younger than halos of the same mass living in voids. This agrees with theoretical work and results from dark-matter simulations. Our work is the first direct observational evidence for strong tidal interactions suppressing the accretion of low mass haloes in regions  where such interactions are more likely - such as knots.

\end{enumerate}

We note that the spectral quality of GAMA data is too low to allow us to confirm the age dependence of haloes with GE with stellar ages. Our spectral mocks show that we currently sit just below the needed S/N in order to be able to make such comparisons, and that we require either higher S/N spectra or larger galaxy numbers.  We hope that future datasets, such as the Bright Galaxy Survey planned as part of Dark Energy Spectroscopic Instrument (DESI), will help us make progress on that front. Finally, we also remark on the fact that integrated stellar populations in and by themselves may simply never prove to be sufficient to fully explore the galaxy-halo connection in sufficient detail. Work as the one detailed here hints at the possibility that stellar and gas kinematics and spacially-resolved SFHs might help us understand the details of this complex relationship - we leave such explorations for future work.

  
\section{Acknowledgements}
RT would like to thank Risa Wechsler for motivating conversations on this topic. RT acknowledges support from the Science and Technology Facilities Council via an Ernest Rutherford Fellowship (grant number ST/K004719/1). VW and EE acknowledge support of the European Research Council via the award of a starting grant (SEDMorph; P.I. V.~Wild).  PAT (ORCID 0000-0001-6888-6483) acknowledges support from the Science and Technology Facilities Council (grant number ST/L000652/1).

GAMA is a joint European-Australasian project based around a spectroscopic campaign using the Anglo-Australian Telescope. The GAMA input catalogue is based on data taken from the Sloan Digital Sky Survey and the UKIRT Infrared Deep Sky Survey. Complementary imaging of the GAMA regions is being obtained by a number of independent survey programmes including GALEX MIS, VST KiDS, VISTA VIKING, WISE, Herschel-ATLAS, GMRT and ASKAP providing UV to radio coverage. GAMA is funded by the STFC (UK), the ARC (Australia), the AAO, and the participating institutions. The GAMA website is http://www.gama-survey.org/

The Millennium Simulation databases used in this paper and the web application providing online access to them were constructed as part of the activities of the German Astrophysical Virtual Observatory (GAVO).

Funding for the SDSS and SDSS-II has been provided by the Alfred P. Sloan Foundation, the Participating Institutions, the National Science Foundation, the U.S. Department of Energy, the National Aeronautics and Space Administration, the Japanese Monbukagakusho, the Max Planck Society, and the Higher Education Funding Council for England. The SDSS Web Site is http://www.sdss.org/.

The SDSS is managed by the Astrophysical Research Consortium for the Participating Institutions. The Participating Institutions are the American Museum of Natural History, Astrophysical Institute Potsdam, University of Basel, University of Cambridge, Case Western Reserve University, University of Chicago, Drexel University, Fermilab, the Institute for Advanced Study, the Japan Participation Group, Johns Hopkins University, the Joint Institute for Nuclear Astrophysics, the Kavli Institute for Particle Astrophysics and Cosmology, the Korean Scientist Group, the Chinese Academy of Sciences (LAMOST), Los Alamos National Laboratory, the Max-Planck-Institute for Astronomy (MPIA), the Max-Planck-Institute for Astrophysics (MPA), New Mexico State University, Ohio State University, University of Pittsburgh, University of Portsmouth, Princeton University, the United States Naval Observatory, and the University of Washington.

\bibliography{bib_me}

\begin{thebibliography}{}

\bibitem[\protect\citeauthoryear{{Alonso}, {Eardley} \& {Peacock}}{{Alonso}
  et~al.}{2015}]{AlonsoEtAl14}
{Alonso} D.,  {Eardley} E.,    {Peacock} J.~A.,  2015, \mnras, 447, 2683

\bibitem[\protect\citeauthoryear{{Alpaslan} et~al.,}{{Alpaslan}
  et~al.}{2016}]{AlpaslanEtAl16}
{Alpaslan} M.,  et~al., 2016, \mnras, 457, 2287

\bibitem[\protect\citeauthoryear{{Bernardi} et~al.,}{{Bernardi}
  et~al.}{2016}]{Bernardi2016}
{Bernardi} M.,  et~al., 2016, \mnras, 455, 4122

\bibitem[\protect\citeauthoryear{{Blanton} \& {Berlind}}{{Blanton} \&
  {Berlind}}{2007}]{Blanton2007}
{Blanton} M.~R.,  {Berlind} A.~A.,  2007, \apj, 664, 791

\bibitem[\protect\citeauthoryear{{Borzyszkowski} et~al.,}{{Borzyszkowski}
  et~al.}{2016}]{Borzyszkowski16}
{Borzyszkowski} M.,  et~al., 2016, arXiv:1610.04231

\bibitem[\protect\citeauthoryear{{Brouwer} et~al.,}{{Brouwer}
  et~al.}{2016}]{BrouwerEtAl16}
{Brouwer} M.~M.,  et~al., 2016, arXiv e-prints: 1604.07233

\bibitem[\protect\citeauthoryear{{Brown} et~al.,}{{Brown}
  et~al.}{2008}]{Brown2008}
{Brown} M.~J.~I.,  et~al., 2008, \apj, 682, 937

\bibitem[\protect\citeauthoryear{{Bruzual} \& {Charlot}}{{Bruzual} \&
  {Charlot}}{2003}]{Bruzual2003}
{Bruzual} G.,  {Charlot} S.,  2003, \mnras, 344, 1000

\bibitem[\protect\citeauthoryear{{Charlot} \& {Fall}}{{Charlot} \&
  {Fall}}{2000}]{CharlotFall00}
{Charlot} S.,  {Fall} S.~M.,  2000, \apj, 539, 718

\bibitem[\protect\citeauthoryear{{Chaves-Montero} et~al.,}{{Chaves-Montero}
  et~al.}{2016}]{ChavesMontero2016}
{Chaves-Montero} J.,  et~al., 2016, \mnras, 460, 3100

\bibitem[\protect\citeauthoryear{{Christodoulou} et~al.,}{{Christodoulou}
  et~al.}{2012}]{Christodoulou2012}
{Christodoulou} L.,  et~al., 2012, \mnras, 425, 1527

\bibitem[\protect\citeauthoryear{{Conroy}, {Gunn} \& {White}}{{Conroy}
  et~al.}{2009}]{Conroy2009}
{Conroy} C.,  {Gunn} J.~E.,    {White} M.,  2009, \apj, 699, 486

\bibitem[\protect\citeauthoryear{{Conroy}, {Wechsler} \& {Kravtsov}}{{Conroy}
  et~al.}{2006}]{Conroy2006}
{Conroy} C.,  {Wechsler} R.~H.,    {Kravtsov} A.~V.,  2006, \apj, 647, 201

\bibitem[\protect\citeauthoryear{{Cooray}}{{Cooray}}{2006}]{Cooray2006}
{Cooray} A.,  2006, \mnras, 365, 842

\bibitem[\protect\citeauthoryear{{Crain} et~al.,}{{Crain}
  et~al.}{2015}]{Crain2015}
{Crain} R.~A.,  et~al., 2015, \mnras, 450, 1937

\bibitem[\protect\citeauthoryear{{Cresswell} \& {Percival}}{{Cresswell} \&
  {Percival}}{2009}]{Cresswell2009}
{Cresswell} J.~G.,  {Percival} W.~J.,  2009, \mnras, 392, 682

\bibitem[\protect\citeauthoryear{{Dalal} et~al.,}{{Dalal}
  et~al.}{2008}]{Dalal2008}
{Dalal} N.,  et~al., 2008, \apj, 687, 12

\bibitem[\protect\citeauthoryear{{Darvish} et~al.,}{{Darvish}
  et~al.}{2014}]{DarvishEtAl14}
{Darvish} B.,  et~al., 2014, \apj, 796, 51

\bibitem[\protect\citeauthoryear{{de Jong} et~al.,}{{de Jong}
  et~al.}{2013}]{deJong2013}
{de Jong} J.~T.~A.,  et~al., 2013, The Messenger, 154, 44

\bibitem[\protect\citeauthoryear{{de la Torre} \& {Peacock}}{{de la Torre} \&
  {Peacock}}{2013}]{delaTorre2013b}
{de la Torre} S.,  {Peacock} J.~A.,  2013, \mnras, 435, 743

\bibitem[\protect\citeauthoryear{{De Lucia} et~al.,}{{De Lucia}
  et~al.}{2006}]{deLucia2006}
{De Lucia} G.,  et~al., 2006, \mnras, 366, 499

\bibitem[\protect\citeauthoryear{{Driver} et~al.,}{{Driver}
  et~al.}{2009}]{Driver2009}
{Driver} S.~P.,  et~al., 2009, Astronomy and Geophysics, 50, 12

\bibitem[\protect\citeauthoryear{{Driver} et~al.,}{{Driver}
  et~al.}{2011}]{Driver2011}
{Driver} S.~P.,  et~al., 2011, \mnras, 413, 971

\bibitem[\protect\citeauthoryear{{Eardley} et~al.,}{{Eardley}
  et~al.}{2015}]{Eardley2015}
{Eardley} E.,  et~al., 2015, \mnras, 448, 3665

\bibitem[\protect\citeauthoryear{{Faltenbacher} \& {White}}{{Faltenbacher} \&
  {White}}{2010}]{Faltenbacher2010}
{Faltenbacher} A.,  {White} S.~D.~M.,  2010, \apj, 708, 469

\bibitem[\protect\citeauthoryear{{Fisher} \& {Faltenbacher}}{{Fisher} \&
  {Faltenbacher}}{2016}]{Fisher2016}
{Fisher} J.~D.,  {Faltenbacher} A.,  2016, arXiv e-prints: 1603.06955

\bibitem[\protect\citeauthoryear{{Gao}, {Springel} \& {White}}{{Gao}
  et~al.}{2005}]{Gao2005}
{Gao} L.,  {Springel} V.,    {White} S.~D.~M.,  2005, \mnras, 363, L66

\bibitem[\protect\citeauthoryear{{Guo} et~al.,}{{Guo}  et~al.}{2013}]{Guo2013}
{Guo} H.,  et~al., 2013, \apj, 767, 122

\bibitem[\protect\citeauthoryear{{Hahn} et~al.,}{{Hahn}
  et~al.}{2009}]{Hahn2009}
{Hahn} O.,  et~al., 2009, \mnras, 398, 1742

\bibitem[\protect\citeauthoryear{{Han} et~al.,}{{Han}
  et~al.}{2015}]{HanEtAl15}
{Han} J.,  et~al., 2015, \mnras, 446, 1356

\bibitem[\protect\citeauthoryear{{Hearin}}{{Hearin}}{2015}]{Hearin2015a}
{Hearin} A.~P.,  2015, \mnras, 451, L45

\bibitem[\protect\citeauthoryear{{Hearin}, {Watson} \& {van den
  Bosch}}{{Hearin} et~al.}{2015}]{Hearin2015b}
{Hearin} A.~P.,  {Watson} D.~F.,    {van den Bosch} F.~C.,  2015, \mnras, 452,
  1958

\bibitem[\protect\citeauthoryear{{Henriques} et~al.,}{{Henriques}
  et~al.}{2015}]{Henriques2015}
{Henriques} B.~M.~B.,  et~al., 2015, \mnras, 451, 2663

\bibitem[\protect\citeauthoryear{{Hopkins} et~al.,}{{Hopkins}
  et~al.}{2013}]{Hopkins2013}
{Hopkins} A.~M.,  et~al., 2013, \mnras, 430, 2047

\bibitem[\protect\citeauthoryear{{Jing}, {Mo} \& {B{\"o}rner}}{{Jing}
  et~al.}{1998}]{Jing1998}
{Jing} Y.~P.,  {Mo} H.~J.,    {B{\"o}rner} G.,  1998, \apj, 494, 1

\bibitem[\protect\citeauthoryear{{Kaiser}}{{Kaiser}}{1984}]{Kaiser84}
{Kaiser} N.,  1984, \apjl, 284, L9

\bibitem[\protect\citeauthoryear{{Kravtsov}, {Berlind}, {Wechsler}, {Klypin},
  {Gottl{\"o}ber}, {Allgood} \& {Primack}}{{Kravtsov}
  et~al.}{2004}]{Kravtsov2004}
{Kravtsov} A.~V.,  {Berlind} A.~A.,  {Wechsler} R.~H.,  {Klypin} A.~A.,
  {Gottl{\"o}ber} S.,  {Allgood} B.,    {Primack} J.~R.,  2004, \apj, 609, 35

\bibitem[\protect\citeauthoryear{{Lacerna} \& {Padilla}}{{Lacerna} \&
  {Padilla}}{2011}]{LacernaPadilla12001}
{Lacerna} I.,  {Padilla} N.,  2011, \mnras, 412, 1283

\bibitem[\protect\citeauthoryear{{Lacerna}, {Padilla} \& {Stasyszyn}}{{Lacerna}
  et~al.}{2014}]{Lacerna2014}
{Lacerna} I.,  {Padilla} N.,    {Stasyszyn} F.,  2014, \mnras, 443, 3107

\bibitem[\protect\citeauthoryear{{Lemson} \& {Virgo Consortium}}{{Lemson} \&
  {Virgo Consortium}}{2006}]{Lemson2006}
{Lemson} G.,  {Virgo Consortium} t.,  2006, arXiv e-prints: 0608019

\bibitem[\protect\citeauthoryear{{Li}, {Mo} \& {Gao}}{{Li}
  et~al.}{2008}]{Li2008}
{Li} Y.,  {Mo} H.~J.,    {Gao} L.,  2008, \mnras, 389, 1419

\bibitem[\protect\citeauthoryear{{Lim}, {Mo}, {Wang} \& {Yang}}{{Lim}
  et~al.}{2015}]{Lim2015}
{Lim} S.,  {Mo} H.,  {Wang} H.,    {Yang} X.,  2015, ArXiv e-prints: 1502.01256

\bibitem[\protect\citeauthoryear{{Lin} et~al.,}{{Lin}  et~al.}{2015}]{Lin2015}
{Lin} Y.-T.,  et~al., 2015, arXiv e-prints: 1504.07632

\bibitem[\protect\citeauthoryear{{Liske} et~al.,}{{Liske}
  et~al.}{2015}]{Liske2015}
{Liske} J.,  et~al., 2015, \mnras, 452, 2087

\bibitem[\protect\citeauthoryear{{Manera} et~al.,}{{Manera}
  et~al.}{2013}]{Manera2013}
{Manera} M.,  et~al., 2013, MNRAS, 428, 1036

\bibitem[\protect\citeauthoryear{{Manera} et~al.,}{{Manera}
  et~al.}{2015}]{Manera2015}
{Manera} M.,  et~al., 2015, \mnras, 447, 437

\bibitem[\protect\citeauthoryear{{Mathis}, {Mezger} \& {Panagia}}{{Mathis}
  et~al.}{1983}]{Mathis83}
{Mathis} J.~S.,  {Mezger} P.~G.,    {Panagia} N.,  1983, \aap, 128, 212

\bibitem[\protect\citeauthoryear{{Mo}, {Jing} \& {White}}{{Mo}
  et~al.}{1996}]{Mo1996}
{Mo} H.~J.,  {Jing} Y.~P.,    {White} S.~D.~M.,  1996, \mnras, 282, 1096

\bibitem[\protect\citeauthoryear{{Norberg} et~al.,}{{Norberg}
  et~al.}{2001}]{Norberg2001}
{Norberg} P.,  et~al., 2001, \mnras, 328, 64

\bibitem[\protect\citeauthoryear{{Norberg} et~al.,}{{Norberg}
  et~al.}{2002}]{Norberg2002}
{Norberg} P.,  et~al., 2002, \mnras, 332, 827

\bibitem[\protect\citeauthoryear{Peacock \& Smith}{Peacock \&
  Smith}{2000}]{Peacock2000}
Peacock J.~A.,  Smith R.~E.,  2000, \mnras, 318, 1144

\bibitem[\protect\citeauthoryear{{Robotham} et~al.,}{{Robotham}
  et~al.}{2011}]{Robotham2011}
{Robotham} A.~S.~G.,  et~al., 2011, \mnras, 416, 2640

\bibitem[\protect\citeauthoryear{{Ross} \& {Brunner}}{{Ross} \&
  {Brunner}}{2009}]{Ross2009}
{Ross} A.~J.,  {Brunner} R.~J.,  2009, \mnras, 399, 878

\bibitem[\protect\citeauthoryear{{Ross}, {Tojeiro} \& {Percival}}{{Ross}
  et~al.}{2011}]{Ross2011}
{Ross} A.~J.,  {Tojeiro} R.,    {Percival} W.~J.,  2011, \mnras, 413, 2078

\bibitem[\protect\citeauthoryear{{Schaye} et~al.,}{{Schaye}
  et~al.}{2015}]{Schaye2015}
{Schaye} J.,  et~al., 2015, \mnras, 446, 521

\bibitem[\protect\citeauthoryear{{Seljak}}{{Seljak}}{2000}]{Seljak2000}
{Seljak} U.,  2000, \mnras, 318, 203

\bibitem[\protect\citeauthoryear{{Shamshiri} et~al.,}{{Shamshiri}
  et~al.}{2015}]{ShamshiriEtAl15}
{Shamshiri} S.,  et~al., 2015, \mnras, 451, 2681

\bibitem[\protect\citeauthoryear{{Sheth} \& {Tormen}}{{Sheth} \&
  {Tormen}}{2004}]{ShethTormen2004}
{Sheth} R.~K.,  {Tormen} G.,  2004, \mnras, 350, 1385

\bibitem[\protect\citeauthoryear{{Skibba} et~al.,}{{Skibba}
  et~al.}{2009}]{Skibba2009}
{Skibba} R.~A.,  et~al., 2009, \mnras, 399, 966

\bibitem[\protect\citeauthoryear{Springel et~al.,}{Springel
  et~al.}{2005}]{Springel2005}
Springel V.,  et~al., 2005, Nature, 435, 629

\bibitem[\protect\citeauthoryear{{Swanson} et~al.,}{{Swanson}
  et~al.}{2008}]{Swanson2008}
{Swanson} M.~E.~C.,  et~al., 2008, \mnras, 385, 1635

\bibitem[\protect\citeauthoryear{{Taylor} et~al.,}{{Taylor}
  et~al.}{2011}]{Taylor2011}
{Taylor} E.~N.,  et~al., 2011, \mnras, 418, 1587

\bibitem[\protect\citeauthoryear{{Tinker}, {Wetzel} \& {Conroy}}{{Tinker}
  et~al.}{2011}]{Tinker2011}
{Tinker} J.,  {Wetzel} A.,    {Conroy} C.,  2011, ArXiv e-prints: 1107.5046

\bibitem[\protect\citeauthoryear{{Tinker} et~al.,}{{Tinker}
  et~al.}{2008}]{Tinker2008b}
{Tinker} J.~L.,  et~al., 2008, \apj, 686, 53

\bibitem[\protect\citeauthoryear{{Tinker} et~al.,}{{Tinker}
  et~al.}{2010}]{Tinker2010}
{Tinker} J.~L.,  et~al., 2010, \apj, 724, 878

\bibitem[\protect\citeauthoryear{{Tojeiro} et~al.,}{{Tojeiro}
  et~al.}{2007}]{Tojeiro2007}
{Tojeiro} R.,  et~al., 2007, \mnras, 381, 1252

\bibitem[\protect\citeauthoryear{{Tojeiro} et~al.,}{{Tojeiro}
  et~al.}{2009}]{Tojeiro2009}
{Tojeiro} R.,  et~al., 2009, \apjs, 185, 1

\bibitem[\protect\citeauthoryear{{Tojeiro} et~al.,}{{Tojeiro}
  et~al.}{2011}]{Tojeiro2011}
{Tojeiro} R.,  et~al., 2011, \mnras, 413, 434

\bibitem[\protect\citeauthoryear{{Tojeiro} et~al.,}{{Tojeiro}
  et~al.}{2012}]{Tojeiro2012}
{Tojeiro} R.,  et~al., 2012, \mnras, 424, 136

\bibitem[\protect\citeauthoryear{{van den Bosch}, {Yang} \& {Mo}}{{van den
  Bosch} et~al.}{2003}]{Bosch2003}
{van den Bosch} F.~C.,  {Yang} X.,    {Mo} H.~J.,  2003, \mnras, 340, 771

\bibitem[\protect\citeauthoryear{{Wake} et~al.,}{{Wake}
  et~al.}{2008}]{Wake2008}
{Wake} D.~A.,  et~al., 2008, \mnras, 387, 1045

\bibitem[\protect\citeauthoryear{{Wake} et~al.,}{{Wake}
  et~al.}{2011}]{Wake2011}
{Wake} D.~A.,  et~al., 2011, \apj, 728, 46

\bibitem[\protect\citeauthoryear{{Wang} et~al.,}{{Wang}
  et~al.}{2011}]{Wang2011}
{Wang} H.,  et~al., 2011, \mnras, 413, 1973

\bibitem[\protect\citeauthoryear{{Wang} et~al.,}{{Wang}
  et~al.}{2013}]{Wang2013}
{Wang} L.,  et~al., 2013, \mnras, 433, 515

\bibitem[\protect\citeauthoryear{{Wang} et~al.,}{{Wang}
  et~al.}{2008}]{Wang2008}
{Wang} Y.,  et~al., 2008, \apj, 687, 919

\bibitem[\protect\citeauthoryear{{Wechsler} et~al.,}{{Wechsler}
  et~al.}{2006}]{Wechsler2006}
{Wechsler} R.~H.,  et~al., 2006, \apj, 652, 71

\bibitem[\protect\citeauthoryear{{White} et~al.,}{{White}
  et~al.}{2007}]{White2007}
{White} M.,  et~al., 2007, \apjl, 655, L69

\bibitem[\protect\citeauthoryear{{White}, {Tinker} \& {McBride}}{{White}
  et~al.}{2014}]{White2014}
{White} M.,  {Tinker} J.~L.,    {McBride} C.~K.,  2014, \mnras, 437, 2594

\bibitem[\protect\citeauthoryear{{White} \& {Rees}}{{White} \&
  {Rees}}{1978}]{White1978}
{White} S.~D.~M.,  {Rees} M.~J.,  1978, \mnras, 183, 341

\bibitem[\protect\citeauthoryear{{Yang} et~al.,}{{Yang}
  et~al.}{2007}]{Yang2007}
{Yang} X.,  et~al., 2007, \apj, 671, 153

\bibitem[\protect\citeauthoryear{{Yang}, {Mo} \& {van den Bosch}}{{Yang}
  et~al.}{2006}]{Yang2006}
{Yang} X.,  {Mo} H.~J.,    {van den Bosch} F.~C.,  2006, \apj, 638, L55

\bibitem[\protect\citeauthoryear{{York} et~al.,}{{York}
  et~al.}{2000}]{York2000}
{York} D.~G.,  et~al., 2000, {The Astronomical Journal}, 120, 1579

\bibitem[\protect\citeauthoryear{{Zehavi} et~al.,}{{Zehavi}
  et~al.}{2005}]{Zehavi2005}
{Zehavi} I.,  et~al., 2005, \apj, 621, 22

\bibitem[\protect\citeauthoryear{{Zehavi} et~al.,}{{Zehavi}
  et~al.}{2011}]{Zehavi2011}
{Zehavi} I.,  et~al., 2011, \apj, 736, 59

\bibitem[\protect\citeauthoryear{{Zentner}}{{Zentner}}{2007}]{Zentner2007}
{Zentner} A.~R.,  2007, International Journal of Modern Physics D, 16, 763

\bibitem[\protect\citeauthoryear{{Zentner} et~al.,}{{Zentner}
  et~al.}{2016}]{Zentner2016}
{Zentner} A.~R.,  et~al., 2016, ArXiv e-prints: 1606.07817

\bibitem[\protect\citeauthoryear{{Zentner}, {Hearin} \& {van den
  Bosch}}{{Zentner} et~al.}{2014}]{Zentner2014}
{Zentner} A.~R.,  {Hearin} A.~P.,    {van den Bosch} F.~C.,  2014, \mnras, 443,
  3044

\bibitem[\protect\citeauthoryear{{Zheng}, {Coil} \& {Zehavi}}{{Zheng}
  et~al.}{2007}]{Zheng2007}
{Zheng} Z.,  {Coil} A.~L.,    {Zehavi} I.,  2007, \apj, 667, 760

\end{thebibliography}

\appendix
\section{Dust modelling effects} \label{sec:appendix_dust}

\begin{figure*}
\includegraphics[scale=0.45]{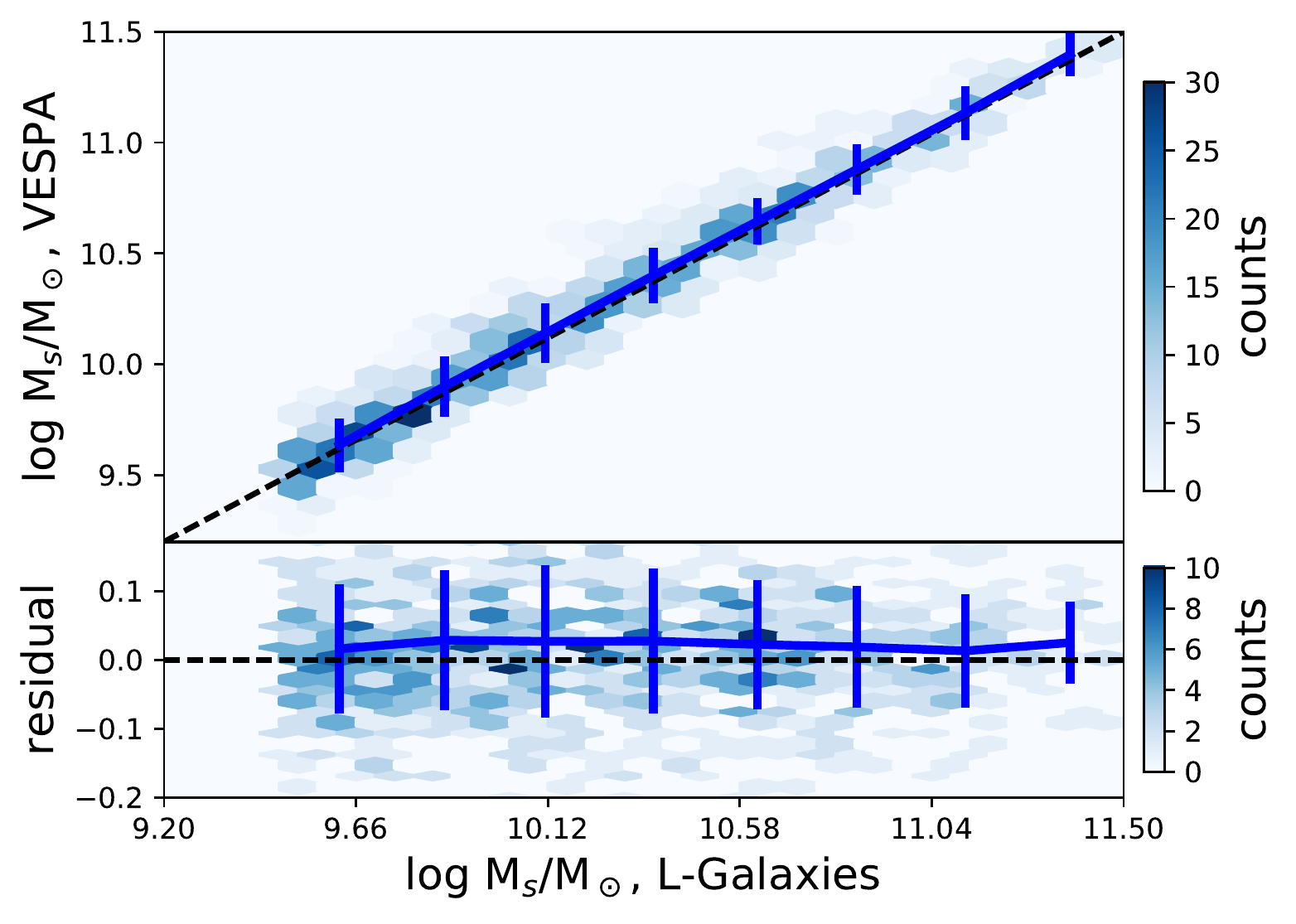}
\includegraphics[scale=0.45]{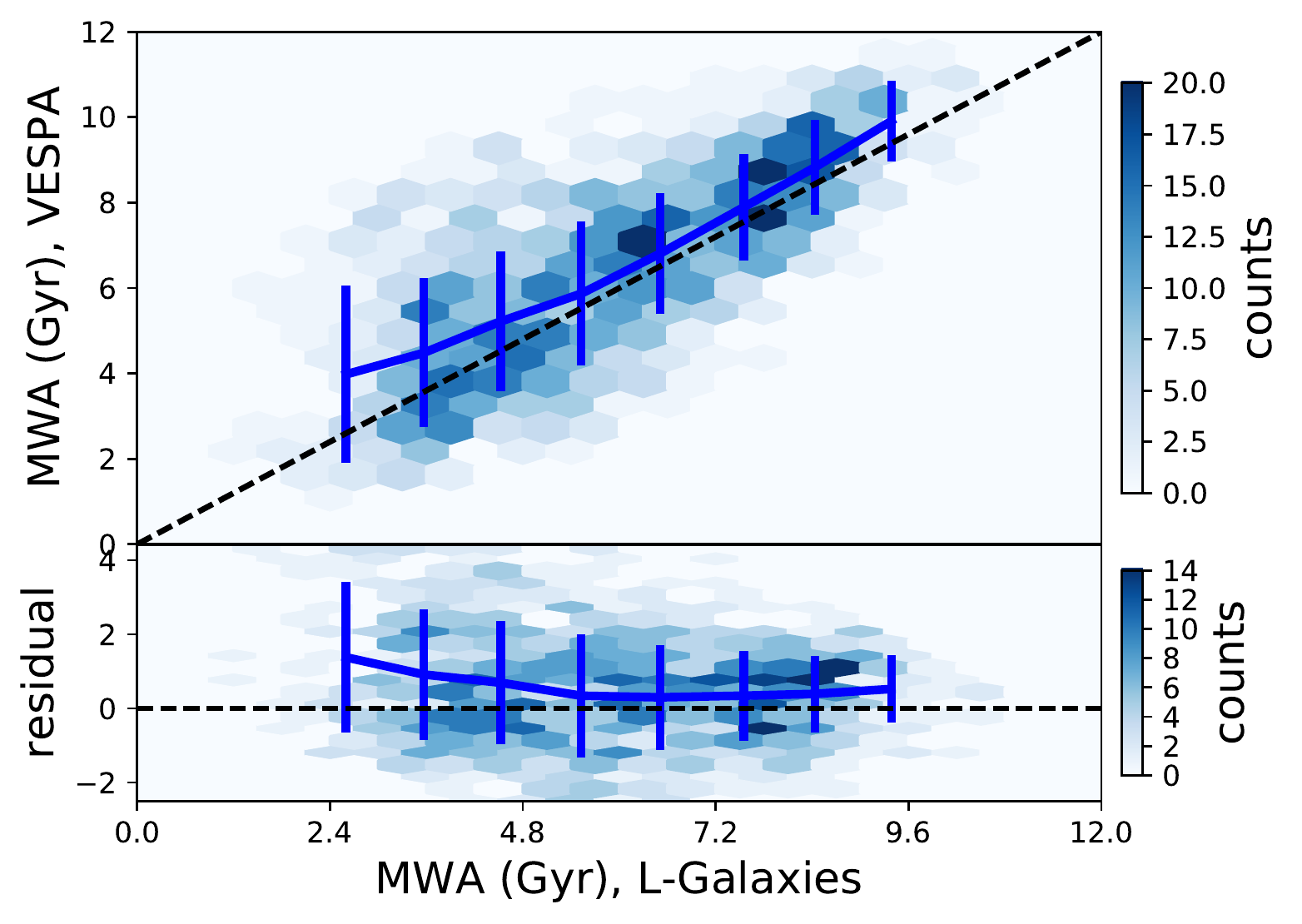}
\includegraphics[scale=0.45]{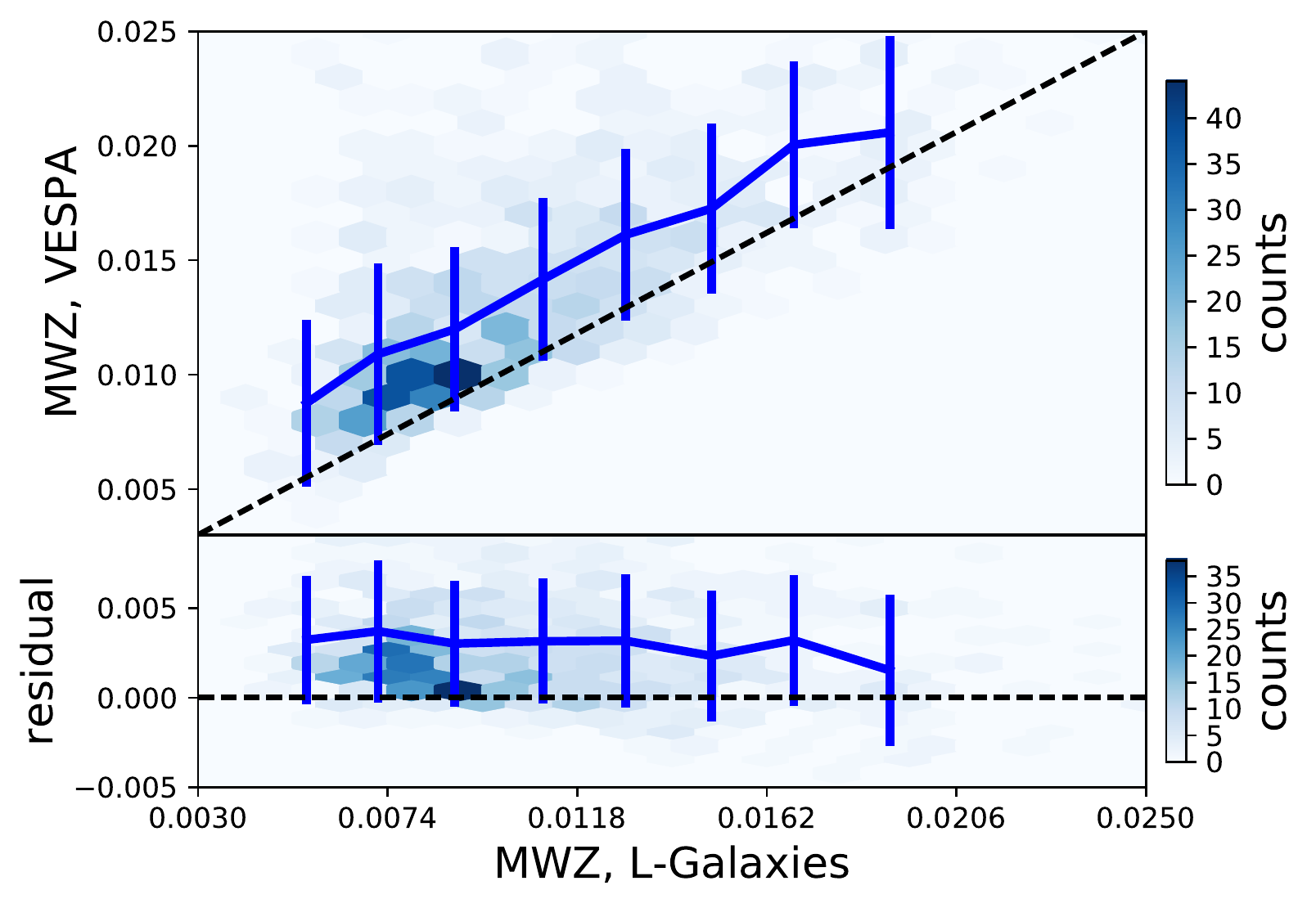}
\includegraphics[scale=0.45]{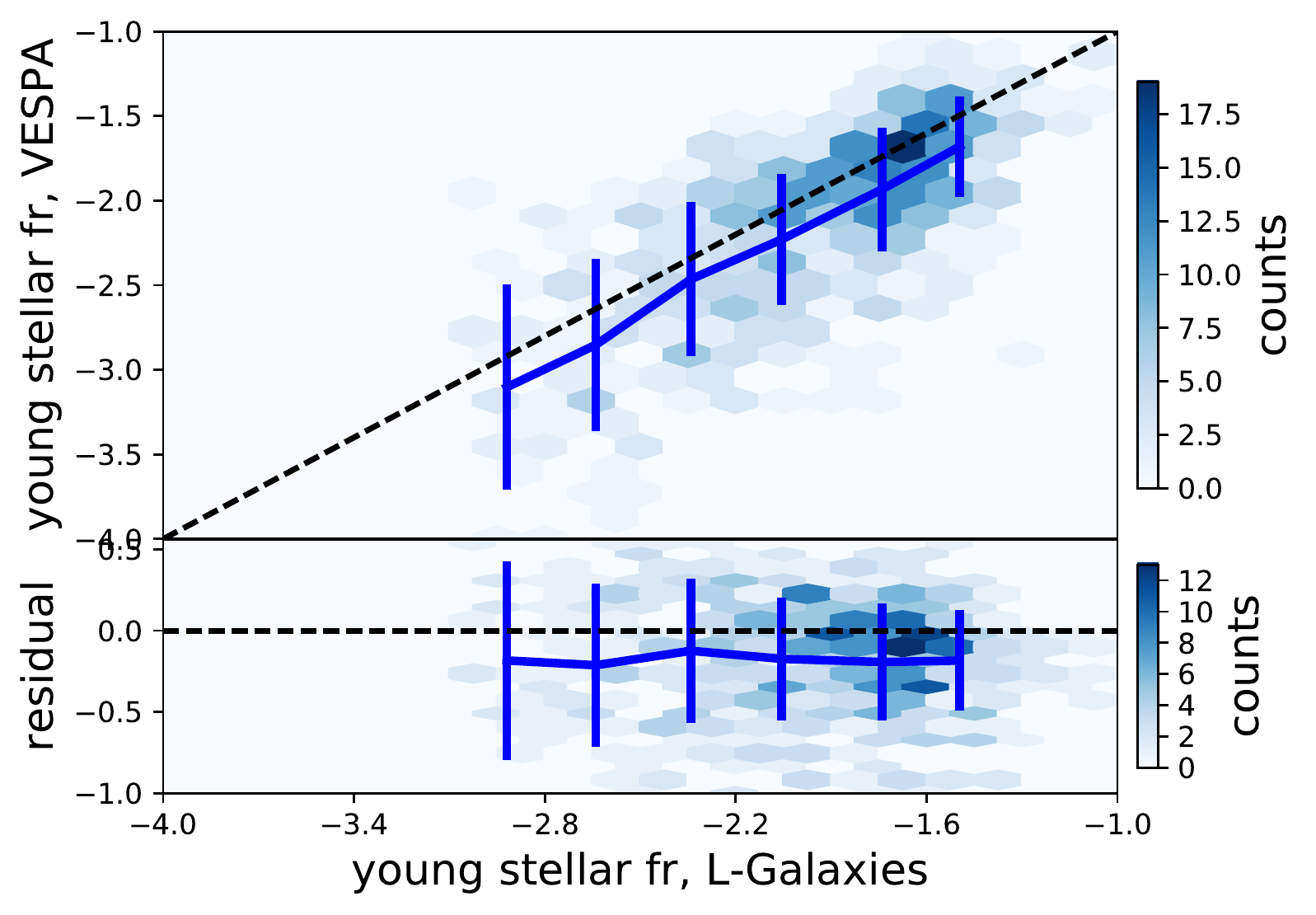}
\caption{Same as Fig.~\ref{fig:mocks_dust}, but using mock galaxies without any dust attenuation. Comparatively to Fig.~\ref{fig:mocks_dust}, the offset it stellar masses disappears, and the offset in young mass fraction and mass-weighted age is much reduced. The offset in metallicity remains, though now it is constant. }
\label{fig:mocks_nodust}
\end{figure*}

\begin{figure*}
\includegraphics[scale=0.45]{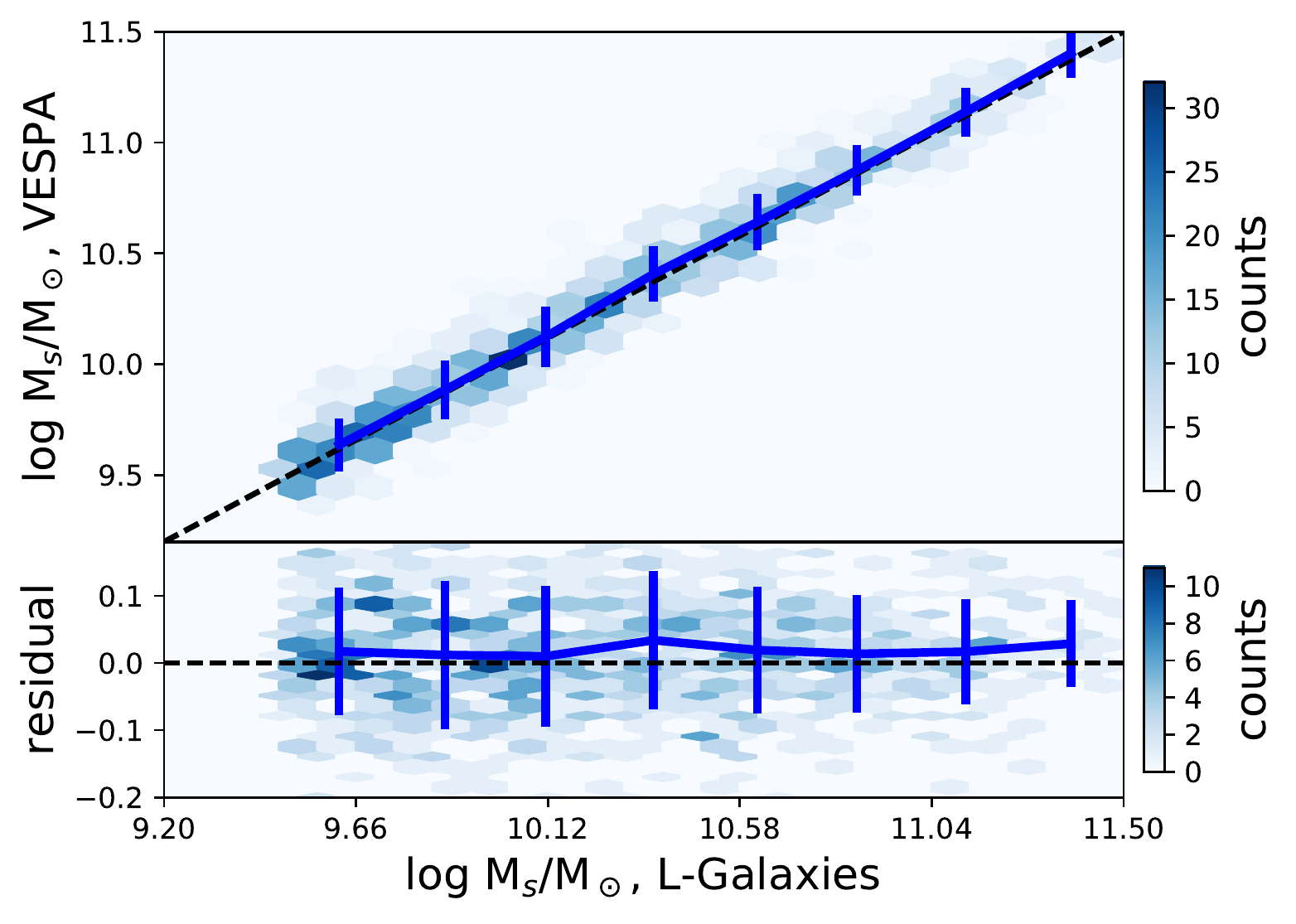}
\includegraphics[scale=0.45]{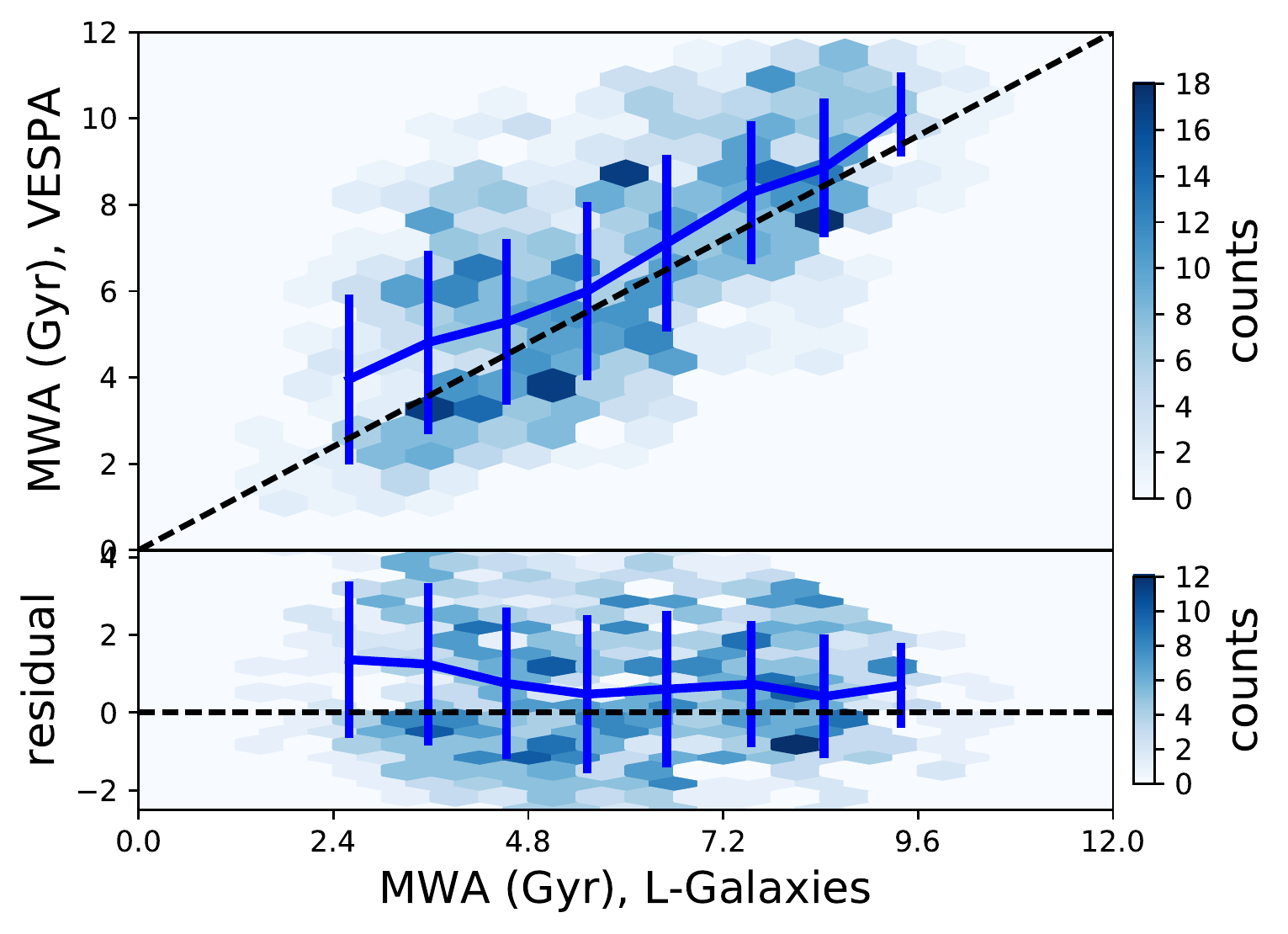}
\includegraphics[scale=0.45]{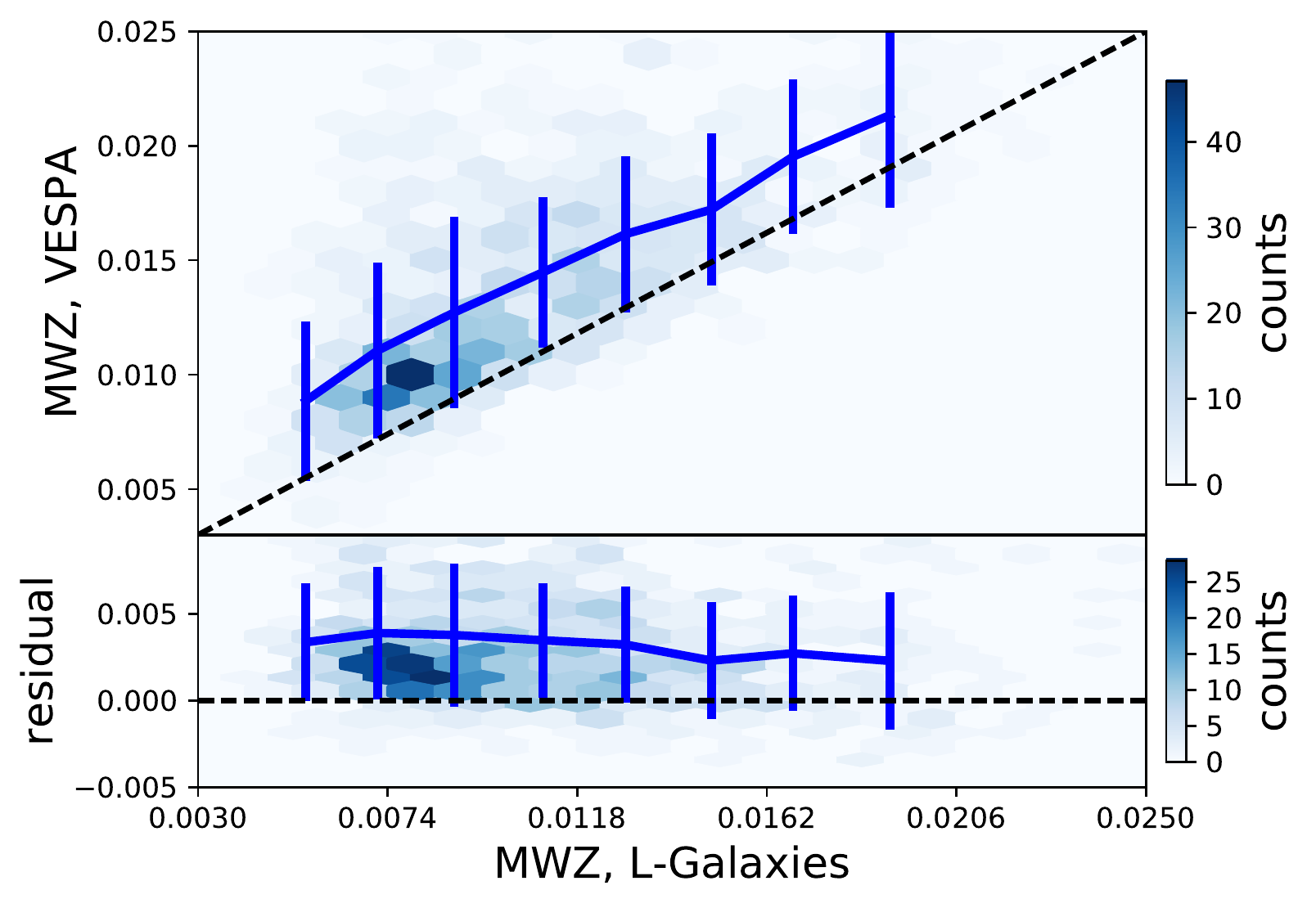}
\includegraphics[scale=0.45]{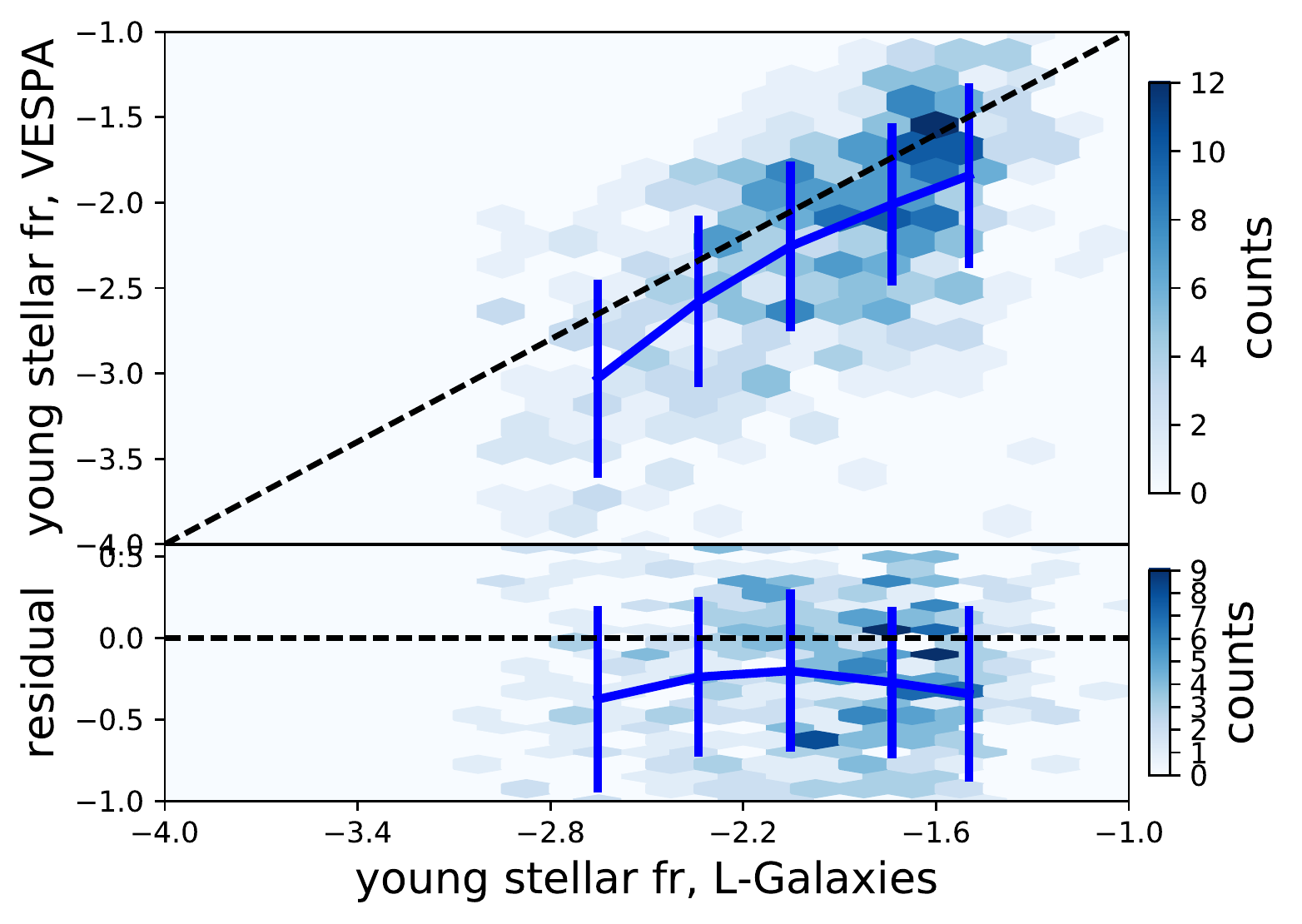}
\caption{ Same as Fig.~\ref{fig:mocks_dust}, but using a dust attenuation model and geometry that matches that assumed by VESPA exactly. The difference with respect to \ref{fig:mocks_nodust} is only mild, showing that if the dust could be modeled exactly it would not impart significant extra biases on the recovered parameters.}
\label{fig:mocks_dustsimple}
\end{figure*}

Here we show the results from analysing mocks with either no dust attenuation applied (Fig.~\ref{fig:mocks_nodust}), or with a model and geometry that matches what is assumed by VESPA exactly (Fig.~\ref{fig:mocks_dustsimple}). We leave a thorough exploration of these issues to a forthcoming paper, and here we show only the results that directly help the interpretation of Fig.~\ref{fig:mocks_dust}. Of the biases seen in that figure, we see that the bias in stellar mass is completely removed in the case of no dust or when the dust model is known. Similarly, the offset in mass-weighted age and fraction of young stars is much reduced. We note that the offset in metallicity remains, indicating that is more likely cause by the poor S/N or is indeed by an intrinsic limitation of VESPA. However, the tilt seen in the residuals of mass-weighted metallicity in Fig.~\ref{fig:mocks_dust} disappears. 

\appendix
\section{BC03 results}

\begin{figure*}
\includegraphics[scale=0.26]
                        {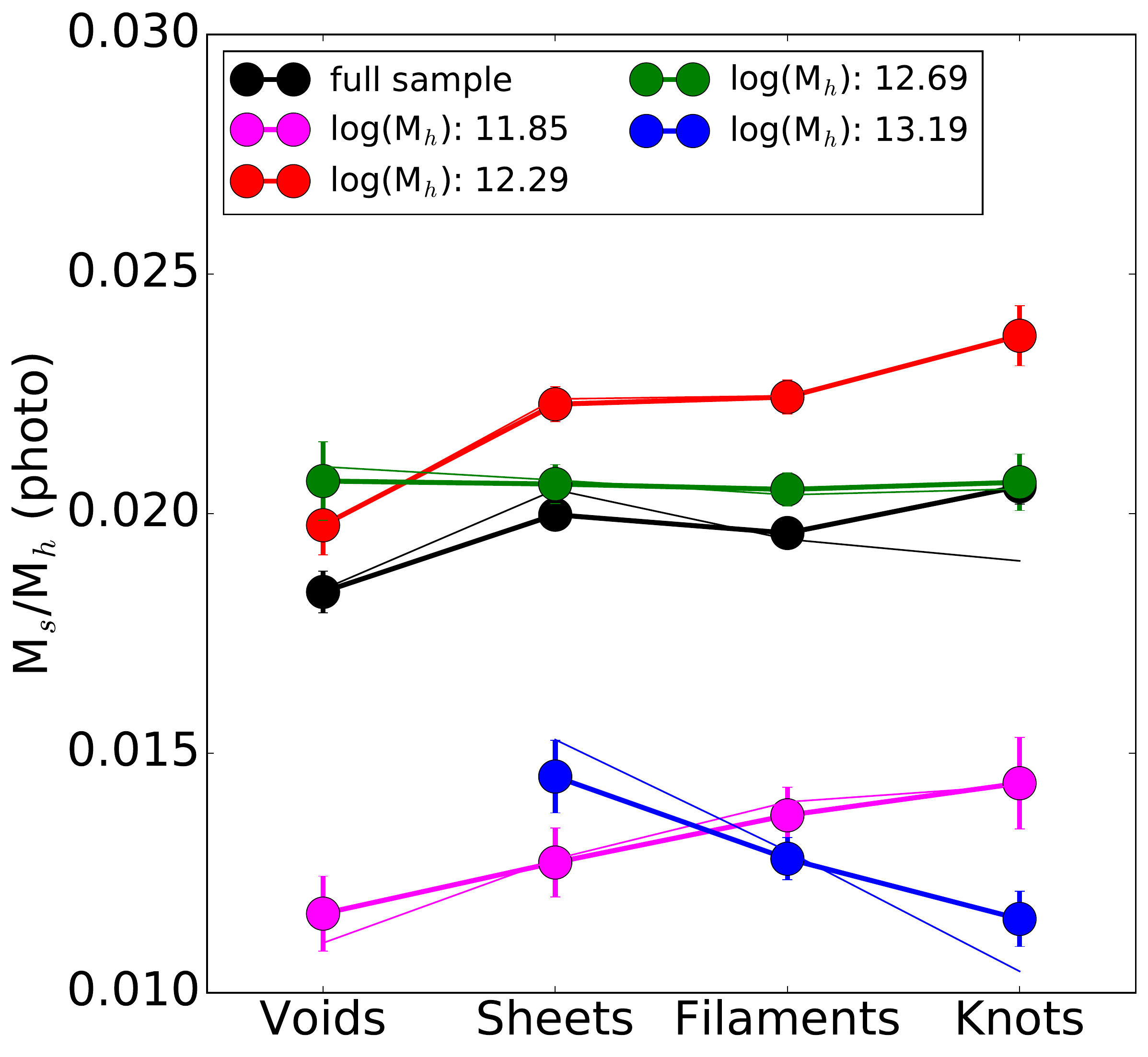}
\includegraphics[scale=0.26]
                        {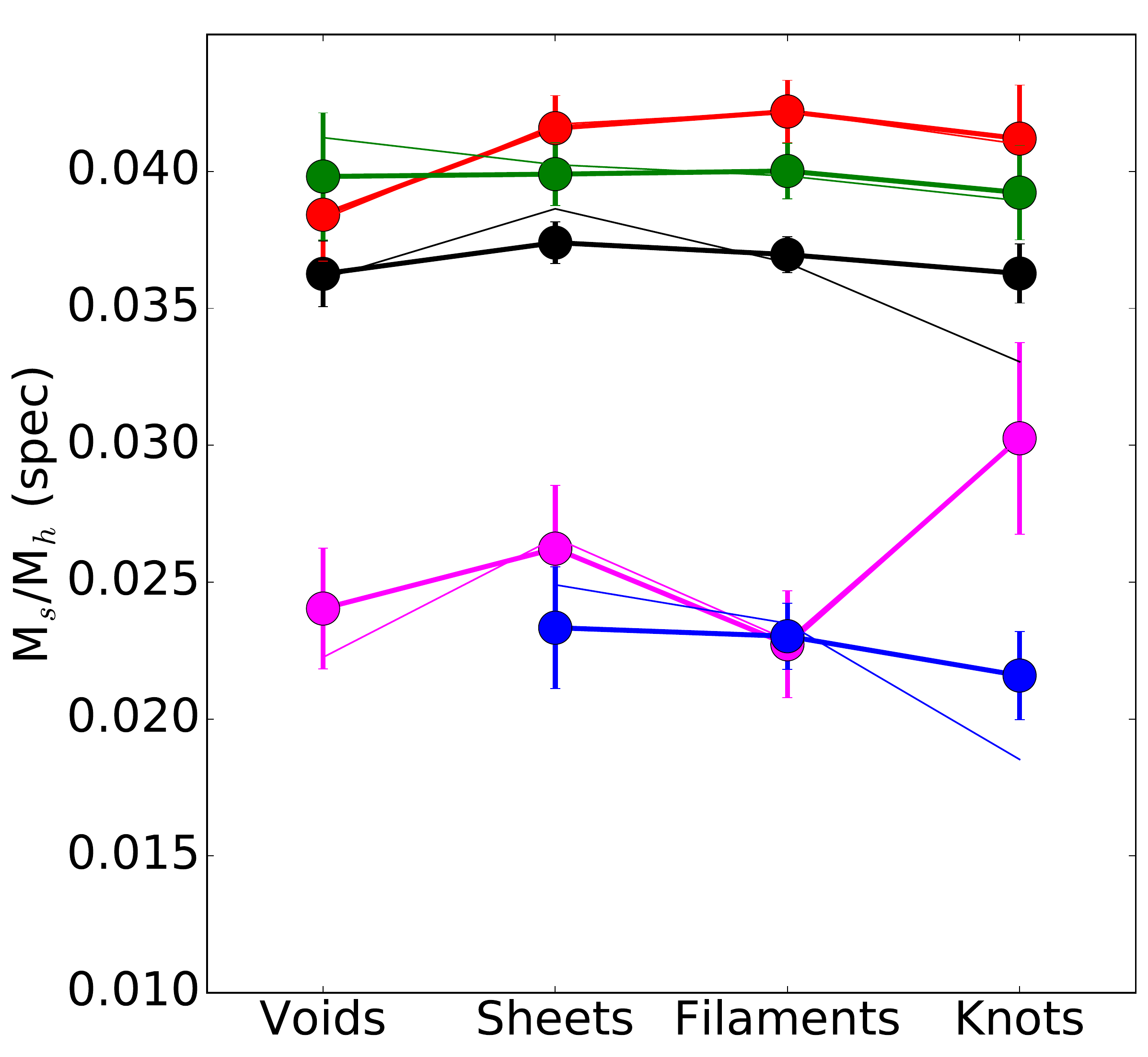}

\includegraphics[scale=0.26]
                        {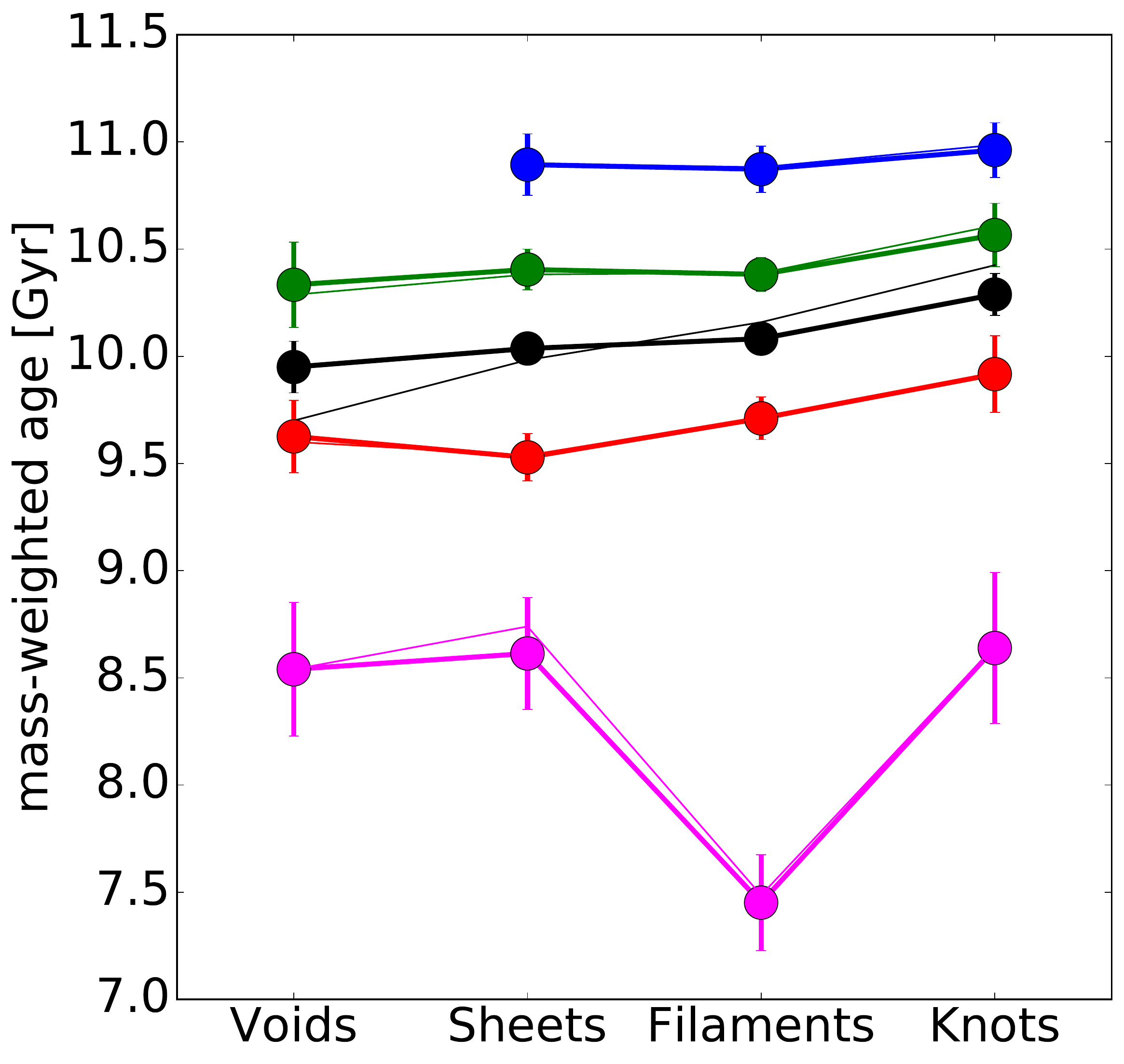}
\includegraphics[scale=0.26]
                        {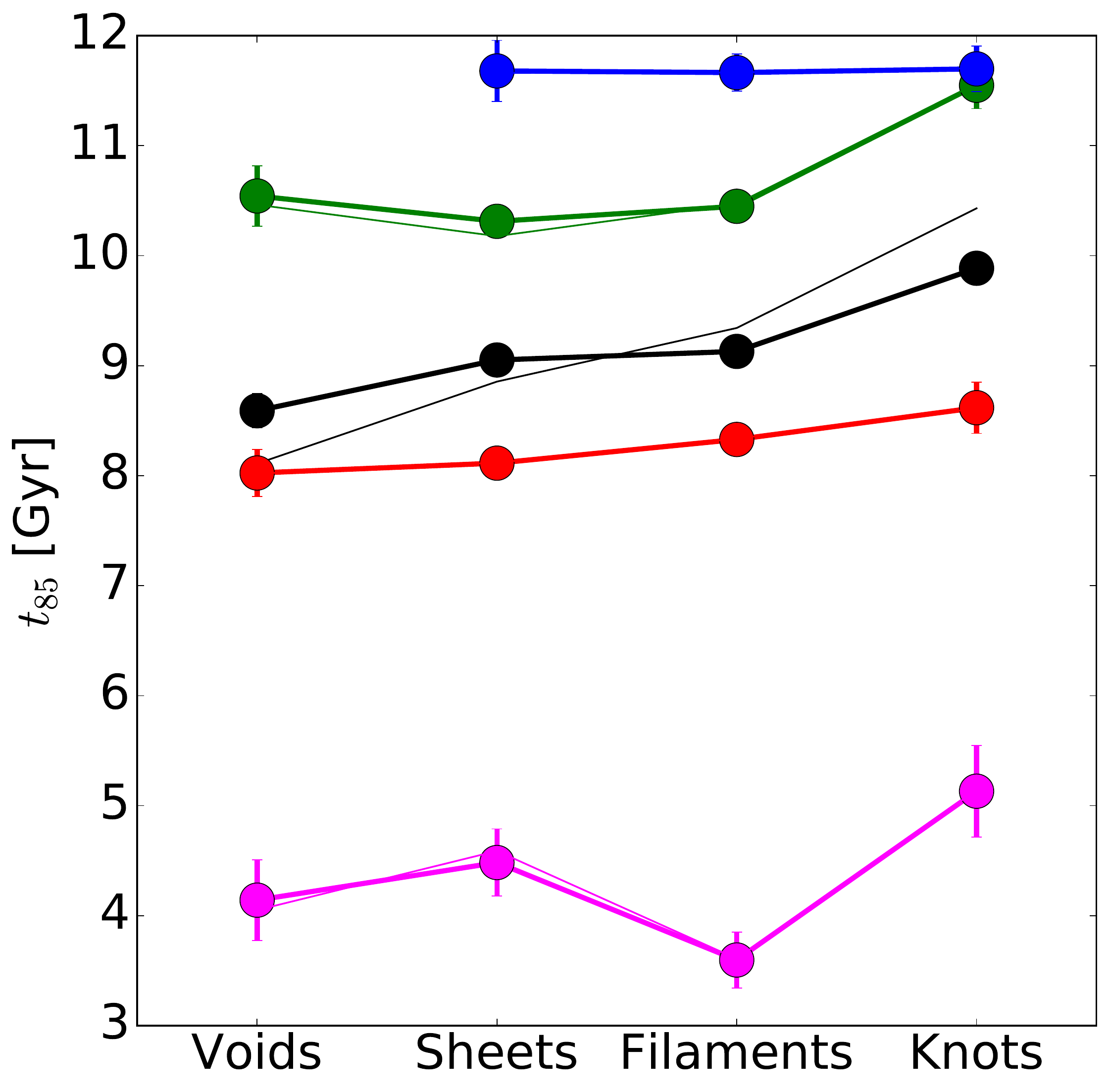}
\includegraphics[scale=0.26]
                        {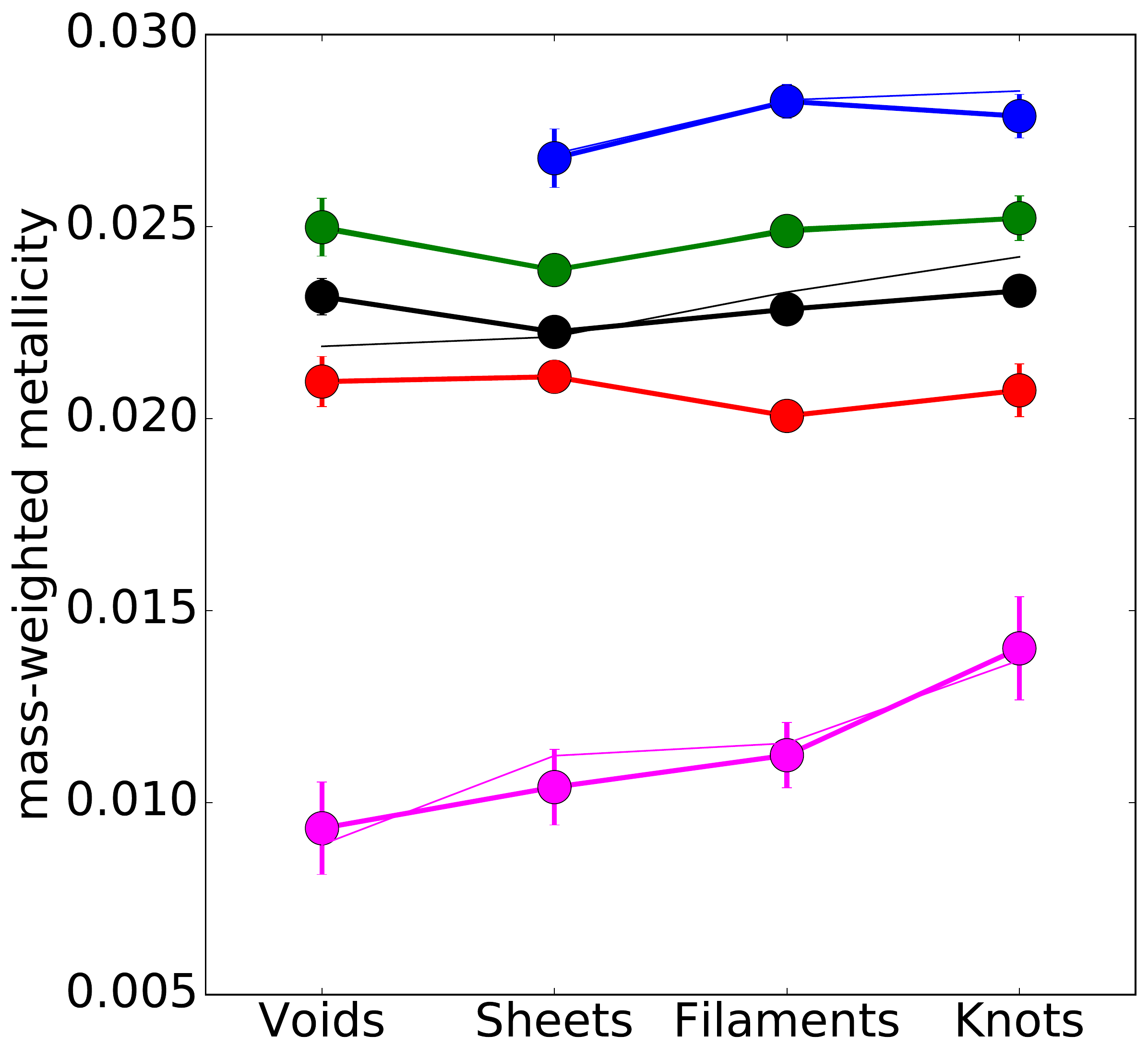}
\includegraphics[scale=0.26]
                        {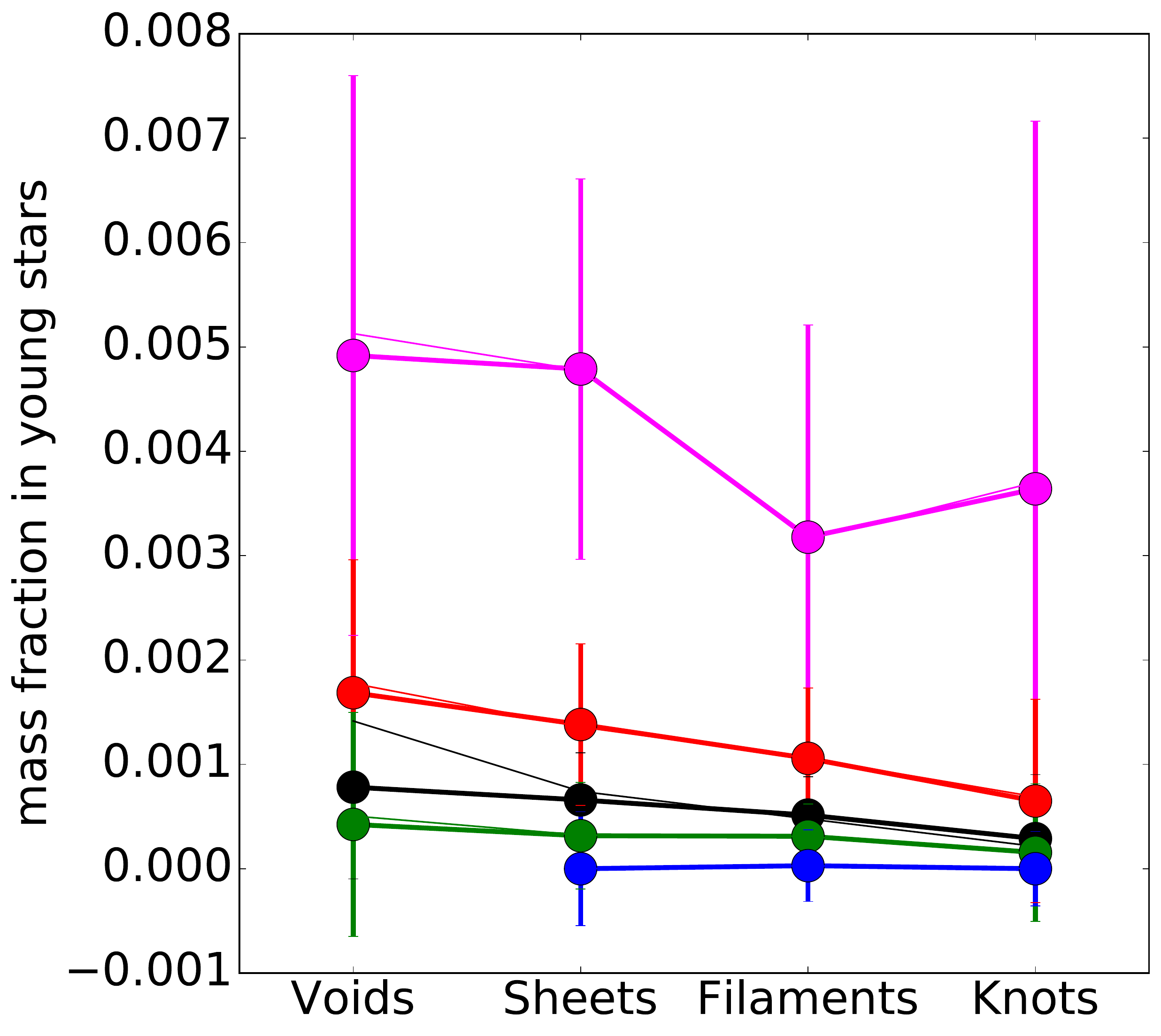}
\caption{Same as Fig.~\ref{fig:GAMA_res}, but obtained using the BC03 population models. The qualitative trends are the same, in line to expectations from Fig.~\ref{fig:GAMA_res_Mh}, but there are offsets with respect to the results obtained using FSPS models. Note that the top left panel is the same as in Fig.~\ref{fig:GAMA_res}, as that panel is independent of the VESPA analysis.}
\label{fig:GAMA_res_BC03}
\end{figure*}

Fig.~\ref{fig:GAMA_res_BC03} shows the same as Fig.~\ref{fig:GAMA_res}, but obtained using the BC03 population models; it demonstrates our conclusions relating to mass-weighted age, mass-weighted metallicity, young mass fraction and time at which 85\% of the stars had formed are robust to the two stellar population models we consider. Note that M$_s$/M$_h$ shown is independent of the VESPA analysis.

\end{document}